\documentclass[12pt,preprint]{aastex}

\def \kms{km s$^{-1}$}
\def \cm3{cm$^{-3}$}
\def \coi{$^{12}$CO}
\def \coii{$^{13}$CO}
\def \hii{H {\small II}}

\shorttitle{Arecibo 6.7 GHz Methanol Maser Survey -- II}

\begin{document}
\shortauthors{Pandian, \& Goldsmith}

\title{The Arecibo Methanol Maser Galactic Plane Survey--II: Statistical and Multi-wavelength Counterpart Analysis}

\author{Jagadheep D. Pandian \altaffilmark{1,2} and Paul F. Goldsmith \altaffilmark{3}}

\altaffiltext{1}{Max-Planck-Institut f\"{u}r Radioastronomie, Auf dem H\"{u}gel 69, 53121 Bonn, Germany; jpandian@mpifr-bonn.mpg.de}
\altaffiltext{2}{Department of Astronomy, Cornell University, Ithaca, NY 14853}
\altaffiltext{3}{Jet Propulsion Laboratory, California Institute of Technology, Pasadena, CA 91109; Paul.F.Goldsmith@jpl.nasa.gov}

\begin{abstract}
We present an analysis of the properties of the 6.7 GHz methanol maser sample detected in the Arecibo Methanol Maser Galactic Plane Survey. The distribution of the masers in the Galaxy, and statistics of their multi-wavelength counterparts is consistent with the hypothesis of 6.7 GHz maser emission being associated with massive young stellar objects. Using the detection statistics of our survey, we estimate the minimum number of methanol masers in the Galaxy to be 1275. The $l-v$ diagram of the sample shows the tangent point of the Carina--Sagittarius spiral arm to be around 49.6\degr, and suggests occurrence of massive star formation along the extension of the Crux--Scutum arm. A Gaussian component analysis of the maser spectra shows the mean line-width to be 0.38 \kms~which is more than a factor of two larger than what has been reported in the literature. We also find no evidence that faint methanol masers have different properties than those of their bright counterparts.
\end{abstract}


\section{Introduction}

The $5_1-6_0$ A$^+$ line of methanol at 6.7 GHz is the strongest of the Class II methanol masers. Theoretical models (e.g. \citealt{crag05}), and observational studies (e.g. \citealt{elli06, mini05}) suggest that the masers are associated with early phases of massive star formation. The fact that 6.7 GHz methanol masers have not been detected to date towards late type stars or low-mass star forming regions, makes them unique compared to their OH and H$_2$O counterparts. This, in addition to their high brightness, makes 6.7 GHz methanol masers powerful tools to identify sites of massive star formation across the Galaxy, which in turn can probe Galactic structure at great distances.

However, some puzzles remain. \citet{szym00b} found that many relatively weak methanol masers were associated with IRAS sources with colors different from those expected for ultracompact \hii~regions. They further found that many faint masers had very narrow linewidths. This raises the question of whether faint methanol masers have properties different from that of bright masers. Recently, we reported on the sensitive blind survey for 6.7 GHz methanol masers carried out using the Arecibo\footnote{The Arecibo Observatory is part of the National Astronomy and Ionosphere Center, which is operated by Cornell University under a cooperative agreement with the National Science Foundation.} radio telescope (\citealt{pand07}; hereafter Paper I). The methanol maser sample from this survey is ideal for probing some of the questions above. Here, we present an analysis of the properties of the methanol masers discovered in the survey.

\section{The Sample}

The methanol maser sample comes from the Arecibo Methanol Maser Galactic Plane Survey (AMGPS) which is discussed in detail in \citet{pand07}. The sample has 86 methanol masers, 77 of which have peak flux densities above the completeness limit of 0.27 Jy. The follow up observations in AMGPS collected high and low velocity resolution data (0.04 \kms~and 0.14 \kms~respectively after Hanning smoothing) for each methanol maser simultaneously. Sources weaker than $\sim 0.5$ Jy were typically observed for two minutes on source, while the stronger sources were observed for one minute on source. The rms noise in the spectra are typically 17 mJy and 8 mJy for the high and low velocity resolutions respectively for the two minute observations, and 24 mJy and 12 mJy respectively for the one minute observations.

\section{Distribution in the Galaxy}

The AMGPS is fully sampled only for $35.2\degr \leq l \leq 53.7\degr$ and $|b|\leq 0.41\degr$. Hence, to consider the distribution of methanol masers in the Galaxy, all sources that lie outside these coordinates must be excluded along with the sources that lie below the completeness limit. This gives a sample of 72 methanol masers.

\begin{figure}[!htb]
\begin{center}
\includegraphics{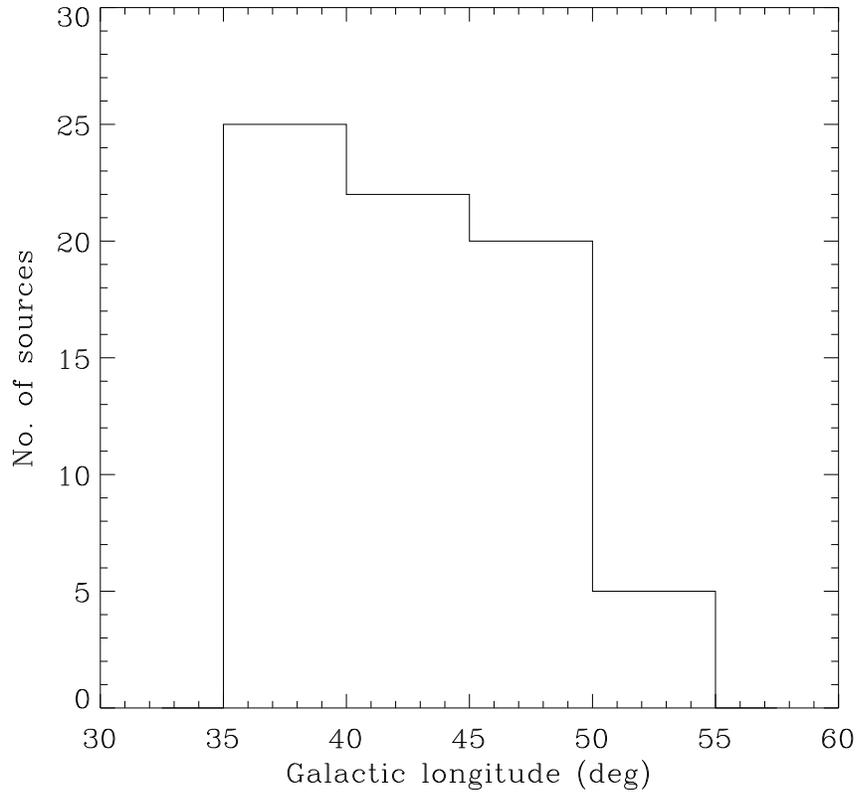}
\caption[Distribution of methanol masers as a function of Galactic longitude.]{Distribution of methanol masers as a function of Galactic longitude. Note that the 35\degr--40\degr~bin and the 50\degr--55\degr~bin will be slightly affected by incomplete coverage.}\label{longdist}
\end{center}
\end{figure}

Figure \ref{longdist} shows the distribution of methanol masers as a function of Galactic longitude. Note the sharp decline in the number of methanol masers beyond a longitude of 50\degr. This is due to the tangent point of the Carina--Sagittarius spiral arm being crossed at $l \sim 49.6\degr$. The increase in the number of methanol masers at lower Galactic longitudes is most likely due to the molecular ring in the Galaxy.

\begin{figure}[!htb]
\begin{center}
\includegraphics{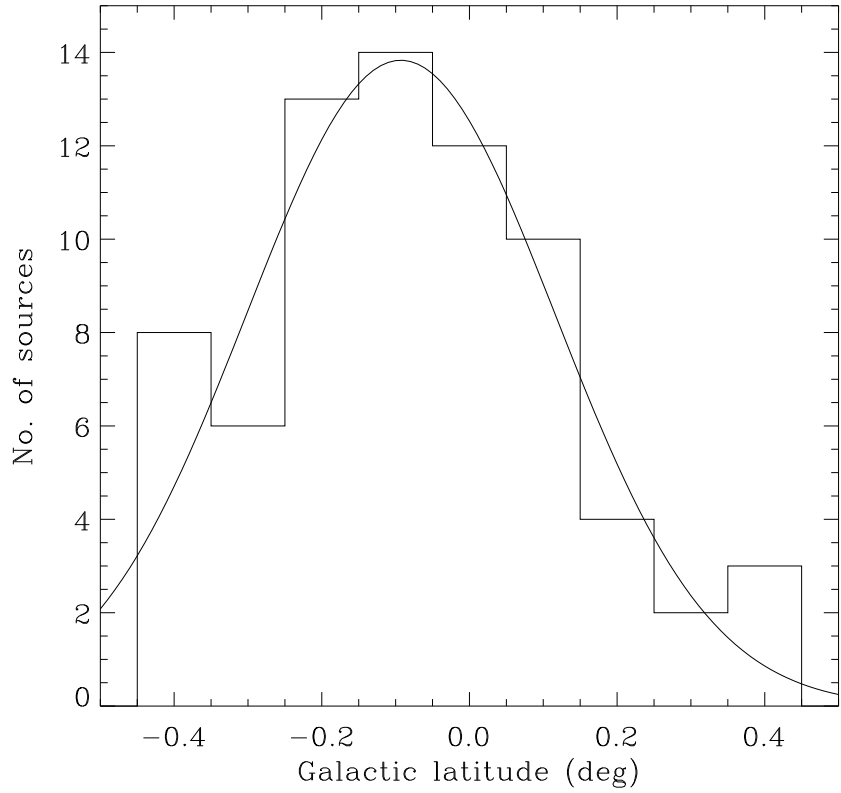}
\caption[Distribution of methanol masers as a function of Galactic latitude.]{Distribution of methanol masers as a function of Galactic latitude. A Gaussian fit to the distribution is overlaid. The Gaussian fit has a mean at $-0.09\degr$ and a full width to half maximum of 0.49\degr.}\label{latdist}
\end{center}
\end{figure}

The distribution of methanol masers as a function of Galactic latitude is shown in Figure \ref{latdist}. A Gaussian fit to the distribution has a mean at a latitude of $-0.09\degr$ and a full width at half maximum of 0.49\degr. The width of the distribution is comparable to that of embedded massive stars determined from the IRAS point source catalog \citep{wood89}. The parameters fitted to the distribution are very similar to those obtained by \citet{pest05} for their general catalog of methanol masers (which contains over 500 sources). Over half of the sources in the general catalog were detected in targeted surveys towards tracers of massive star formation such as ultracompact \hii~regions and OH masers. Thus, the close agreement of the distribution of the AMGPS sources (which come from a blind survey) with that of \citet{pest05} gives further evidence for the association of 6.7 GHz methanol masers with massive star formation.

The equatorial plane of the Galaxy was defined by the International Astronomical Union (IAU) based on primarily neutral hydrogen measurements of the Galaxy \citep{iau59}. The measurement of the mid-plane of the distribution of massive stars at a Galactic latitude of --0.09\degr~could be due to an uncertainty in the definition of the Galactic plane from the HI observations, especially since the Sun is located $\sim 15.5$ pc above the plane \citep{hamm95}, or a difference in the plane defined by HI and young massive stars. It is to be noted that most models of the warp in the Galactic disk keep the mid-plane close to $b = 0$ or move it to positive Galactic latitudes in the Galactic longitude range probed by AMGPS (e.g. \citealt{vig05}). Hence, a warped disk cannot explain the observed distribution of methanol masers in AMGPS.

\section{Implications for Galactic Structure}
Since massive stars form preferentially along spiral density waves which compress molecular gas to high densities, 6.7 GHz methanol masers are potential probes of the spiral structure of the Galaxy. Unfortunately, distances to methanol masers are not known precisely, except for a few regions (e.g. the distance to W3OH was measured through parallax by \citealt{xu06}). Distances are mostly calculated kinematically by assuming that the observed radial velocity of the source arises from its circular motion around the Galaxy. Then, using a model for the rotation curve of the Galaxy (e.g. \citealt{clem85,bran93}), one can derive an estimate for the distance to the source. The uncertainty in the distance is large since the source is usually not in perfect circular rotation around the Galactic Center, deviations from which arise due to random motions, spiral shocks, expansion such as at edges of \hii~regions and supernova explosions, etc. To compound these difficulties, the distance to the source from the Sun is double valued for certain radial velocities in the inner regions of the Galaxy (positive radial velocities for $0\degr\leq l\leq 90\degr$, negative radial velocities for $270\degr\leq l \leq 360\degr$).

Although there are techniques to distinguish between the two values of the kinematic distance (called near and far distance), we do not at present have the data required to resolve the distance ambiguity for our methanol maser sample. Hence, we use an $l-v$ diagram to see what information methanol masers provide regarding the structure of our Galaxy. It is to be noted that the radial velocity of a source can have significant uncertainties when determined from maser emission which can span a velocity range as high as 20 \kms. To overcome this difficulty, we measured \coii~($J=2-1$) spectra (velocity resolution $\sim 1.4$ \kms; 1$\sigma$ sensitivity $\sim 0.04$ K) towards all our sources using the Heinrich Hertz Submillimeter Telescope operated by the Arizona Radio Observatory. We chose \coii~rather than \coi~as it has narrower linewidths, and is likely to display self-absorbed profiles as is occasionally seen with \coi. For each source, the LSR velocity of CO emission within 5 \kms~of the methanol maser emission was adopted as its systemic velocity. For a few sources, there were two \coii~peaks within the velocity range of the maser emission; the velocities of the two peaks were typically equidistant from the peak of maser emission, and occurred near the extremities of the maser emission range. In these cases, the velocity of peak maser emission was adopted for the systemic velocity of the source. In a majority of cases, the \coii~peak was within 5 \kms~of the methanol maser peak emission.

\begin{figure}[!htb]
\begin{center}
\includegraphics[width=\textwidth]{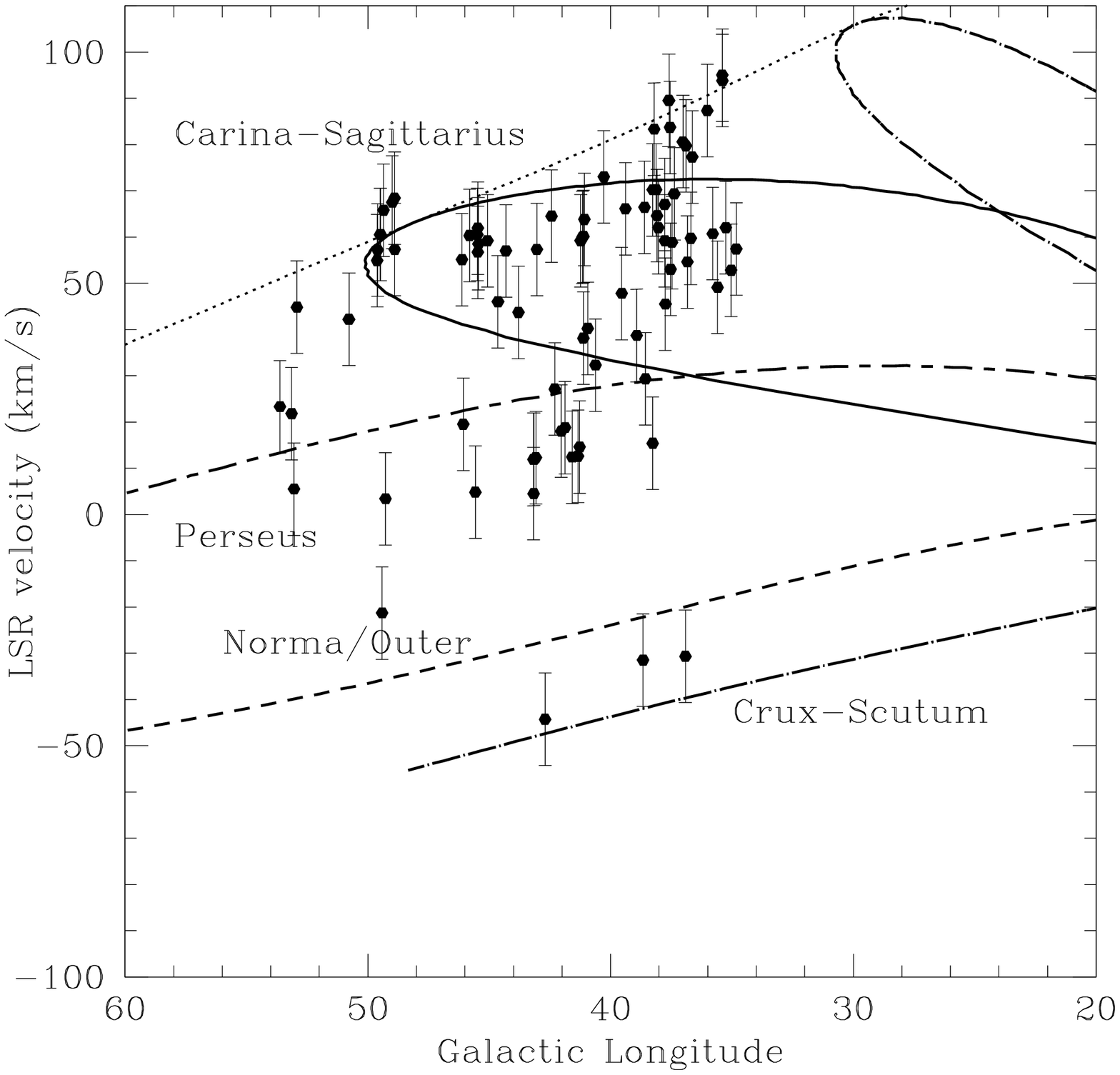}
\caption{The methanol maser sample of AMGPS on an $l-v$ diagram. The loci of spiral arms from the model of \citet{vall95} are overplotted. The dotted line shows the tangent point velocity as a function of Galactic longitude.}\label{lvvallee}
\end{center}
\end{figure}

\begin{figure}[!htb]
\begin{center}
\includegraphics[width=\textwidth]{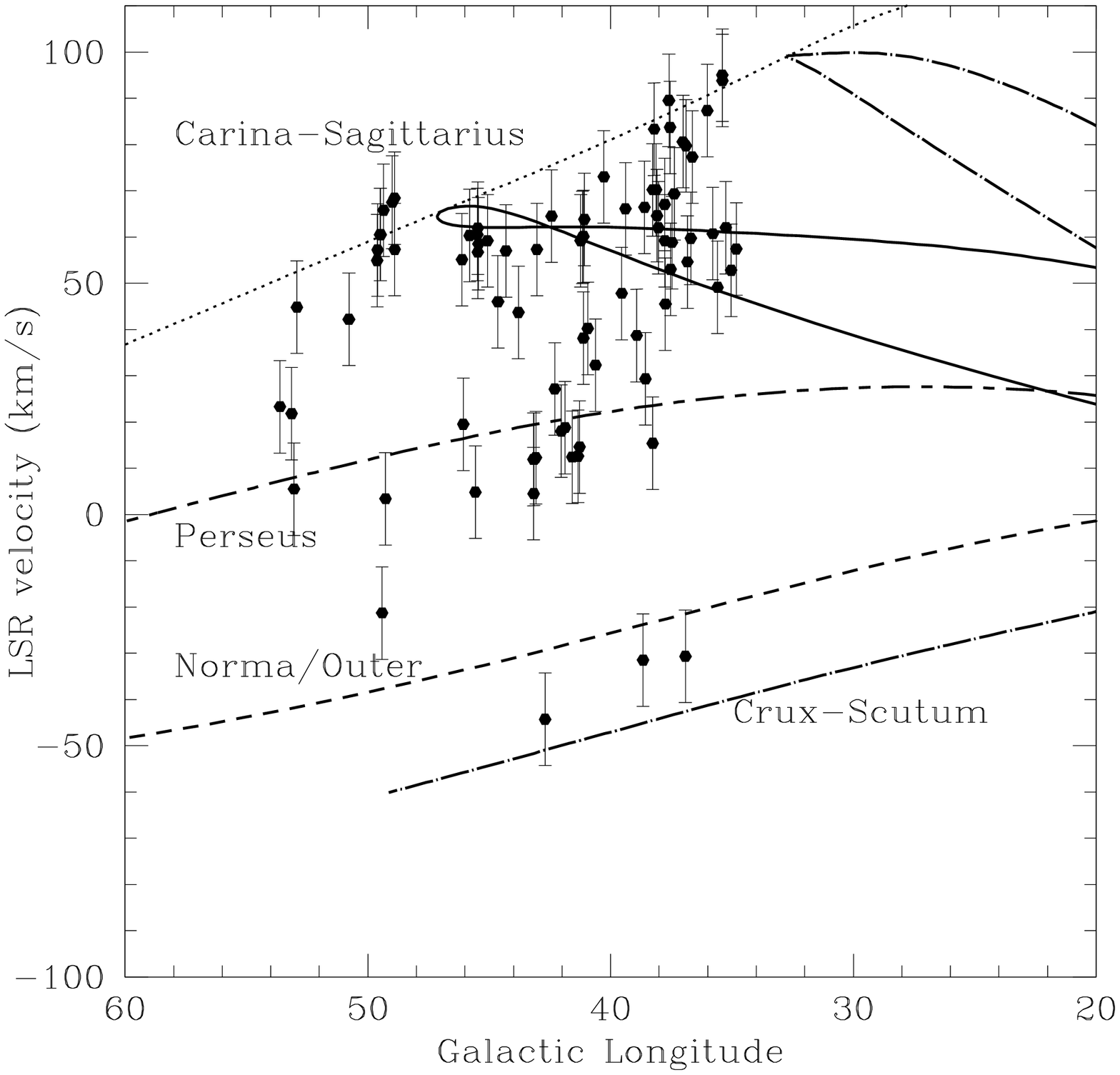}
\caption{The methanol maser sample of AMGPS on an $l-v$ diagram. The loci of spiral arms from the NE2001 model of \citet{cord02} are overplotted. The dotted line shows the tangent point velocity as a function of Galactic longitude.}\label{lvne2001}
\end{center}
\end{figure}

Two constructs of the $l-v$ diagram are shown in Figures \ref{lvvallee} and \ref{lvne2001} respectively. The observed radial velocities are assumed to have deviations from the values observed from pure circular rotation by up to $\pm 10$ km s$^{-1}$ due to the various phenomena described above. Figure \ref{lvvallee} shows the loci of various spiral arms in the model of \citet{vall95}. Figure \ref{lvne2001} shows the spiral arm loci for the NE2001 model of \citet{cord02}. The spiral arm loci are calculated using the rotation curve of \citet{clem85}.

The concentration of sources around $l \sim 49.6\degr$ seems to arise from the tangent point of the Carina--Sagittarius spiral arm. The model of \citet{vall95} fits this feature reasonably well, while the tangent point in the model of \citet{cord02} occurs at approximately 1\degr~lower Galactic longitude which does not fit our data. It is to be noted that the spiral arm model of \citet{cord02} is derived from ionized gas, which is likely to be offset from the location of young massive stellar objects in a spiral arm. As the molecular gas orbiting the Galaxy encounters a spiral density wave, it gets compressed thereby initiating star formation. Since massive stars form on a relatively short timescale, the region produces \hii~regions by the time it exits the density wave. Thus, for trailing spiral arms, deeply embedded massive young stellar objects are more likely to be found at the inner (concave) edge of the spiral arm, while ionized gas is more likely to be found along the outer (convex) edge of the spiral arm. Thus, one would expect the tangent point of a spiral arm to be located at larger Galactic longitudes as seen in ionized gas, compared to early tracers of massive star formation such as 6.7 GHz methanol masers. This makes the discrepancy between the tangent point of the Carina--Sagittarius spiral arm as seen from our data, and the NE2001 model especially puzzling.

It can also be seen that a significant fraction of the methanol masers do not lie near spiral arm loci. It is possible that this is due to significant deviations of the velocities of spiral arm loci from circular motion due to spiral shocks. The other possibility is that the spiral arm model of the Galaxy needs revision. It is also evident from comparing Fig. \ref{lvvallee} and Fig. \ref{lvne2001} that deviations of spiral arms from simple log spiral models can result in significant changes in their loci in $l-v$ diagrams. Moreover, it is apparent from images of external galaxies that massive star formation takes place along spurs of spiral arms in addition to the arms themselves. A combination of these effects could be responsible for the apparent lack of correspondence between spiral arm models and the locations of methanol masers.

The other interesting feature is the observation of three sources (36.92+0.48, 38.66+0.08 and 42.70--0.15) at radial velocities lower than $-30$ \kms~(with respect to the local standard of rest). These points lie beyond the Outer arm and within the extension of the Crux--Scutum arm (which can also be seen to the upper right in Figures \ref{lvvallee} and \ref{lvne2001}). The kinematic distances to these sources are over 17 kpc from the Sun, which is among the farthest known distance to a methanol maser. Thus, sensitive surveys for 6.7 GHz methanol masers can detect structures beyond those detected by other tracers of massive star formation.

\section{The Number of Methanol Masers in the Galaxy}

The number of 6.7 GHz methanol masers in the Galaxy has been estimated by \cite{van05}. The procedure followed is to model the distribution of young massive star forming regions as a function of Galactic longitude using a star formation law, initial mass function, and a spiral arm model of the Galaxy. This distribution is then scaled such that the number of maser sources in a chosen longitude bin coincides with the statistics from an existing survey. Since all surveys are flux limited, a minimum number of methanol masers can be derived by summing over all longitude bins. Thus, the minimum number of methanol masers was determined by \citet{van05} to be 845, by scaling the model distribution to the observations between Galactic longitudes of $330^\circ$ and $340^\circ$ (which includes a blind survey by \citealt{casw96} and is partially covered by the blind survey of \citealt{elli96}).

\begin{figure}[!htb]
\begin{center}
\includegraphics{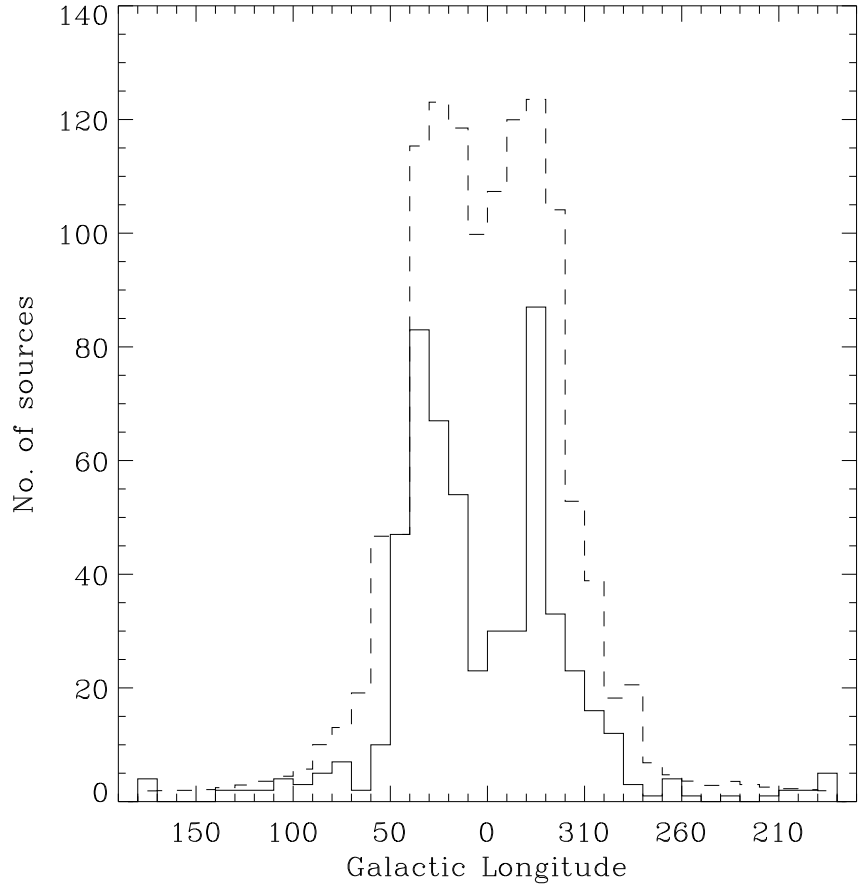}
\caption{The longitude distribution of all 6.7 GHz methanol masers published to date is shown in solid lines. The dashed line shows the expected distribution of methanol masers using one of the simulations of \citet{van05}, and normalized to the observed distribution between longitudes of 40\degr~and 50\degr.}\label{longdistsimul}
\end{center}
\end{figure}

Since our survey is almost an order of magnitude more sensitive than that of \citet{casw96} and \citet{elli96} (and has more uniform sensitivity than that of \citealt{casw96}), our observations can be used to revise the minimum number of methanol masers in the Galaxy. AMGPS detected 47 methanol masers between longitudes of 40\degr~and 50\degr, 44 of which are brighter than 0.27 Jy . An additional three methanol masers have been detected outside the AMGPS survey area in this Galactic longitude range in previous surveys (all of which are much brighter than 0.27 Jy). Thus there are at least 47 methanol masers brighter than 0.27 Jy, and 50 methanol masers detected to date, between Galactic longitudes of 40\degr~and 50\degr. Using these statistics to normalize the longitude distribution of \citet{van05}, the number of methanol masers in the Galaxy brighter than 0.27 Jy is at least 1200. Note that this number should be taken as a lower limit as the region $|b|>0.41\degr$ is incompletely sampled by AMGPS, and is thus not a complete census of methanol masers brighter than 0.27 Jy. Further, considering the entire sample that is not flux limited to 0.27 Jy, the minimum number of methanol masers in the Galaxy is 1275. The observed distribution of methanol masers and the expected distribution normalized to our statistics is shown in Fig. \ref{longdistsimul}.

The total number of methanol masers in the Galaxy is also estimated through Monte Carlo simulations of \citet{van05}. The simulations involved assigning luminosities to a simulated population of methanol masers, and determining the completeness limit for a survey as a function of the limiting flux density. Since assigning luminosities involved calculating kinematic distances which often suffers from an ambiguity between near and far distance, the completeness limits were computed as a function of $P_{\mbox{\scriptsize near}}$, the probability that a maser is at the near distance. Using this technique, the total number of methanol masers in the Galaxy was estimated to be $1200\pm 84$. 

\begin{figure}[!htb]
\begin{center}
\includegraphics{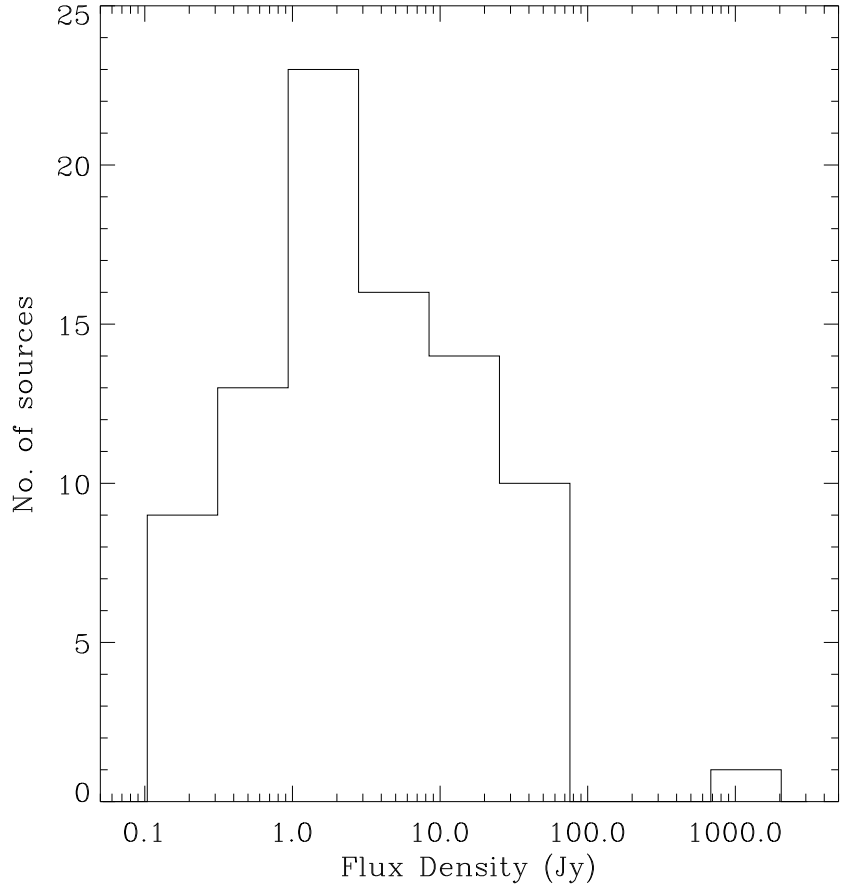}
\caption{The distribution of flux densities of methanol masers discovered in the AMGPS. The turn-over in the number of sources at lower flux densities is likely to be real since only the lowest bin is affected by incompleteness.}\label{fluxdist}
\end{center}
\end{figure}

From the simulations described above, the completeness limit for a survey with a flux density limit of 0.27 Jy is $\sim 0.89-0.93$ for $P_{\mbox{\scriptsize near}} = 0.5-0.9$. This translates to a total number of methanol masers in the Galaxy to be 1290--1348, which is slightly above the number calculated by \citet{van05}. In light of this discussion, it is interesting to note the turn-over in the distribution of methanol masers as a function of flux density (Fig. \ref{fluxdist}). Note that only the first bin in this plot is affected by incompleteness. The plot shows that the peak of the distribution occurs for flux densities between $\sim 0.9$ and 3 Jy. This indicates that the rate of detection of a survey that is much deeper than AMGPS will not be much higher than the detection rate of AMGPS itself. This information will be useful for planning future methanol maser surveys.

\section{Line Properties: Gaussian Analysis}

The spectra of 6.7 GHz methanol masers are in general very complex, with several velocity components blended together. This makes Gaussian component analysis a daunting task. We used a modified version of the interactive IDL routine ``xgaussfit'' available from the FUSE IDL Tools website\footnote{http://fuse.pha.jhu.edu/analysis/fuse\_idl\_tools.html} for our Gaussian analysis. We modified the routine so that it could fit up to 25 Gaussians simultaneously (as opposed to 8 Gaussians in the original version), and also incorporated the $\chi^2$ metric into the interactive widget to determine the quality of a given fit.

\begin{figure}[!htb]
\begin{center}
\includegraphics[width=0.85\textwidth]{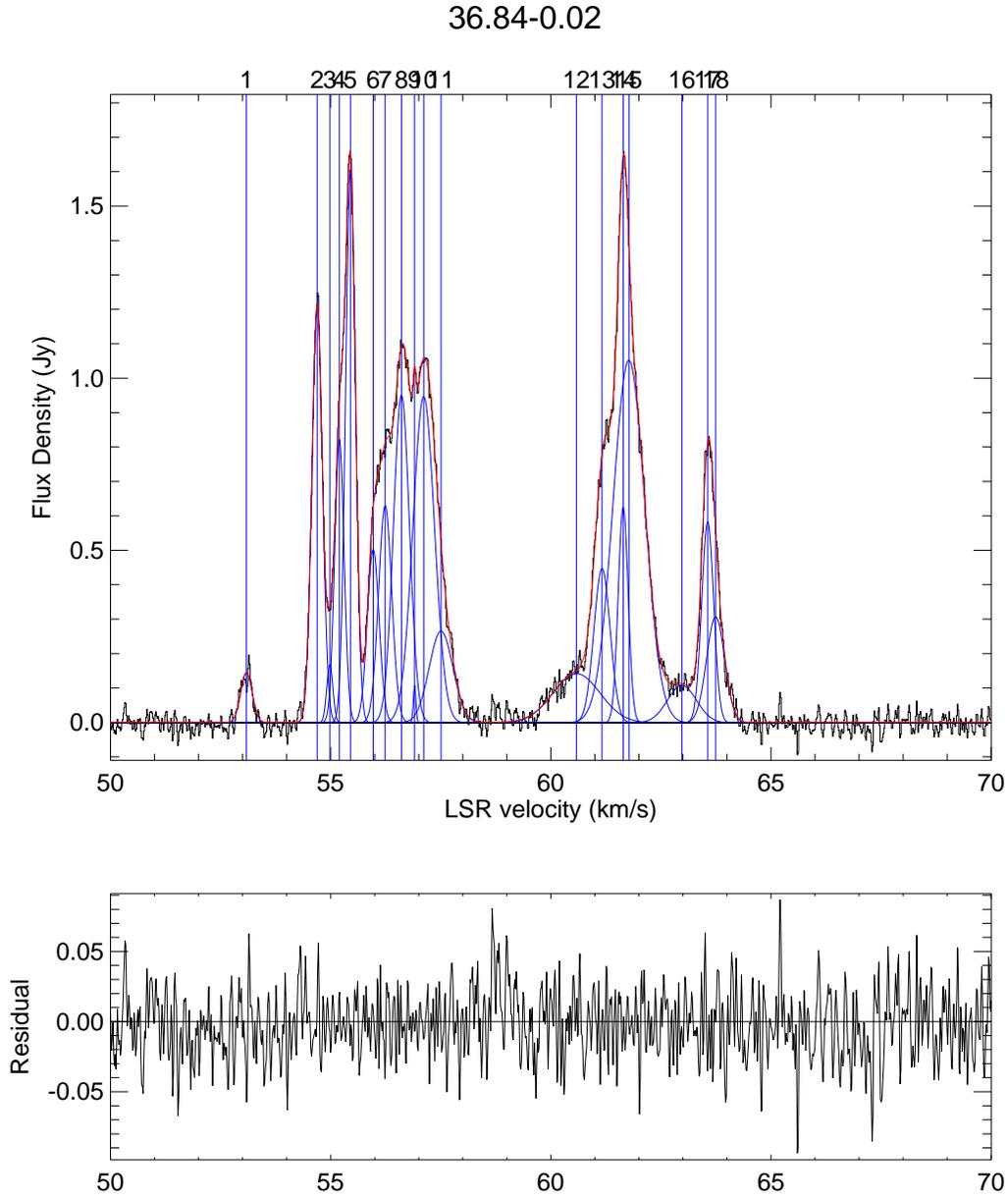}
\caption{An example of a complex spectrum, and an attempted Gaussian component analysis. The dotted lines show the individual Gaussians, and the solid line shows the overall fit. A similar fit can be achieved by using slightly fewer or more components, and by different placement of components. The lower panel shows the residuals to the fit. A color version of this figure is available electronically.}\label{complexspectrum}
\end{center}
\end{figure}

We found it extremely difficult to get a ``good'' fit to complex spectra. Moreover, we found that fits obtained using generic fitting routines are not unique. This is primarily because there are no {\it a priori} constraints on the number of Gaussian components in the spectrum. This translates to an uncertainty in the statistics of the component linewidths. Even if we restrict the number of Gaussians from $\chi^2$ arguments, there are many arrangements of the different components to achieve a similar fit. An example of a complex spectrum, and an attempted fit is shown in Figure \ref{complexspectrum}.

Due to the difficulty in obtaining good fits to complex sources, we restrict our analysis to relatively simple sources, and have at present carried out the Gaussian component analysis for 49 out of the 86 sources detected in AMGPS. We are exploring partial deconvolution techniques to fit the remaining 37 sources. A technique that shows promise is the deconvolution of a Gaussian (through division in the Fourier domain) with a width smaller than that of the narrowest feature -- this has the effect of artificially reducing the linewidth of each feature in the spectrum thereby separating the features that were originally blended together. We are also obtaining data at high spatial resolution using MERLIN for our sources. The MERLIN observations are expected to resolve each source into a number of masing spots, which will enable us to unambiguously determine the number of velocity components per source, and their properties such as linewidths.

\begin{figure}[!htb]
\begin{center}
\includegraphics[width=0.9\textwidth]{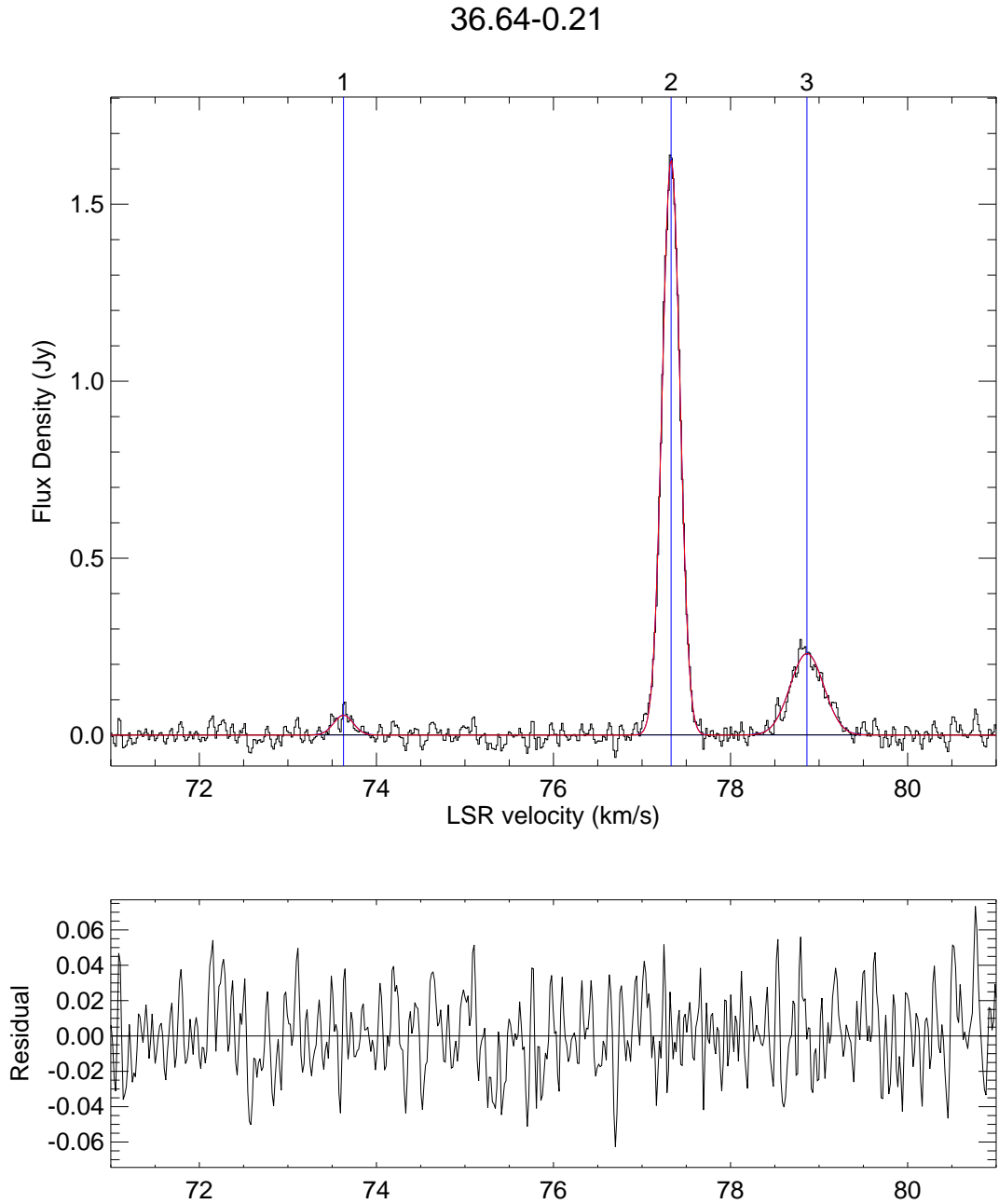}
\caption{Gaussian component fits to the 49 relatively simple sources in the AMGPS sample. The top panel shows the spectrum, with overlays of individual Gaussians in dotted lines, and the fit in solid line. The bottom panel shows the residual. The Gaussians are numbered in the same order as in Table 1. Fits for four sources in order of inreasing complexity are shown here. Color versions of this figure, and the fits for all the remaining sources are available electronically.}\label{gausscompfigures}
\end{center}
\end{figure}

\begin{figure}[!htb]
\begin{center}
\includegraphics{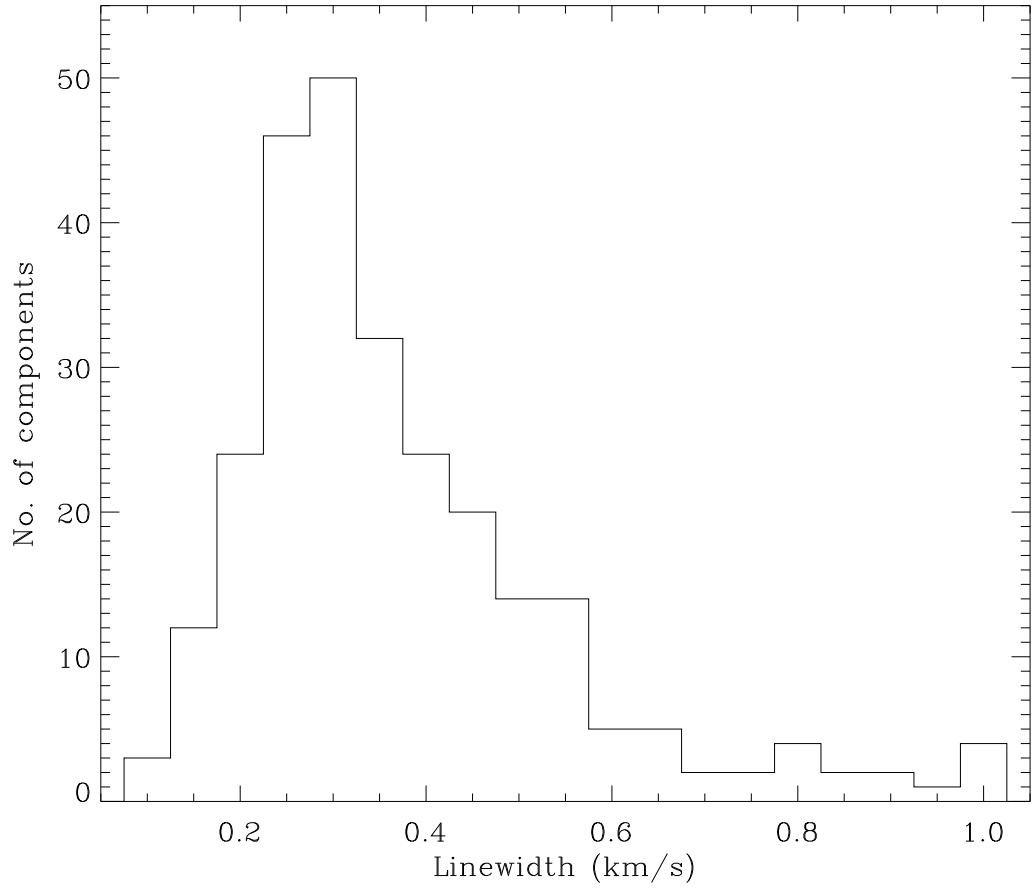}
\caption[The distribution of FWHM linewidths.]{The distribution of FWHM linewidths. There are three components that have linewidths greater than 1 \kms, that are not shown in the figure.}\label{lwhist}
\end{center}
\end{figure}

Our Gaussian component analysis for the 49 relatively simple sources yielded a total of 266 components. The parameters of the best multiple Gaussian fit for each source are shown in Table 1, and images of individual fits are shown in Figure \ref{gausscompfigures}. The first page of Table 1, and fit images for four sources in order of increasing complexity are shown here, while the rest of the material is available in the eletronic version of the paper. The distribution of full width at half maximum (FWHM) linewidths binned into 0.05 \kms~bins is shown in Figure \ref{lwhist}. The mean linewidth is found to be 0.38 \kms, and the median value is 0.33 \kms.

\citet{szym00b} determined the mean FWHM of Gaussian components to be 0.17 \kms, and the median to be 0.14 \kms. These values are a factor of two lower than the values that we determine from our analysis. Unlike the claim of \citet{szym00b}, we find no evidence for faint sources to have narrow linewidths. To make sure that our results are not biased by our excluding complex sources, we obtained sample fits for a few complex sources. While these fits are not likely to be unique, it is unlikely that the number of components used in the fit are off by more than 20\% based on the $\chi^2$ of the fit. The distribution of these components was found to be similar to that of Figure \ref{lwhist}, and the values of mean and median linewidths were similar to those quoted above. We note that our result is consistent with the measurements of \citet{ment91} and \citet{casw95}. It is also curious that our histogram has the same shape as that of \citet{szym00b}, the primary difference being the scaling of the linewidth axis by a factor of two\footnote{Private communication with the author revealed that the discrepancy arose from a mistake of \citet{szym00b} reporting half width at half maximum as FWHM linewidth.}.

\section{Counterparts at Other Wavelengths}

A multi-wavelength analysis of the astronomical sources responsible for pumping 6.7 GHz methanol masers is required to understand the relation between the maser emission and its environment, which is surmised to be high-mass star formation. Hence, it is of interest to search for counterparts to the masers in catalogs published in the literature, as well as undertake new observations at wavelengths where data is not available. The catalogs of interest at present are 2MASS ($J$, $H$ and $K$ bands), GLIMPSE (3.6, 4.5, 5.8 and 8.0 \micron), MSX (8.28, 12.13, 14.65 and 21.3 \micron), IRAS (12, 25, 60 and 100 \micron) and NVSS (1.4 GHz continuum).

A question that must be addressed in this regard is the search radius for a counterpart, keeping in mind that a large search radius may preclude a direct association between the source and the maser. Since the metric governing association of two sources is the physical separation between them, the angle of separation on the sky will be a function of the distance to the source. This is however, not practical since distances to most methanol masers are not known. Considering a fiducial distance to a massive star forming region as $\sim 3$ kpc, an angular separation of 5\arcsec~translates to a physical separation of 15,000 AU. Since warm dust heated by the central star is required for the maser pump, a search radius of 5\arcsec~can be taken as a conservative upper limit for possible association.

It is also important to consider the uncertainty in the position of the methanol maser itself. This is primarily related to the pointing error of the telescope. The root mean square (rms) pointing error of the Arecibo radio telescope is $\sim 5\arcsec$ in both azimuth and zenith angle. This has been verified by both our measurements of pointing using continuum sources, and independent measurements by the observatory staff. Hence, the overall rms pointing error in right ascension and declination can be taken to be $\sim 7\arcsec$. Although the pointing error is to some extent dependent on the azimuth angle and zenith angle, the overall statistical distribution of pointing errors is a Gaussian to good approximation. The final radius of the error circle defining the uncertainty in the position of a methanol maser, defined at the 95\% confidence level, is $\sim 18\arcsec$. Hence, the search radius for counterparts is taken to be 23\arcsec.

The large search radius, resulting from the pointing error precludes a counterpart analysis with the 2MASS and GLIMPSE catalogs. The relatively high angular resolution of these surveys coupled with their sensitivity and the numerous stars visible in near infrared wavelengths imply that a large number of point sources (50 sources is not uncommon) from these catalogs will fall within the error circle surrounding the methanol maser. Thus, these catalogs are excluded from our analysis until the methanol maser positions are refined to sub-arcsecond accuracy.

The counterparts to the methanol maser sample from AMGPS in IRAS, MSX and NVSS catalogs are summarized in Table 2. Sources in the well known star forming regions W49N and W51 are excluded since the density of methanol masers in these regions is high enough that a given IRAS, MSX or NVSS source will lie within the search radius of more than one maser. The source 41.87--0.10 is also excluded due to the uncertainty in its position -- this source was detected at the edge of a data cube, and due to source variability could not be detected in follow-up mapping as explained in Paper I.

\subsection{IRAS Counterparts}

From Table 2, it is clear that only 26 out of 76 sources have possible IRAS counterparts. This is similar to the result of \citet{elli05} who found that 30 out of a statistically complete sample of 68 Class II methanol masers had IRAS counterparts within 30\arcsec. Furthermore, 52 out of 76 sources have IRAS sources within 1 \arcmin. This is larger than results from the previous blind surveys of \citet{elli96} and \citet{szym02}, who found that about half of the methanol masers had IRAS sources within 1\arcmin. This discrepancy can be explained by the fact that methanol masers occur in clusters reflecting the clustered nature of massive star formation. Thus, a survey with higher sensitivity will detect fainter methanol masers in a given cluster, leading to a larger fraction of masers being in close proximity to infrared sources associated with the clusters.

\begin{figure}[!htb]
\begin{center}
\includegraphics[width=\textwidth]{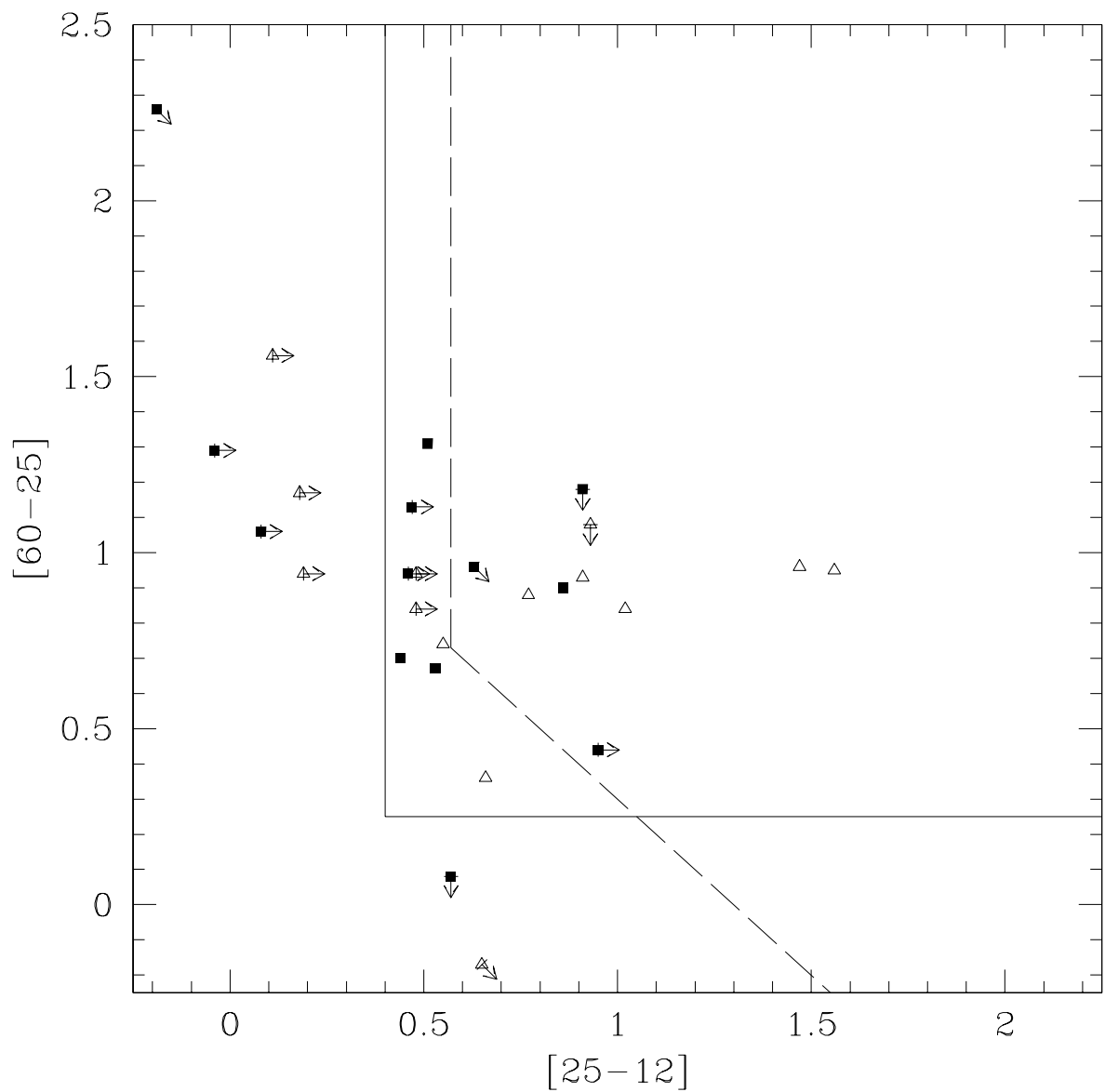}
\caption{The color-color diagram for possible IRAS counterparts of 6.7 GHz methanol masers discovered in AMGPS. The colors for previous detections are indicated by open triangles, while those of new detections are indicated by filled squares. Color limits pointing at 45\degr~indicate upper limits in both [25--12] color and [60--25] colors. Sources to the right of the dashed lines satisfy WC89 colors for ultracompact \hii~regions, while sources to the right of the bold lines satisfy HM89 criteria for the same.}\label{irascolors}
\end{center}
\end{figure}

The color-color diagram of the IRAS sources within 23\arcsec~(the one associated with W49 region included) is shown in Figure \ref{irascolors}. The colors for previous detections are shown in open triangles, while those of new detections are shown in filled squares. The two sets of criteria usually used to identify embedded massive stars (primarily ultracompact \hii~regions) are those of \citet{wood89} and \citet{hugh89}. These criteria will be abbreviated as WC89 and HM89 respectively. In Figure \ref{irascolors}, sources to the right of the dashed lines satisfy WC89, while sources to the right of the solid lines satisfy HM89. HM89 imposes an additional constraint on the 100 \micron~flux of the source, which is not shown in the figure. 

Only six sources satisfy WC89 criteria, which exclude sources with poorly determined colors (lower or upper limits on any color). A larger number of sources satisfy HM89 criteria, although the requirements on the 100 \micron~flux eliminates some of the sources that lie on the right of the solid lines. It can be seen in Figure \ref{irascolors} that a few sources do not satisfy WC89 or HM89 criteria at all. It is not clear whether these are due to projection effects, or whether a small population of methanol masers are associated with IRAS sources which do not satisfy color criteria for embedded massive stars.

Ten of the new detections in Fig. \ref{irascolors} (filled squares) are weaker than 5 Jy, while most of the previous detections (open triangles) are stronger than 5 Jy. The lack of segregation between the colors of new and old detections is contrary to the result of \citet{szym00b} who found that IRAS sources associated with weak methanol masers populate the upper left part Fig. \ref{irascolors}.

It is clear that the number of IRAS sources associated with methanol masers keeps diminishing as the positions of methanol masers are measured with greater accuracy, and the search radius for counterparts is reduced (although the large error in the IRAS source positions will dictate the search radius beyond a certain point). Hence, a number of IRAS sources that are reported as associations with methanol masers in the literature (including some cases in Table 2) are not likely to be the sources exciting the maser emission. Thus, the primary utility of IRAS sources satisfying color criteria like WC89 and HM89 may be to locate sites of massive star formation, where there are potential sources with conditions favourable for excitation of 6.7 GHz methanol masers -- targeted searches towards IRAS sources are best done by mapping a region around an individual source, or by using a telescope with a large beam.

As surmised by previous work (e.g. \citealt{elli96}), one of the reasons for the poor correspondence between methanol masers and IRAS sources could be severe source confusion in crowded fields of the Galactic plane, which resulted in many sources being not included in the point source catalog. The relatively coarse resolution of the IRAS satellite compounds problems in the context of clustered massive star formation.

\subsection{MSX Counterparts}

Only 41 out of 76 methanol masers have possible MSX point source \citep{egan03} counterparts. Visually inspecting MSX image fields, four sources are clearly associated with MSX dark clouds, while the association with dark clouds is more uncertain for an additional four sources. The lack of point source counterparts for a majority of sources is a bit surprising, especially because there are no obvious dark clouds at the sites of a majority of masers lacking point source counterparts. However, a number of masers lie close to a bright point source suggesting that a deeper search at higher resolution could detect mid-infrared emission associated with the masers. The fact that about half of the methanol masers have no associated MSX point sources, has implications on the completeness of massive young stellar object (MYSO) samples selected through MSX colors (e.g. \citealt{lums02}), although there are semantics associated with the definition of a MYSO. Our results are also consistent with those of \citet{elli05} who found that methanol masers (with sub-arcsecond positions) were generally offset from MSX point sources suggesting that the latter were associated with other objects in the star forming region.

\subsection{Counterparts in cm Wavelengths}

Seven out of 76 sources have NVSS point sources \citep{cond98} located within 23\arcsec. The emission at 1.4 GHz is thought to come from thermal bremsstrahlung (or free-free emission) in the ionized regions surrounding the massive star. The optical depth of a plasma due to free-free emission, under the \citet{alten60} approximation is
\begin{equation}
\tau_\nu \sim \frac{0.08235}{\nu^{2.1}T_e^{1.35}}~\int n_e^2dl
\end{equation}
where the frequency, $\nu$ is in GHz, $T_e$ is the electron temperature and $\int n_e^2dl$ is the emission measure in cm$^{-6}$ pc. If the \hii~region is optically thick, the emission scales as the square of the frequency. Thus, the intensity of radiation at low frequencies can be a small fraction of the intensity at a frequency corresponding to an optical depth of unity. This, compounded by the compactness of the source, makes it undetectable at low frequencies. The emission measure of a \hii~region decreases with age (as it expands), and hence younger \hii~regions are optically thick at higher frequencies. Thus, the lack of NVSS counterparts for a majority of methanol masers strongly suggests that the masers are primarily associated with phases of massive star formation which occur prior to the formation of an ultracompact \hii~region.

In addition to the NVSS, the AMGPS target region was covered in a 20 cm survey using the VLA in the B--configuration, and sub-region ($l < 42\degr$ where $l$ is the Galactic longitude) was also surveyed at 6 cm in the C--configuration \citep{whit05}. This data is useful for determining the spectral index of sources detected at both wavelengths, since both surveys have the same angular resolution. We found that six AMGPS sources had possible counterparts at 6 cm, out of which four had no counterpart at 20 cm. The latter four sources were relatively weak at 6 cm (integrated flux density 4--15 mJy), which coupled with the flux density limit of 13.8 mJy at 20 cm, places only weak constraints on the spectral index (defined by $S_\nu \propto \nu^\alpha$) of the source, as tabulated in Table 4. However, this is consistent with the picture of \hii~regions associated with methanol masers being very young and optically thick at centimeter wavelengths. It is to be noted that one of the masers, 37.77--0.22 had emission components within 23\arcsec~of the maser listed in the source catalog, but was excluded as the images seemed showed the maser to lie close to a cometary tail of a source that is centered at a greater distance.

\subsection{Other Star Formation Tracers}

While sources that are farther than 23\arcsec~from the methanol maser are unlikely to be directly associated with the maser emission, it is of interest to determine whether any IRAS, MSX, NVSS sources, or other tracers of star formation (e.g. OH, H$_2$O masers, NH$_3$ emission) are detected within 1\arcmin~of the maser. A search radius of 1\arcmin~is chosen in part to conform with previous studies (e.g. \citealt{pest05}), and in part to ensure that any star formation tracers detected would be associated with the same complex that contains the massive young stellar object responsible for the methanol maser emission. The results of this search, carried out through the VizieR service \citep{ochs00}, are summarized for all sources in Table 3.

It is apparent from Table 3 that a few sources have no tracers of star formation detected to date within 1\arcmin. It is possible that these sources are at a very early stage of evolution where the methanol maser emission has just turned on, and thus have counterparts primarily in the form of continuum emission at millimeter and submillimeter wavelengths. Observations at these wavelengths are required to understand these ``isolated'' methanol masers.

\section{Conclusions}
The properties of 6.7 GHz methanol masers detected in the AMGPS are consistent with their being associated with early phases of massive star formation. The mean linewidth of a methanol maser feature is around 0.38 \kms. There does not seem to be any distinction between the properties of bright and faint methanol masers. The clustering of methanol masers around a Galactic longitude of 49.6\degr~and the sharp decline in the number of sources at larger Galactic longitudes is likely a result of the tangency of the Carina-Sagittarius spiral arm at that longitude. The total number of methanol masers in the Galaxy is estimated to be between 1290 and 1350, most of which reside in the inner Galaxy within $\pm 50\degr$ of the Galactic center.

\acknowledgments
We are very grateful to Kiriaki Xiluri, Gene Lauria, and the operators Teresa Jiles, Bob Moulton and Patrick Fimbres at the Heinrich Hertz Submillimeter Telescope (SMT) for assistance in obtaining the \coii~spectra for our sample. The SMT is operated by the Arizona Radio Observatory (ARO), Steward Observatory, University of Arizona. We thank Johan van der Walt for providing data regarding his simulations on the distribution of methanol masers in the Galaxy, and the completeness limits of surveys as a function of flux density. We also thank the referee for comments that improved the paper. This work was supported in part by the Jet Propulsion Laboratory, California Institute of Technology. This research has made use of NASA's Astrophysics Data System.

\clearpage

\begin{deluxetable}{cccc}
\tabletypesize{\footnotesize}
\tablecolumns{4}
\tablewidth{0pc}
\tablecaption{The parameters of Gaussian components for 49 sources for which the analysis has been carried out. The columns show the amplitude, $S$ in Jy, center, $v_c$ in \kms~and the FWHM $\Delta v_{FWHM}$ in \kms~for each component.}\label{gaussiancomptable}
\tablehead{
\colhead{No.} & \colhead{$S$ (Jy)} & \colhead{$v_c$ (\kms)} & \colhead{$\Delta v_{FWHM}$ (\kms)}}
\startdata
%\hline \\[-3mm]
\multicolumn{4}{c}{36.64--0.21} \\[1mm]
\hline \\[-3mm]
1 & 0.06 & 73.63 & 0.26 \\
2 & 1.63 & 77.33 & 0.23 \\
3 & 0.23 & 78.86 & 0.44 \\[2mm]
\hline \\[-3mm]
\multicolumn{4}{c}{36.90--0.41} \\[1mm]
\hline \\[-3mm]
1 & 0.06 & 81.65 & 0.14 \\
2 & 0.06 & 83.38 & 0.79 \\
3 & 0.11 & 84.45 & 0.82 \\
4 & 0.37 & 84.71 & 0.35 \\[2mm]
\hline \\[-3mm]
\multicolumn{4}{c}{38.26--0.20} \\[1mm]
\hline \\[-3mm]
1 & 0.07 & 64.31 & 0.18 \\
2 & 0.13 & 64.72 & 0.78 \\
3 & 0.19 & 65.33 & 0.40 \\
4 & 0.11 & 67.75 & 0.45 \\
5 & 0.36 & 68.81 & 0.41 \\
6 & 0.58 & 69.39 & 0.47 \\
7 & 0.71 & 70.20 & 0.29 \\
8 & 0.07 & 70.80 & 0.41 \\
9 & 0.15 & 72.27 & 0.48 \\
10 & 0.14 & 72.88 & 0.27 \\
11 & 0.57 & 73.16 & 0.39 \\[2mm]
\hline \\[-3mm]
\multicolumn{4}{c}{38.92--0.36} \\[1mm]
\hline \\[-3mm]
1 & 0.34 & 31.15 & 0.34 \\
2 & 0.59 & 31.63 & 0.44 \\
3 & 0.99 & 31.91 & 0.34 \\
4 & 0.73 & 32.26 & 0.32 \\
5 & 0.40 & 32.76 & 0.61 \\
6 & 0.08 & 33.25 & 0.31 \\[2mm]
\multicolumn{4}{c}{34.82+0.35} \\[1mm]
\hline \\[-3mm]
1 & 0.08 & 58.62 & 0.16 \\
2 & 0.23 & 59.67 & 0.28 \\
3 & 0.05 & 59.95 & 0.28 \\[1.5mm]
\hline \\[-3mm]
\multicolumn{4}{c}{35.25--0.24} \\[1mm]
\hline \\[-3mm]
1 & 0.14 & 56.21 & 0.30 \\
2 & 0.07 & 71.38 & 0.18 \\
3 & 0.16 & 71.71 & 0.30 \\
4 & 0.31 & 72.25 & 0.36 \\
5 & 1.24 & 72.39 & 0.21 \\
6 & 0.25 & 72.59 & 0.12 \\
7 & 0.35 & 72.71 & 0.24 \\
8 & 0.22 & 73.08 & 0.33 \\[1.5mm]
\hline \\[-3mm]
\multicolumn{4}{c}{35.39+0.02} \\[1mm]
\hline \\[-3mm]
1 & 0.03 & 91.27 & 0.25 \\
2 & 0.17 & 94.20 & 0.25 \\
3 & 0.17 & 96.92 & 0.38 \\[1.5mm]
\hline \\[-3mm]
\multicolumn{4}{c}{35.40+0.03} \\[1mm]
\hline \\[-3mm]
1 & 0.53 & 89.03 & 0.26 \\
2 & 0.08 & 89.39 & 0.54 \\
3 & 0.17 & 89.96 & 0.46 \\
4 & 0.13 & 90.49 & 0.25 \\[1.5mm]
\hline \\[-3mm]
\multicolumn{4}{c}{36.02--0.20} \\[1mm]
\hline \\[-3mm]
1 & 0.05 & 92.55 & 0.23 \\
2 & 0.13 & 92.96 & 0.31 \\
3 & 0.09 & 93.22 & 0.26 \\[2mm]
\multicolumn{4}{c}{36.92+0.48} \\[1mm]
\hline \\[-3mm]
1 & 0.56 & --36.10 & 0.19 \\
2 & 1.47 & --35.91 & 0.25 \\[2mm]
\hline \\[-3mm]
\multicolumn{4}{c}{37.02--0.03} \\[1mm]
\hline \\[-3mm]
1 & 0.75 & 78.38 & 0.85 \\
2 & 6.37 & 78.43 & 0.35 \\
3 & 2.71 & 78.88 & 0.42 \\
4 & 0.44 & 79.32 & 0.34 \\
5 & 0.32 & 79.93 & 0.82 \\
6 & 0.07 & 80.76 & 0.24 \\
7 & 0.08 & 84.77 & 0.23 \\
8 & 0.10 & 85.19 & 0.25 \\[2mm]
\hline \\[-3mm]
\multicolumn{4}{c}{37.38--0.09} \\[1mm]
\hline \\[-3mm]
1 & 0.04 & 68.16 & 1.33 \\
2 & 0.10 & 70.12 & 0.97 \\
3 & 0.13 & 70.57 & 0.34 \\[2mm]
\hline \\[-3mm]
\multicolumn{4}{c}{37.74--0.12} \\[1mm]
\hline \\[-3mm]
1 & 0.21 & 50.15 & 0.26 \\
2 & 0.84 & 50.29 & 0.22 \\[2mm]
\hline \\[-3mm]
\multicolumn{4}{c}{37.76--0.19} \\[1mm]
\hline \\[-3mm]
1 & 0.06 & 54.33 & 0.25 \\
2 & 0.57 & 55.03 & 0.54 \\
3 & 0.13 & 56.36 & 0.57 \\
4 & 0.19 & 57.12 & 0.64 \\
5 & 0.06 & 58.50 & 0.57 \\
6 & 0.54 & 60.55 & 0.38 \\
7 & 0.12 & 61.87 & 0.37 \\
8 & 0.15 & 62.24 & 1.35 \\
9 & 0.41 & 63.50 & 0.29 \\
10 & 0.39 & 64.67 & 0.43 \\
11 & 0.34 & 65.34 & 0.28 \\
12 & 0.13 & 65.71 & 0.40 \\[2mm]
\hline \\[-3mm]
\multicolumn{4}{c}{37.77--0.22} \\[1mm]
\hline \\[-3mm]
1 & 0.11 & 69.11 & 0.35 \\
2 & 0.71 & 69.57 & 0.31 \\
3 & 0.17 & 69.80 & 0.50 \\[2mm]
\hline \\[-3mm]
\multicolumn{4}{c}{38.08--0.27} \\[1mm]
\hline \\[-3mm]
1 & 0.04 & 62.33 & 0.53 \\
2 & 0.09 & 67.03 & 0.47 \\
3 & 0.57 & 67.50 & 0.23 \\[2mm]
\hline \\[-3mm]
\multicolumn{4}{c}{38.56+0.15} \\[1mm]
\hline \\[-3mm]
1 & 0.06 & 21.95 & 0.15 \\
2 & 0.12 & 23.27 & 0.30 \\
3 & 0.03 & 29.71 & 0.46 \\
4 & 0.07 & 30.97 & 0.36 \\
5 & 0.16 & 31.48 & 0.40 \\[2mm]
\hline \\[-3mm]
\multicolumn{4}{c}{38.60--0.21} \\[1mm]
\hline \\[-3mm]
1 & 0.17 & 61.69 & 0.34 \\
2 & 0.13 & 61.88 & 0.13 \\
3 & 0.40 & 62.49 & 0.20 \\
4 & 0.29 & 62.62 & 0.25 \\
5 & 0.33 & 63.08 & 0.39 \\
6 & 0.05 & 63.77 & 0.44 \\
7 & 0.06 & 68.23 & 0.66 \\
8 & 0.09 & 68.83 & 0.31 \\
9 & 0.34 & 69.04 & 0.28 \\
10 & 0.04 & 69.11 & 1.00 \\[2mm]
\hline \\[-3mm]
\multicolumn{4}{c}{38.66+0.08} \\[1mm]
\hline \\[-3mm]
1 & 1.53 & --31.49 & 0.24 \\
2 & 0.91 & --31.37 & 0.31 \\[2mm]
\hline \\[-3mm]
\multicolumn{4}{c}{39.39--0.14} \\[1mm]
\hline \\[-3mm]
1 & 0.20 & 58.49 & 0.35 \\
2 & 0.14 & 58.89 & 0.28 \\
3 & 0.06 & 59.37 & 0.44 \\
4 & 0.22 & 60.11 & 0.62 \\
5 & 1.02 & 60.40 & 0.25 \\
6 & 0.05 & 64.65 & 0.28 \\
7 & 0.16 & 68.57 & 0.48 \\
8 & 0.16 & 69.41 & 0.58 \\
9 & 0.03 & 70.34 & 0.86 \\
10 & 0.19 & 71.81 & 0.25 \\
11 & 0.07 & 75.36 & 0.29 \\[2mm]
\hline \\[-3mm]
\multicolumn{4}{c}{39.54--0.38} \\[1mm]
\hline \\[-3mm]
1 & 0.19 & 47.72 & 0.45 \\
2 & 0.17 & 48.31 & 0.48 \\
3 & 0.09 & 48.90 & 0.63 \\[2mm]
\hline \\[-3mm]
\multicolumn{4}{c}{40.62--0.14} \\[1mm]
\hline \\[-3mm]
1 & 1.01 & 30.02 & 0.28 \\
2 & 0.70 & 30.23 & 0.31 \\
3 & 0.43 & 30.80 & 0.38 \\
4 & 4.47 & 30.95 & 0.25 \\
5 & 12.89 & 31.11 & 0.22 \\
6 & 4.51 & 31.26 & 0.17 \\
7 & 0.81 & 31.52 & 0.72 \\
8 & 0.07 & 35.72 & 0.22 \\
9 & 1.15 & 36.08 & 0.24 \\
10 & 0.76 & 36.22 & 0.39 \\[2mm]
\hline \\[-3mm]
\multicolumn{4}{c}{40.94--0.04} \\[1mm]
\hline \\[-3mm]
1 & 2.21 & 36.61 & 0.32 \\
2 & 0.78 & 36.84 & 0.27 \\
3 & 0.43 & 37.32 & 0.29 \\
4 & 0.23 & 40.83 & 0.41 \\
5 & 0.72 & 40.97 & 0.22 \\
6 & 0.20 & 41.26 & 0.35 \\
7 & 0.07 & 42.40 & 0.08 \\
8 & 0.14 & 43.63 & 0.22 \\[2mm]
\hline \\[-3mm]
\multicolumn{4}{c}{41.08--0.13} \\[1mm]
\hline \\[-3mm]
1 & 0.80 & 57.52 & 0.30 \\
2 & 0.14 & 58.06 & 0.49 \\
3 & 0.04 & 59.36 & 0.37 \\[2mm]
\hline \\[-3mm]
\multicolumn{4}{c}{41.12--0.11} \\[1mm]
\hline \\[-3mm]
1 & 0.04 & 33.28 & 0.29 \\
2 & 0.05 & 33.52 & 0.13 \\
3 & 0.12 & 35.55 & 0.29 \\
4 & 0.23 & 35.91 & 0.32 \\
5 & 1.13 & 36.54 & 0.32 \\
6 & 0.21 & 37.04 & 0.45 \\[15mm]
\multicolumn{4}{c}{41.12--0.22} \\[1mm]
\hline \\[-3mm]
1 & 0.25 & 55.25 & 0.24 \\
2 & 0.25 & 62.76 & 0.38 \\
3 & 1.96 & 63.45 & 0.47 \\
4 & 0.16 & 63.99 & 0.36 \\
5 & 0.08 & 66.43 & 0.27 \\[2mm]
\hline \\[-3mm]
\multicolumn{4}{c}{41.16--0.20} \\[1mm]
\hline \\[-3mm]
1 & 0.08 & 61.73 & 0.32 \\
2 & 0.10 & 62.16 & 0.33 \\
3 & 0.07 & 62.57 & 0.54 \\
4 & 0.27 & 63.52 & 0.35 \\[2mm]
\hline \\[-3mm]
\multicolumn{4}{c}{41.27+0.37} \\[1mm]
\hline \\[-3mm]
1 & 0.04 & 19.32 & 0.56 \\
2 & 0.22 & 20.04 & 0.46 \\
3 & 0.18 & 20.35 & 0.23 \\[2mm]
\hline \\[-3mm]
\multicolumn{4}{c}{41.58+0.04} \\[1mm]
\hline \\[-3mm]
1 & 0.05 & 10.68 & 0.44 \\
2 & 0.47 & 11.91 & 0.38 \\[2mm]
\hline \\[-3mm]
\multicolumn{4}{c}{41.87--0.10} \\[1mm]
\hline \\[-3mm]
1 & 0.18 & 15.81 & 0.29 \\
2 & 0.06 & 20.70 & 0.40 \\
3 & 0.07 & 23.50 & 0.34 \\[15mm]
\multicolumn{4}{c}{43.08--0.08} \\[1mm]
\hline \\[-3mm]
1 & 0.52 & 10.08 & 0.41 \\
2 & 8.56 & 10.19 & 0.33 \\
3 & 1.30 & 10.49 & 0.26 \\
4 & 0.63 & 10.82 & 0.41 \\
5 & 0.20 & 13.94 & 0.28 \\[2mm]
\hline \\[-3mm]
\multicolumn{4}{c}{43.17--0.00} \\[1mm]
\hline \\[-3mm]
1 & 0.17 & --1.39 & 0.41 \\
2 & 1.87 & --1.22 & 0.20 \\
3 & 1.83 & --1.06 & 0.23 \\
4 & 0.82 & --0.68 & 0.49 \\
5 & 0.05 & 1.29 & 0.22 \\
6 & 0.13 & 1.89 & 0.42 \\
7 & 0.06 & 3.55 & 0.55 \\
8 & 0.18 & 4.02 & 0.31 \\[2mm]
\hline \\[-3mm]
\multicolumn{4}{c}{43.18--0.01} \\[1mm]
\hline \\[-3mm]
1 & 0.07 & 8.18 & 0.29 \\
2 & 0.08 & 9.17 & 0.73 \\
3 & 0.55 & 10.93 & 0.69 \\
4 & 0.54 & 11.23 & 0.46 \\
5 & 0.07 & 13.03 & 0.26 \\
6 & 0.09 & 13.55 & 0.50 \\[2mm]
\hline \\[-3mm]
\multicolumn{4}{c}{44.31+0.04} \\[1mm]
\hline \\[-3mm]
1 & 0.22 & 55.32 & 0.37 \\
2 & 0.65 & 55.77 & 0.39 \\
3 & 0.28 & 56.17 & 0.25 \\[10mm]
\multicolumn{4}{c}{44.64--0.52} \\[1mm]
\hline \\[-3mm]
1 & 0.02 & 47.78 & 0.50 \\
2 & 0.18 & 49.11 & 0.39 \\
3 & 0.47 & 49.37 & 0.30 \\
4 & 0.24 & 49.60 & 0.25 \\[2mm]
\hline \\[-3mm]
\multicolumn{4}{c}{45.07+0.13} \\[1mm]
\hline \\[-3mm]
1 & 0.18 & 57.31 & 0.52 \\
2 & 14.05 & 57.65 & 0.25 \\
3 & 40.50 & 57.79 & 0.32 \\
4 & 8.54 & 58.08 & 0.30 \\
5 & 4.03 & 58.23 & 0.37 \\
6 & 0.16 & 58.78 & 0.48 \\
7 & 0.58 & 59.67 & 0.29 \\[2mm]
\hline \\[-3mm]
\multicolumn{4}{c}{45.44+0.07} \\[1mm]
\hline \\[-3mm]
1 & 0.80 & 49.76 & 0.45 \\
2 & 0.79 & 50.07 & 0.31 \\
3 & 0.31 & 50.40 & 0.20 \\[2mm]
\hline \\[-3mm]
\multicolumn{4}{c}{45.49+0.13} \\[1mm]
\hline \\[-3mm]
1 & 4.98 & 57.09 & 0.19 \\
2 & 3.75 & 57.24 & 0.25 \\
3 & 4.05 & 57.36 & 0.43 \\
4 & 2.53 & 57.71 & 0.31 \\
5 & 0.70 & 57.88 & 0.21 \\
6 & 0.05 & 64.29 & 0.36 \\
7 & 0.04 & 65.20 & 0.65 \\
8 & 0.17 & 65.64 & 0.23 \\
9 & 0.07 & 65.87 & 0.89 \\[5mm]
\multicolumn{4}{c}{45.57--0.12} \\[1mm]
\hline \\[-3mm]
1 & 0.12 & 1.45 & 0.34 \\
2 & 0.31 & 1.64 & 0.33 \\
3 & 0.14 & 3.10 & 0.32 \\
4 & 0.05 & 3.44 & 0.57 \\
5 & 0.22 & 4.17 & 0.23 \\
6 & 0.11 & 9.60 & 0.28 \\[2mm]
\hline \\[-3mm]
\multicolumn{4}{c}{46.07+0.22} \\[1mm]
\hline \\[-3mm]
1 & 0.05 & 22.65 & 0.60 \\
2 & 1.17 & 23.25 & 0.50 \\
3 & 0.68 & 23.64 & 0.40 \\
4 & 0.11 & 24.10 & 0.34 \\
5 & 0.43 & 24.44 & 0.46 \\[2mm]
\hline \\[-3mm]
\multicolumn{4}{c}{48.89--0.17} \\[1mm]
\hline \\[-3mm]
1 & 0.12 & 57.33 & 0.17 \\[2mm]
\hline \\[-3mm]
\multicolumn{4}{c}{48.90--0.27} \\[1mm]
\hline \\[-3mm]
1 & 0.11 & 63.75 & 0.34 \\
2 & 0.04 & 69.22 & 0.18 \\
3 & 0.63 & 71.79 & 0.54 \\
4 & 0.50 & 72.11 & 0.34 \\[2mm]
\hline \\[-3mm]
\multicolumn{4}{c}{48.99--0.30} \\[1mm]
\hline \\[-3mm]
1 & 0.04 & 61.22 & 0.63 \\
2 & 0.10 & 62.67 & 0.24 \\
3 & 0.18 & 63.09 & 0.37 \\
4 & 0.03 & 66.79 & 0.40 \\
5 & 0.14 & 67.26 & 0.32 \\
6 & 0.12 & 67.61 & 0.34 \\
7 & 0.04 & 69.28 & 0.57 \\
8 & 0.04 & 70.47 & 0.25 \\
9 & 0.15 & 71.49 & 0.31 \\
10 & 0.47 & 71.63 & 0.61 \\
11 & 0.08 & 72.39 & 0.37 \\[2mm]
\hline \\[-3mm]
\multicolumn{4}{c}{49.62--0.36} \\[1mm]
\hline \\[-3mm]
1 & 1.21 & 49.26 & 0.32 \\
2 & 0.24 & 49.47 & 0.18 \\
3 & 0.34 & 49.72 & 0.51 \\
4 & 0.42 & 50.10 & 0.32 \\
5 & 0.10 & 50.33 & 0.24 \\
6 & 0.09 & 58.06 & 0.54 \\
7 & 0.15 & 59.15 & 0.89 \\
8 & 0.31 & 59.32 & 0.31 \\
9 & 0.05 & 59.84 & 0.57 \\[2mm]
\hline \\[-3mm]
\multicolumn{4}{c}{50.78+0.15} \\[1mm]
\hline \\[-3mm]
1 & 0.05 & 47.66 & 0.49 \\
2 & 0.98 & 48.71 & 0.47 \\
3 & 1.23 & 48.91 & 0.34 \\
4 & 4.23 & 49.06 & 0.25 \\
5 & 0.26 & 49.35 & 0.23 \\
6 & 1.60 & 49.78 & 0.32 \\
7 & 0.16 & 50.10 & 0.21 \\
8 & 0.12 & 50.56 & 0.54 \\[2mm]
\hline \\[-3mm]
\multicolumn{4}{c}{52.92+0.41} \\[1mm]
\hline \\[-3mm]
1 & 4.96 & 39.09 & 0.21 \\
2 & 1.97 & 39.21 & 0.17 \\
3 & 2.47 & 39.38 & 0.48 \\
4 & 1.82 & 39.67 & 0.24 \\
5 & 0.16 & 40.01 & 0.21 \\
6 & 0.14 & 40.23 & 0.19 \\
7 & 0.16 & 40.50 & 0.32 \\
8 & 0.04 & 41.07 & 0.32 \\
9 & 1.16 & 42.61 & 0.24 \\
10 & 2.83 & 44.59 & 0.35 \\[1.5mm]
\hline \\[-3mm]
\multicolumn{4}{c}{53.04+0.11} \\[1mm]
\hline \\[-3mm]
1 & 1.33 & 9.87 & 0.18 \\
2 & 0.64 & 9.99 & 0.12 \\
3 & 1.62 & 10.13 & 0.20 \\[1.5mm]
\hline \\[-3mm]
\multicolumn{4}{c}{53.14+0.07} \\[1mm]
\hline \\[-3mm]
1 & 0.66 & 23.81 & 0.27 \\
2 & 0.10 & 24.28 & 0.73 \\
3 & 0.93 & 24.56 & 0.24 \\
4 & 0.08 & 24.78 & 0.23 \\
5 & 0.04 & 25.21 & 0.38 \\[1.5mm]
\hline \\[-3mm]
\multicolumn{4}{c}{53.62+0.04} \\[1mm]
\hline \\[-3mm]
1 & 2.67 & 18.49 & 0.17 \\
2 & 6.03 & 18.59 & 0.17 \\
3 & 2.32 & 18.78 & 0.18 \\
4 & 16.69 & 18.95 & 0.24 \\
5 & 3.03 & 19.01 & 0.13 \\
6 & 1.08 & 19.18 & 0.29 \\
\enddata
\end{deluxetable}
%\clearpage

\begin{deluxetable}{ccccccc}
\tabletypesize{\footnotesize}
\tablecaption{IRAS, MSX and NVSS counterparts for the AMGPS sample}\label{multiwave1}
\tablewidth{0pt}
\tablehead{
\colhead{Methanol Maser} & \colhead{IRAS source} & \colhead{sep (\arcsec)} & \colhead{MSX source} & \colhead{sep (\arcsec)} & \colhead{NVSS source} & \colhead{sep (\arcsec)}
}
\startdata
34.82+0.35  & 18511+0146 & 12.1 & G034.8211+00.3519 &  7.9 &               &       \\
35.03+0.35  &            &      & G035.0252+00.3502 & 12.2 &               &       \\
35.25--0.24 & 18539+0153 & 17.2 &                   &      &               &       \\
35.39+0.02  &            &      &                   &      & 185550+021153 & 16.75 \\
35.40+0.03  & 18533+0208 &  7.3 & G035.3988+00.0237 &  4.2 &               &       \\
35.79--0.17 & 18547+0223 &  6.9 & G035.8012-00.1779 & 22.7 &               &       \\
36.70+0.09  & 18554+0319 & 11.5 &                   &      &               &       \\
36.84--0.02 & 18561+0323 & 12.9 &                   &      &               &       \\
36.92+0.48  &            &      & G036.9194+00.4825 & 12.4 &               &       \\
37.38--0.09 &            &      & G037.3815-00.0816 & 13.9 &               &       \\
37.53--0.11 &            &      & G037.5450-00.1118 & 22.0 & 190016+040307 & 19.34 \\
37.55+0.19  & 18566+0408 & 22.8 & G037.5536+00.2008 & 21.6 &               &       \\
37.60+0.42  & 18559+0416 & 13.4 &                   &      &               &       \\
37.74--0.12 &            &      & G037.7391-00.1156 & 10.4 &               &       \\
37.76--0.19 &            &      & G037.7527-00.1933 &  6.0 &               &       \\
37.77--0.22 &            &      & G037.7625-00.2181 & 19.5 &               &       \\
            &            &      & G037.7631-00.2140 & 20.3 &               &       \\
38.03--0.30 & 18593+0419 &  6.3 &                   &      &               &       \\
38.08--0.27 &            &      & G038.0757-00.2652 & 18.8 &               &       \\
38.26--0.08 & 18589+0437 & 15.6 & G038.2577-00.0733 & 17.7 &               &       \\
38.26--0.20 & 18594+0434 &  6.3 & G038.2537-00.1996 & 18.3 &               &       \\
38.66+0.08  &            &      &                   &      & 190135+050739 & 20.83 \\
39.39--0.14 & 19012+0536 &  2.7 & G039.3880-00.1421 &  6.4 &               &       \\
40.28--0.22 &            &      & G040.2816-00.2190 & 15.2 &               &       \\
40.62--0.14 & 19035+0641 & 17.9 & G040.6225-00.1377 & 11.4 &               &       \\
41.08--0.13 &            &      & G041.0723-00.1256 &  7.9 &               &       \\
41.12--0.11 & 19044+0709 &  9.8 & G041.1207-00.1062 &  9.8 &               &       \\
41.12--0.22 &            &      & G041.1195-00.2216 & 11.1 &               &       \\
41.23--0.20 &            &      & G041.2282-00.1966 & 11.5 &               &       \\
41.27+0.37  &            &      & G041.2676+00.3732 & 18.5 &               &       \\
41.34--0.14 &            &      & G041.3477-00.1414 &  7.9 &               &       \\
42.03+0.19  & 19050+0806 & 17.0 & G042.0341+00.1905 & 19.8 &               &       \\
42.43--0.26 & 19074+0814 &  7.8 & G042.4343-00.2597 & 14.6 & 190950+081927 &  4.31 \\
42.70--0.15 & 19075+0832 &  5.5 &                   &      &               &       \\
43.80--0.13 & 19095+0930 & 22.6 & G043.7955-00.1275 & 11.0 &               &       \\
44.31+0.04  &            &      & G044.3103+00.0416 & 15.8 &               &       \\
45.07+0.13  & 19110+1045 & 11.6 & G045.0711+00.1325 & 10.4 &               &       \\
45.47+0.05  &            &      & G045.4658+00.0457 & 20.9 &               &       \\
45.47+0.13  & 19117+1107 &  6.0 & G045.4782+00.1323 &  7.8 & 191408+111229 &  6.07 \\
45.57--0.12 &            &      & G045.5664-00.1193 & 14.8 &               &       \\
45.81--0.36 & 19141+1110 &  8.4 & G045.8059-00.3521 & 20.6 &               &       \\
46.07+0.22  &            &      & G046.0599+00.2240 & 19.6 & 191455+114601 & 11.26 \\
48.99--0.30 & 19201+1400 &  4.4 & G048.9897-00.2992 &  1.6 & 192225+140633 &  7.24 \\
49.27+0.31  & 19184+1432 & 15.5 &                   &      &               &       \\
49.35+0.41  &            &      & G049.3496+00.4150 & 17.8 &               &       \\
49.60--0.25 & 19211+1434 & 11.4 & G049.5993-00.2488 &  3.4 &               &       \\
50.78+0.15  &            &      & G050.7796+00.1520 & 11.4 &               &       \\
52.92+0.41  & 19253+1748 & 12.6 & G052.9217+00.4142 & 13.3 &               &       \\
53.04+0.11  &            &      & G053.0366+00.1110 &  3.2 &               &       \\
53.14+0.07  & 19270+1750 &  4.8 & G053.1417+00.0705 &  1.5 &               &       \\
53.62+0.04  &            &      & G053.6185+00.0376 &  4.3 &               &       \\
\enddata
\tablecomments{Only point sources within 23\arcsec~of the methanol maser are listed above. Sources at larger angular separations are not likely to be responsible for pumping the maser. The methanol masers in W49 and W51, and the source 41.87--0.10 are omitted from this analysis.}
\end{deluxetable}
\clearpage

\begin{deluxetable}{cccccccc}
\tabletypesize{\footnotesize}
\tablecaption{This table indicates whether there are IRAS, MSX or NVSS point sources, or other tracers of star formation such as OH and H$_2$O masers within 1\arcmin~of the methanol masers in AMGPS}
\tablewidth{0pt}
\tablehead{
\colhead{Methanol maser} & \colhead{IRAS} & \colhead{MSX} & \colhead{NVSS} & \colhead{H$_2$O} & \colhead{OH} & \colhead{other methanol} & \colhead{other SF} \\
\colhead{} & \colhead{source?} & \colhead{source?} & \colhead{source?} & \colhead{maser?} &  \colhead{maser?} &  \colhead{masers?} &  \colhead{tracers?} 
}
\startdata
34.82+0.35  & y & y & n & n & n & y & y \\
35.03+0.35  & y & y & y & y & y & y & y \\
35.25--0.24 & y & y & n & n & n & n & n \\
35.39+0.02  & y & y & y & n & n & n & n \\
35.40+0.03  & y & y & y & n & n & n & n \\
35.59+0.06  & n & y & y & n & n & n & y \\
35.79--0.17 & y & y & n & n & y & n & n \\
36.02--0.20 & n & n & n & n & n & n & n \\
36.64--0.21 & y & n & n & n & n & n & n \\
36.70+0.09  & y & y & n & n & y & n & y \\
36.84--0.02 & y & y & n & n & n & n & n \\
36.90--0.41 & n & y & n & n & n & n & n \\
36.92+0.48  & y & y & y & n & n & n & y \\
37.02--0.03 & n & y & y & n & n & n & n \\
37.38--0.09 & n & y & n & n & n & n & n \\
37.47--0.11 & n & y & n & y & n & n & y \\
37.53--0.11 & y & y & y & y & n & n & y \\
37.55+0.19  & y & y & n & y & y & n & y \\
37.60+0.42  & y & n & n & y & n & n & n \\
37.74--0.12 & n & y & n & y & n & n & y \\
37.76--0.19 & y & y & n & n & n & n & y \\
37.77--0.22 & y & y & y & y & n & n & y \\
38.03--0.30 & y & y & y & n & y & n & n \\
38.08--0.27 & y & y & n & n & n & n & n \\
38.12--0.24 & y & y & y & n & n & n & y \\
38.20--0.08 & n & y & n & n & n & n & n \\
38.26--0.08 & y & y & n & n & n & n & n \\
38.26--0.20 & y & y & n & n & n & n & n \\
38.56+0.15  & y & y & y & n & n & n & n \\
38.60--0.21 & n & y & n & n & n & n & y \\
38.66+0.08  & y & y & y & n & n & n & n \\
38.92--0.36 & y & y & n & y & n & n & y \\
39.39--0.14 & y & y & n & n & n & n & y \\
39.54--0.38 & n & n & n & n & n & n & n \\
40.28--0.22 & n & y & n & n & n & n & n \\
40.62--0.14 & y & y & n & y & y & n & y \\
40.94--0.04 & n & n & n & n & n & n & n \\
41.08--0.13 & n & y & n & n & n & n & n \\
41.12--0.11 & y & y & n & n & n & n & n \\
41.12--0.22 & y & y & n & y & n & n & n \\
41.16--0.20 & n & y & n & n & n & n & y \\
41.23--0.20 & y & y & n & n & n & n & n \\
41.27+0.37  & n & y & n & n & n & n & n \\
41.34--0.14 & y & y & n & n & n & n & n \\
41.58+0.04  & n & y & n & n & n & n & n \\
42.03+0.19  & y & y & n & n & n & n & n \\
42.30--0.30 & y & n & n & n & n & n & n \\
42.43--0.26 & y & y & y & y & n & n & y \\
42.70--0.15 & y & n & n & n & n & n & n \\
43.04--0.46 & y & y & n & y & y & y & y \\
43.08--0.08 & n & n & n & n & n & n & n \\
43.80--0.13 & y & y & y & y & y & n & y \\
44.31+0.04  & y & y & n & n & n & n & n \\
44.64--0.52 & n & y & n & n & n & n & n \\
45.07+0.13  & y & y & y & y & y & y & y \\
45.44+0.07  & y & y & y & y & y & n & y \\
45.47+0.05  & y & y & y & y & y & y & y \\
45.47+0.13  & y & y & y & y & y & n & y \\
45.49+0.13  & n & y & n & y & n & n & y \\
45.57--0.12 & y & y & n & n & n & n & n \\
45.81--0.36 & y & y & n & n & n & n & n \\
46.07+0.22  & n & y & y & n & n & n & n \\
46.12+0.38  & n & y & n & n & n & n & n \\
48.89--0.17 & y & y & n & n & n & n & n \\
48.90--0.27 & y & y & n & n & n & n & n \\
48.99--0.30 & y & y & y & y & n & n & n \\
49.27+0.31  & y & y & n & n & n & n & n \\
49.35+0.41  & n & y & n & n & n & n & n \\
49.41+0.33  & y & y & y & n & n & y & n \\
49.60--0.25 & y & y & y & n & n & n & y \\
49.62--0.36 & n & y & n & n & n & n & n \\
50.78+0.15  & y & y & n & n & n & n & n \\
52.92+0.41  & y & y & n & n & n & n & n \\
53.04+0.11  & y & y & n & n & n & n & y \\
53.14+0.07  & y & y & n & n & n & n & n \\
53.62+0.04  & n & y & n & n & n & n & y \\
\enddata
\tablecomments{The methanol masers in W49 and W51, and the source 41.87--0.10 are omitted from this analysis.}
\end{deluxetable}
\clearpage

\begin{deluxetable}{cccc}
\tablecaption{Spectral indices of radio sources potentially associated with 6.7 GHz methanol masers}
\tablewidth{0pt}
\tablehead{
\colhead{Methanol maser} & \colhead{$S_6$ (mJy)} & \colhead{$S_{20}$ (mJy)} & \colhead{$\alpha$}
}
\startdata
37.53--0.11 & 678.23 & 472.81  & +0.30 \\
38.66+0.08  &  33.29 &  19.30  & +0.45 \\
35.03+0.35  &  14.48 & $<13.8$ & $>+0.03$ \\
35.39+0.02  &   8.04 & $<13.8$ & $>-0.45$ \\
39.39--0.14 &   4.33 & $<13.8$ & $>-0.90$ \\
40.62--0.14 &   4.27 & $<13.8$ & $>-0.91$ \\
\enddata
\tablecomments{The columns show the methanol maser, flux densities at 6 cm and 20 cm respectively, and the power law index defined by $S_\nu \propto \nu^\alpha$.}
\end{deluxetable}
\clearpage

\clearpage
{\includegraphics[width=0.9\textwidth]{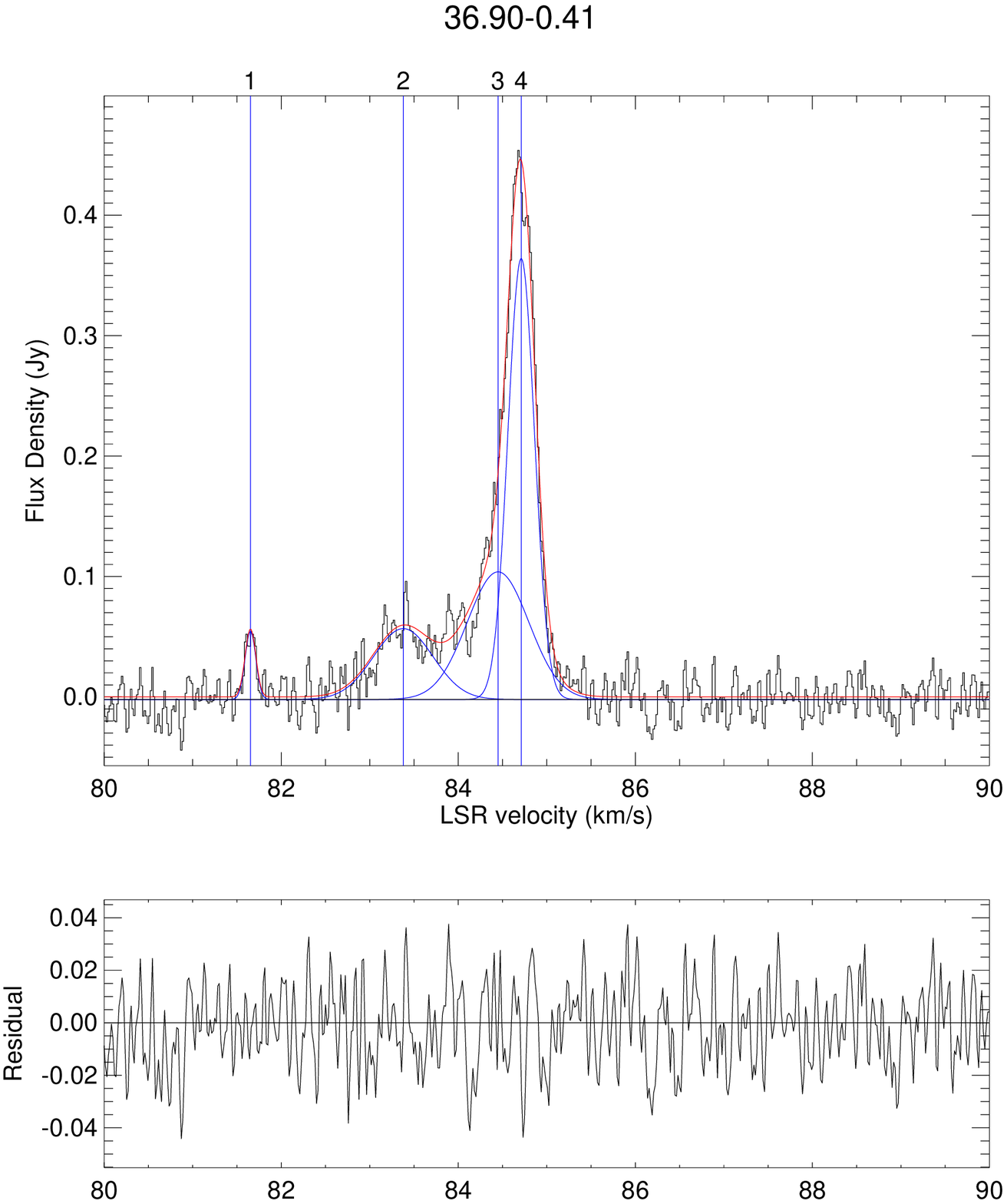}}\\
\centerline{Fig. 8. --- Continued.}
{\includegraphics[width=0.9\textwidth]{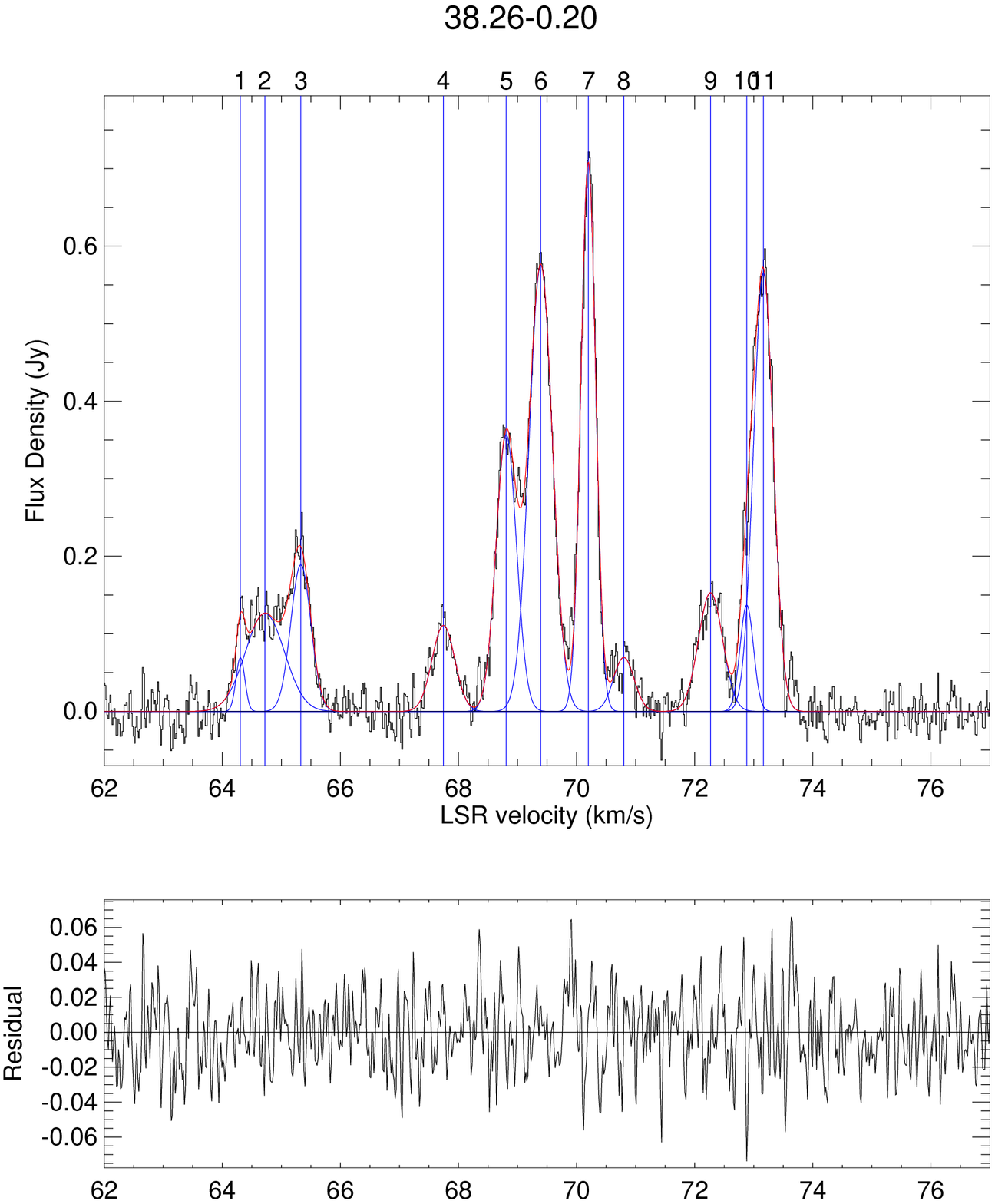}}\\
\centerline{Fig. 8. --- Continued.}
{\includegraphics[width=0.9\textwidth]{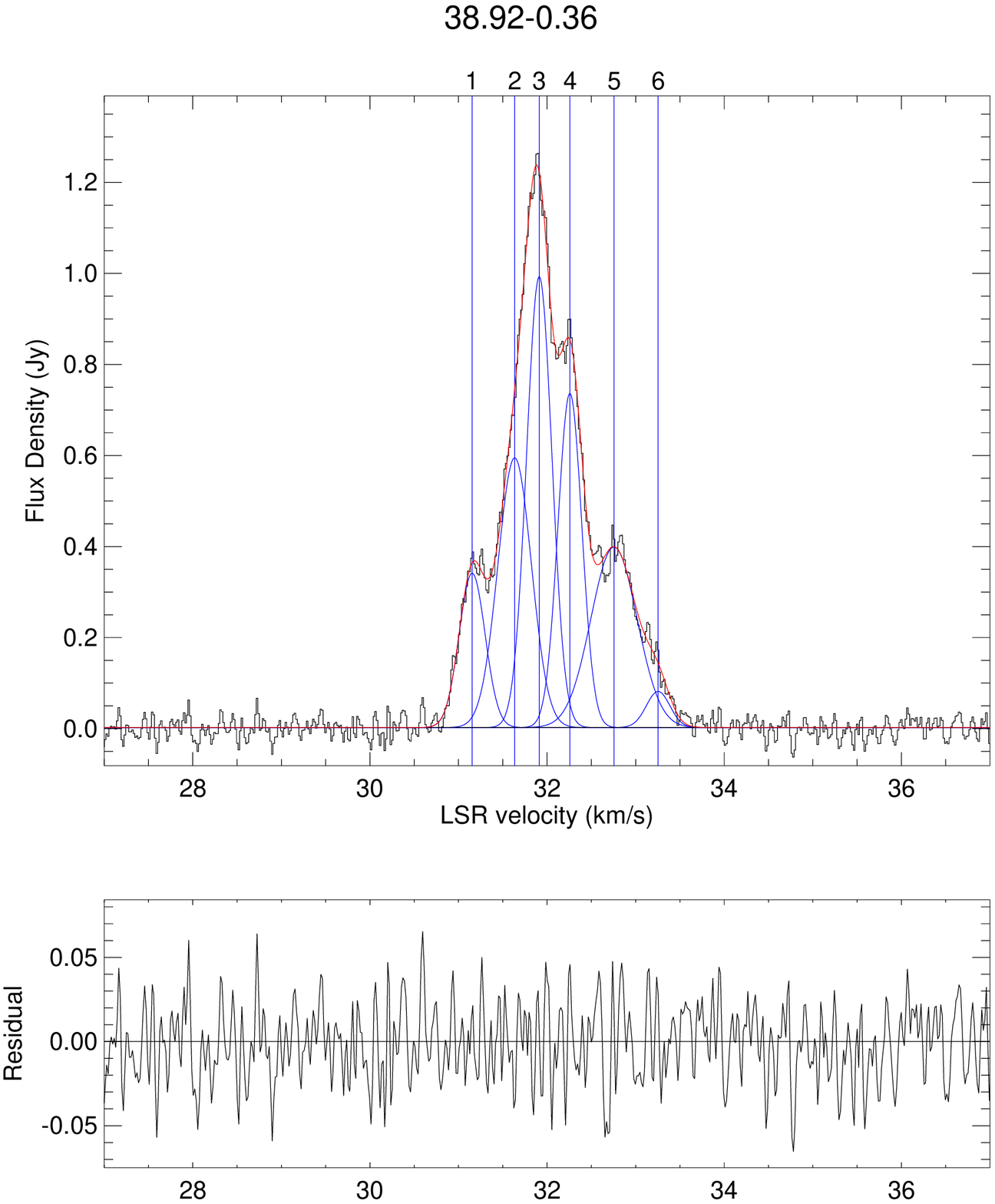}}\\
\centerline{Fig. 8. --- Continued.}
{\includegraphics[width=0.9\textwidth]{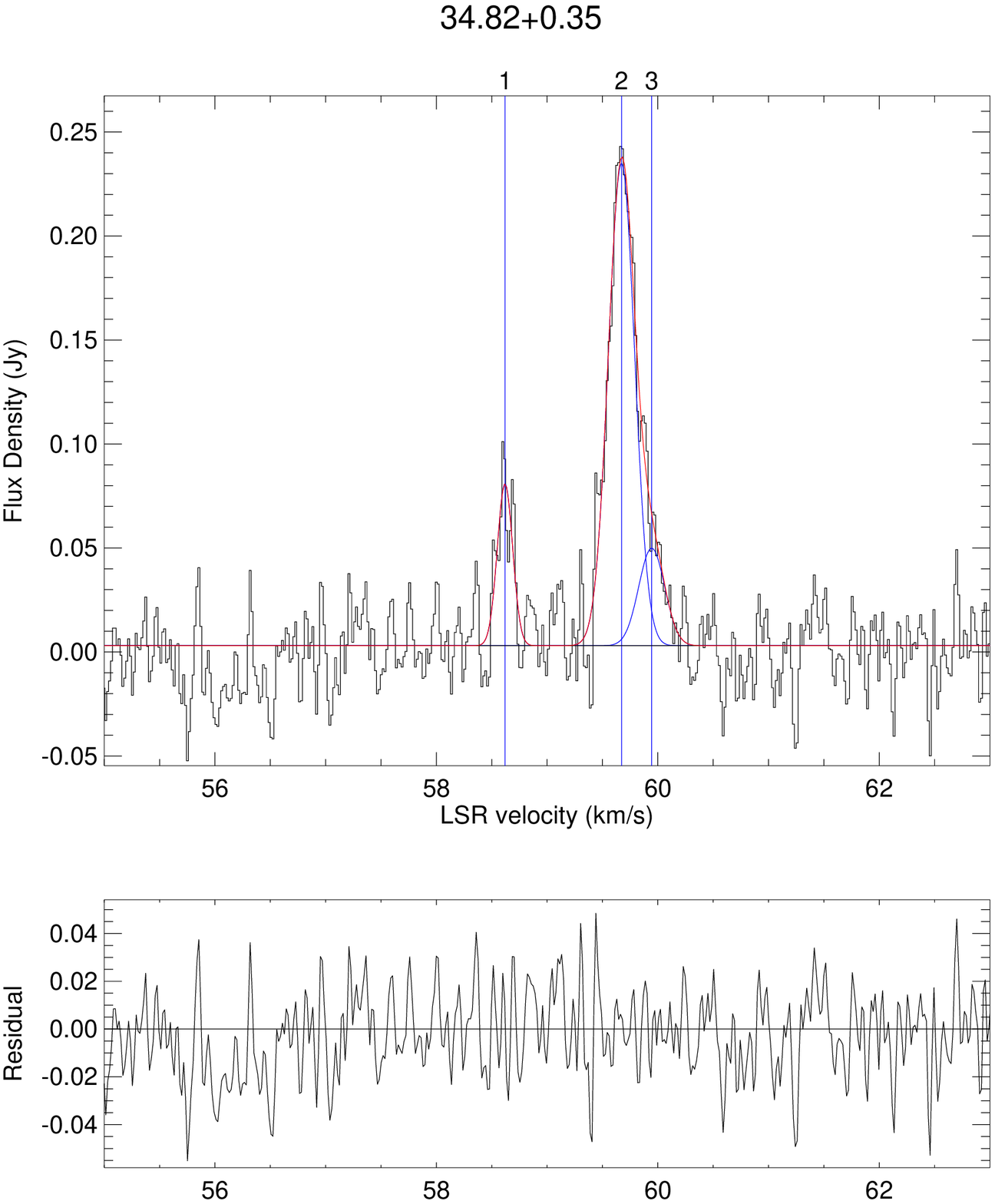}}\\
\centerline{Fig. 8. --- Continued.}
{\includegraphics[width=0.9\textwidth]{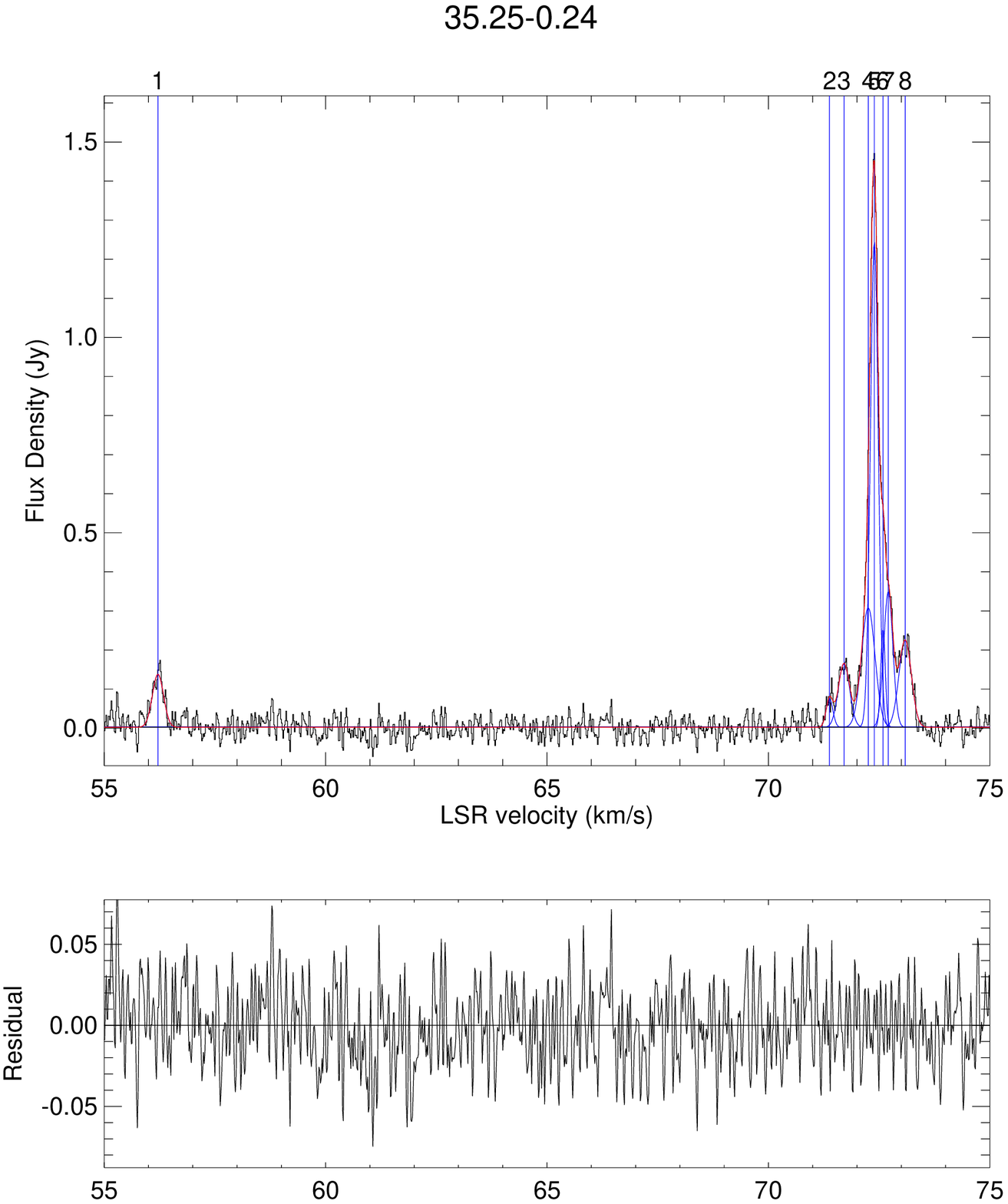}}\\
\centerline{Fig. 8. --- Continued.}
{\includegraphics[width=0.9\textwidth]{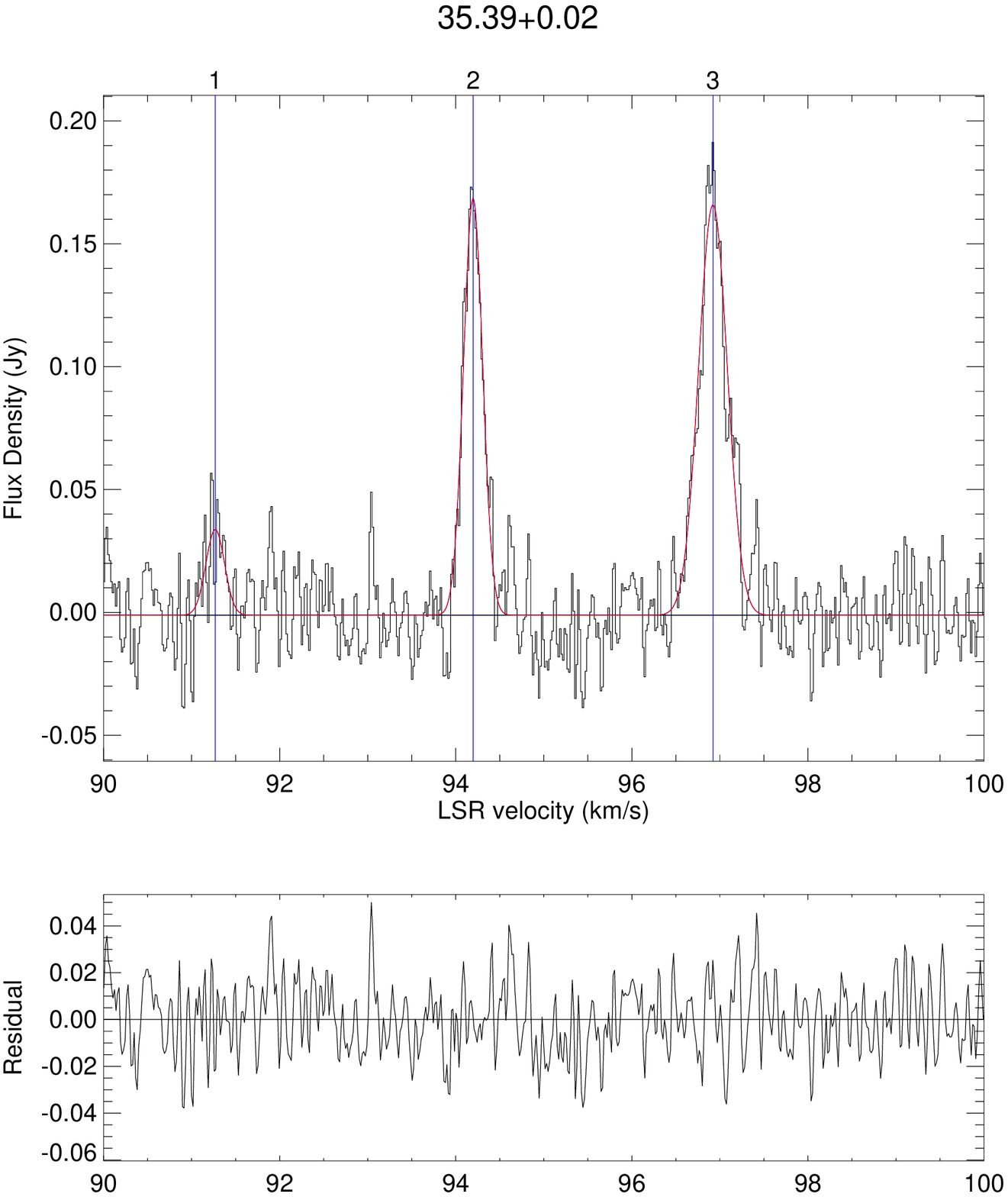}}\\
\centerline{Fig. 8. --- Continued.}
{\includegraphics[width=0.9\textwidth]{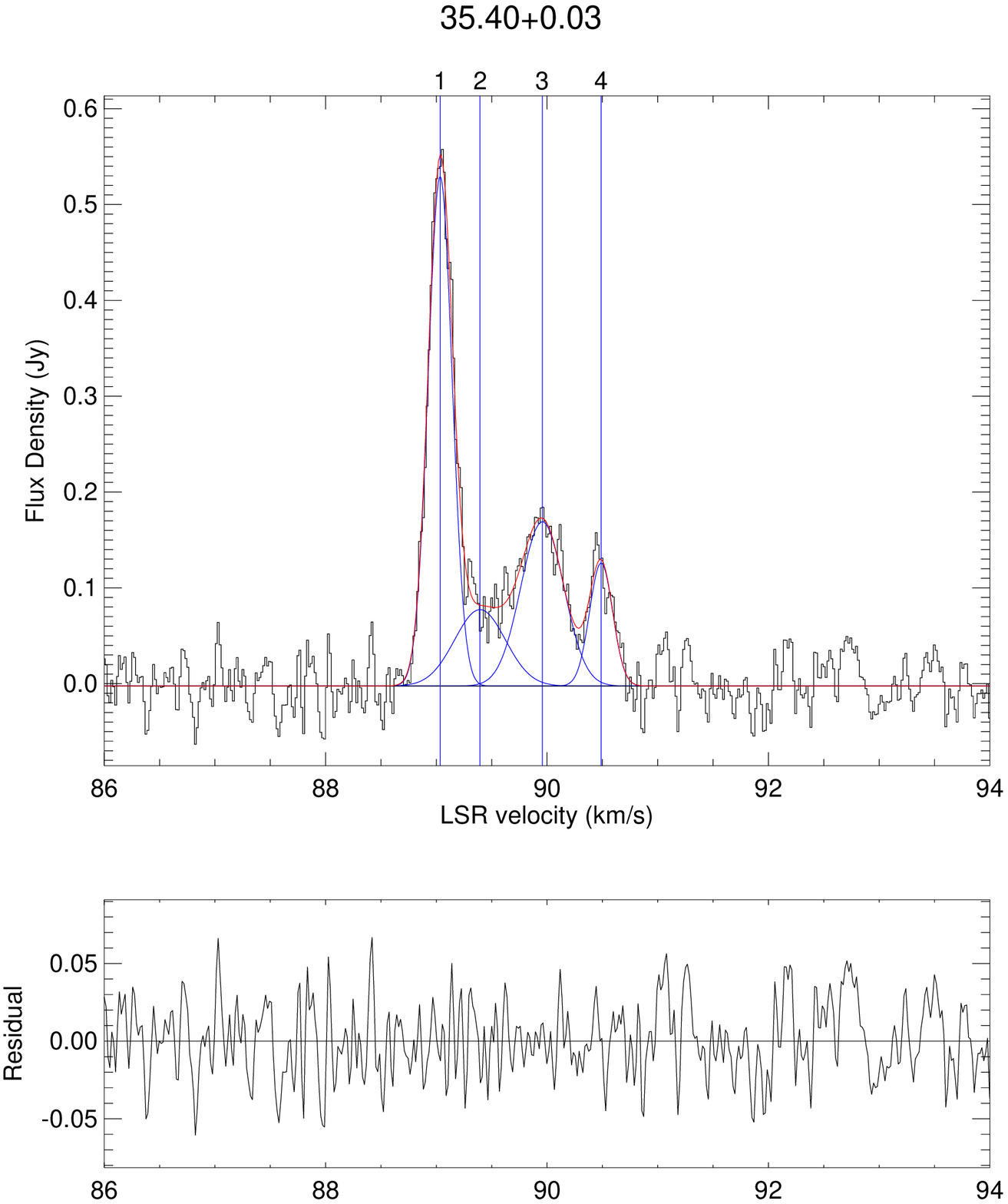}}\\
\centerline{Fig. 8. --- Continued.}
{\includegraphics[width=0.9\textwidth]{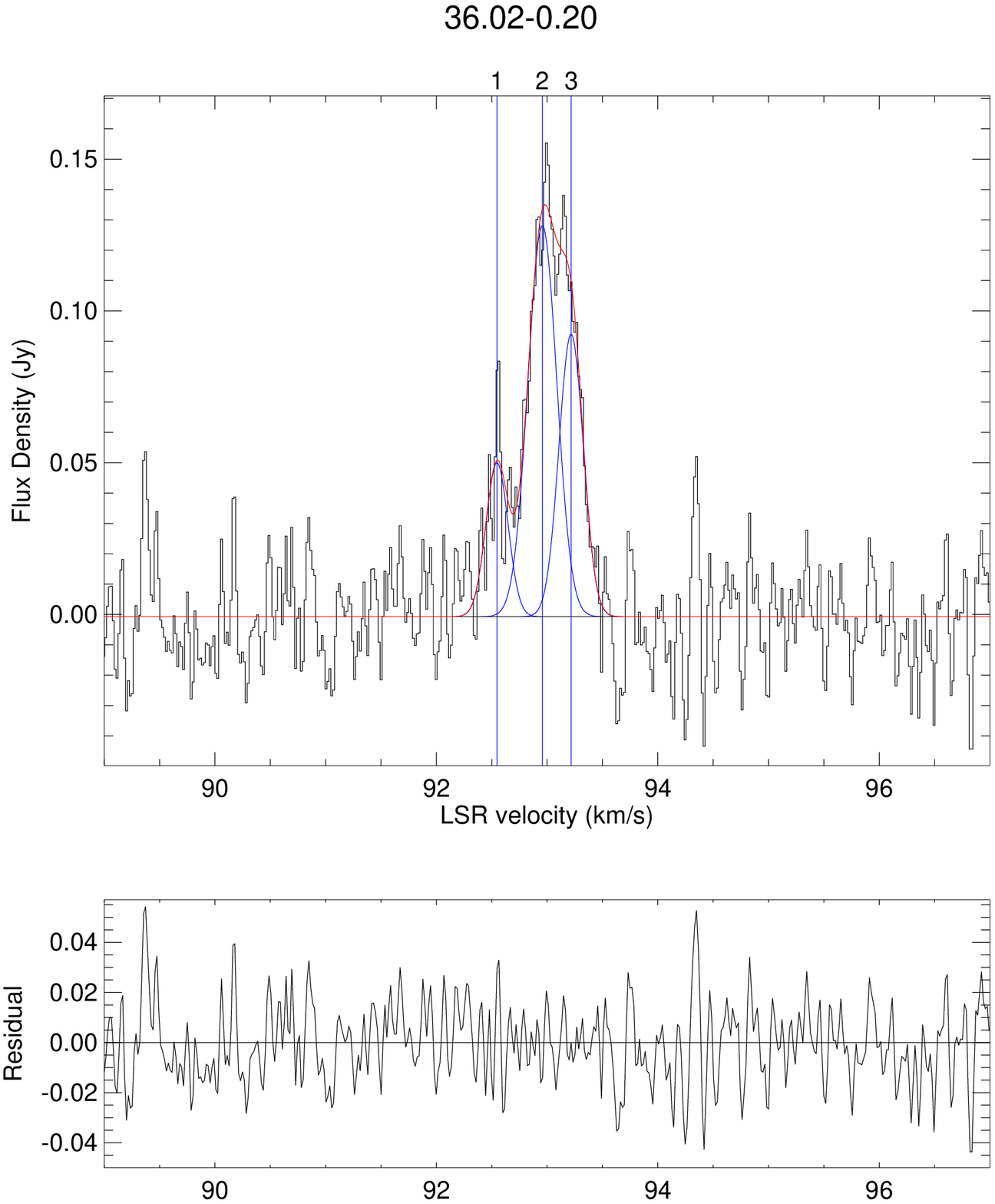}}\\
\centerline{Fig. 8. --- Continued.}
{\includegraphics[width=0.9\textwidth]{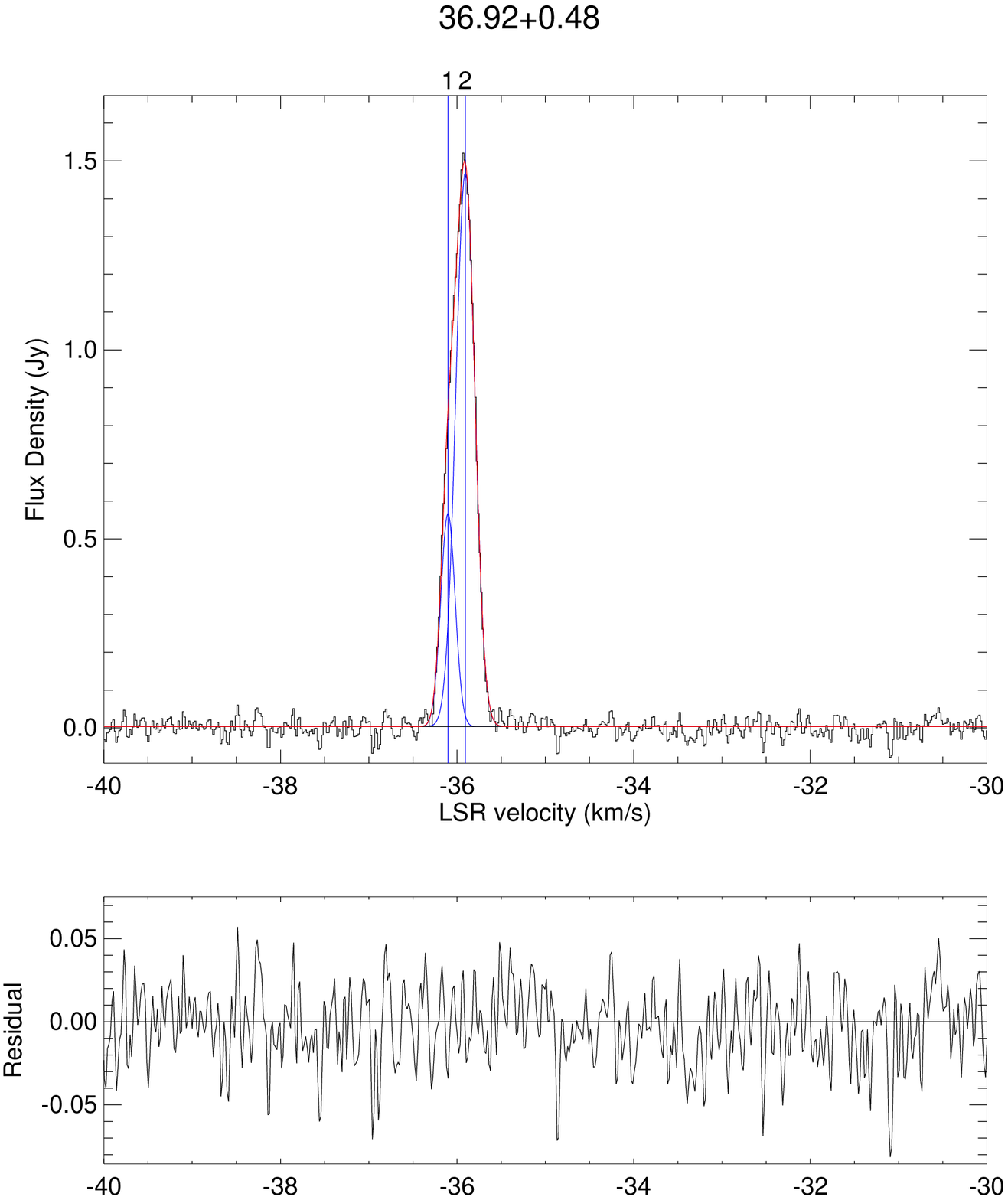}}\\
\centerline{Fig. 8. --- Continued.}
{\includegraphics[width=0.9\textwidth]{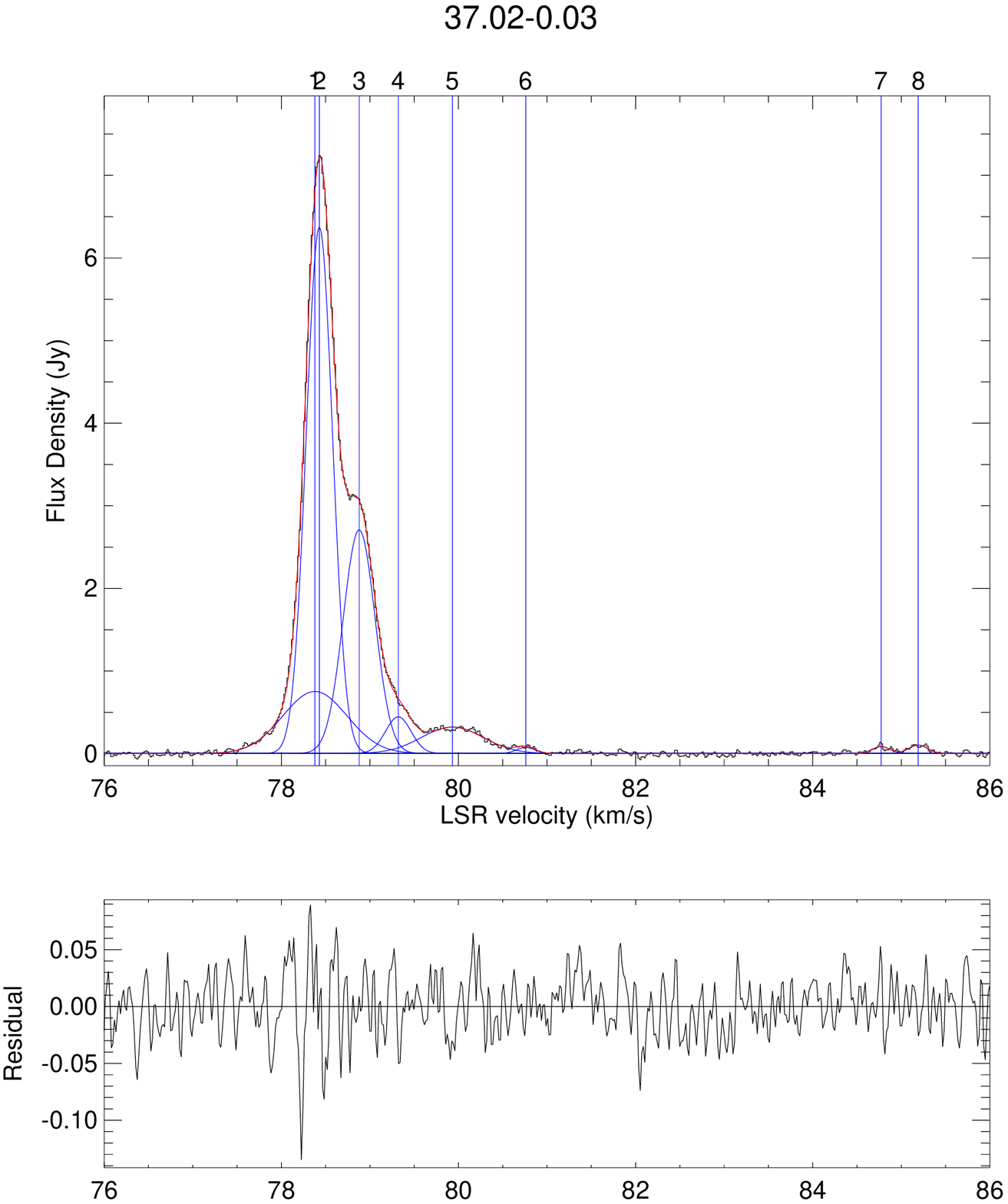}}\\
\centerline{Fig. 8. --- Continued.}
{\includegraphics[width=0.9\textwidth]{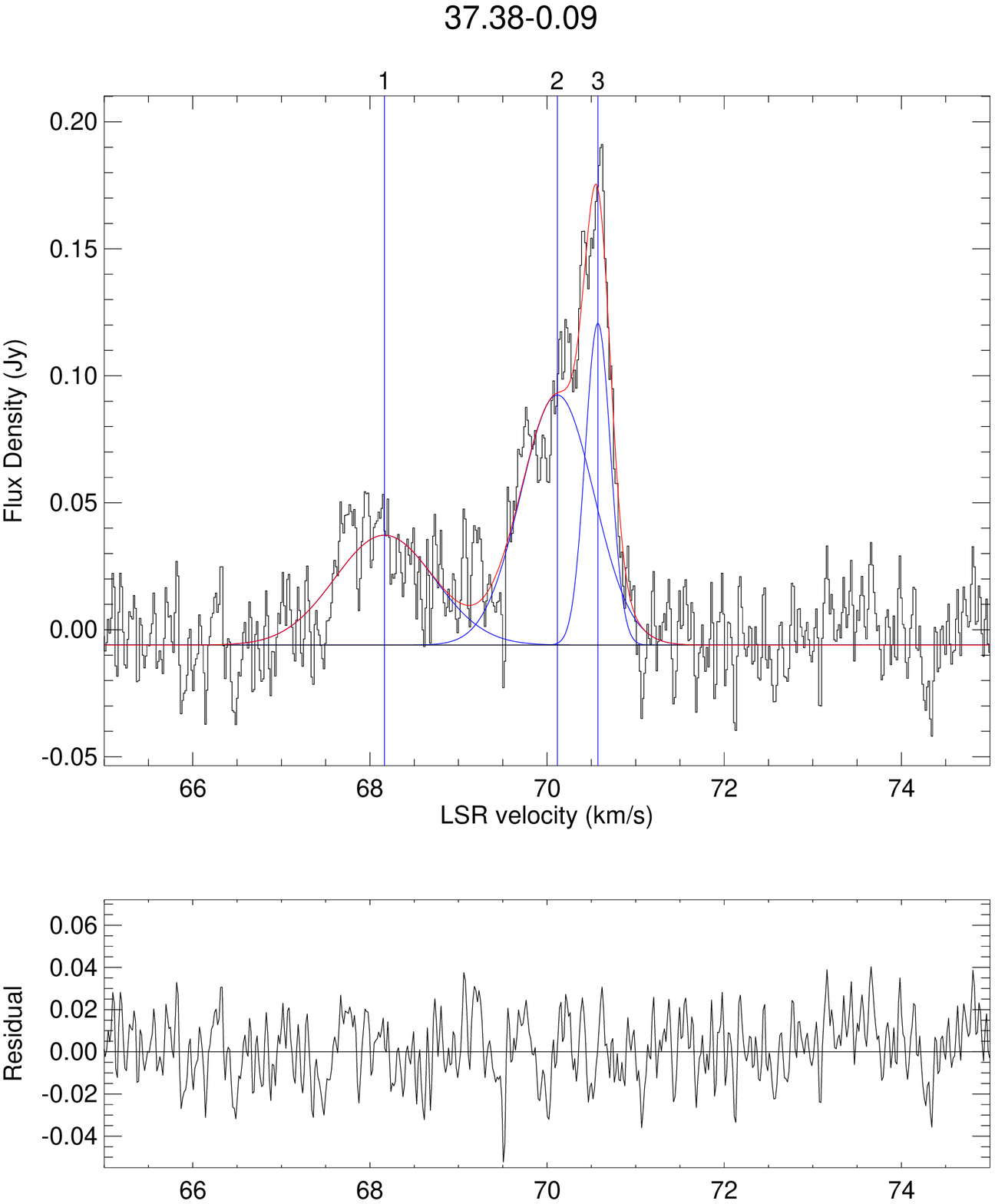}}\\
\centerline{Fig. 8. --- Continued.}
{\includegraphics[width=0.9\textwidth]{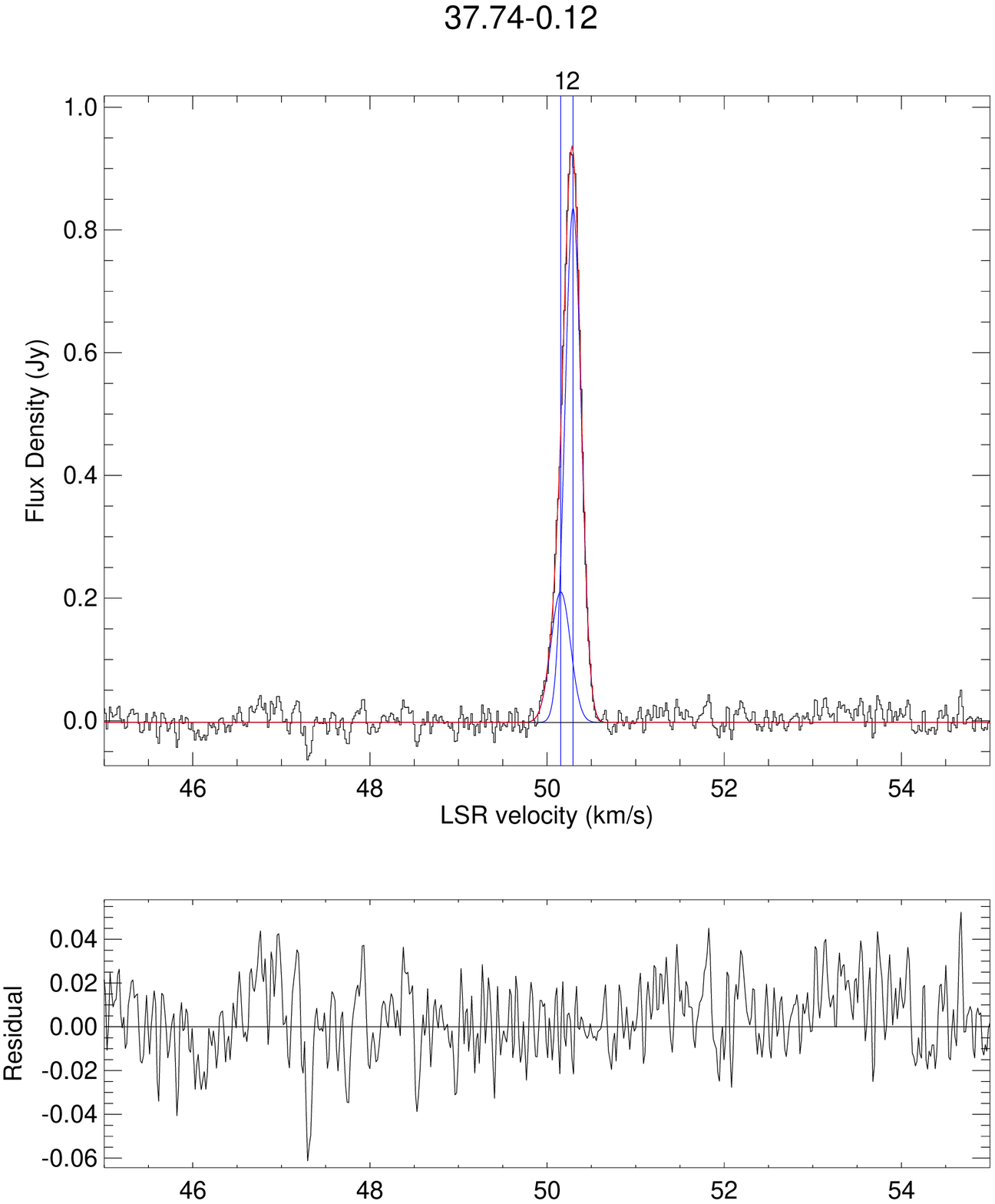}}\\
\centerline{Fig. 8. --- Continued.}
{\includegraphics[width=0.9\textwidth]{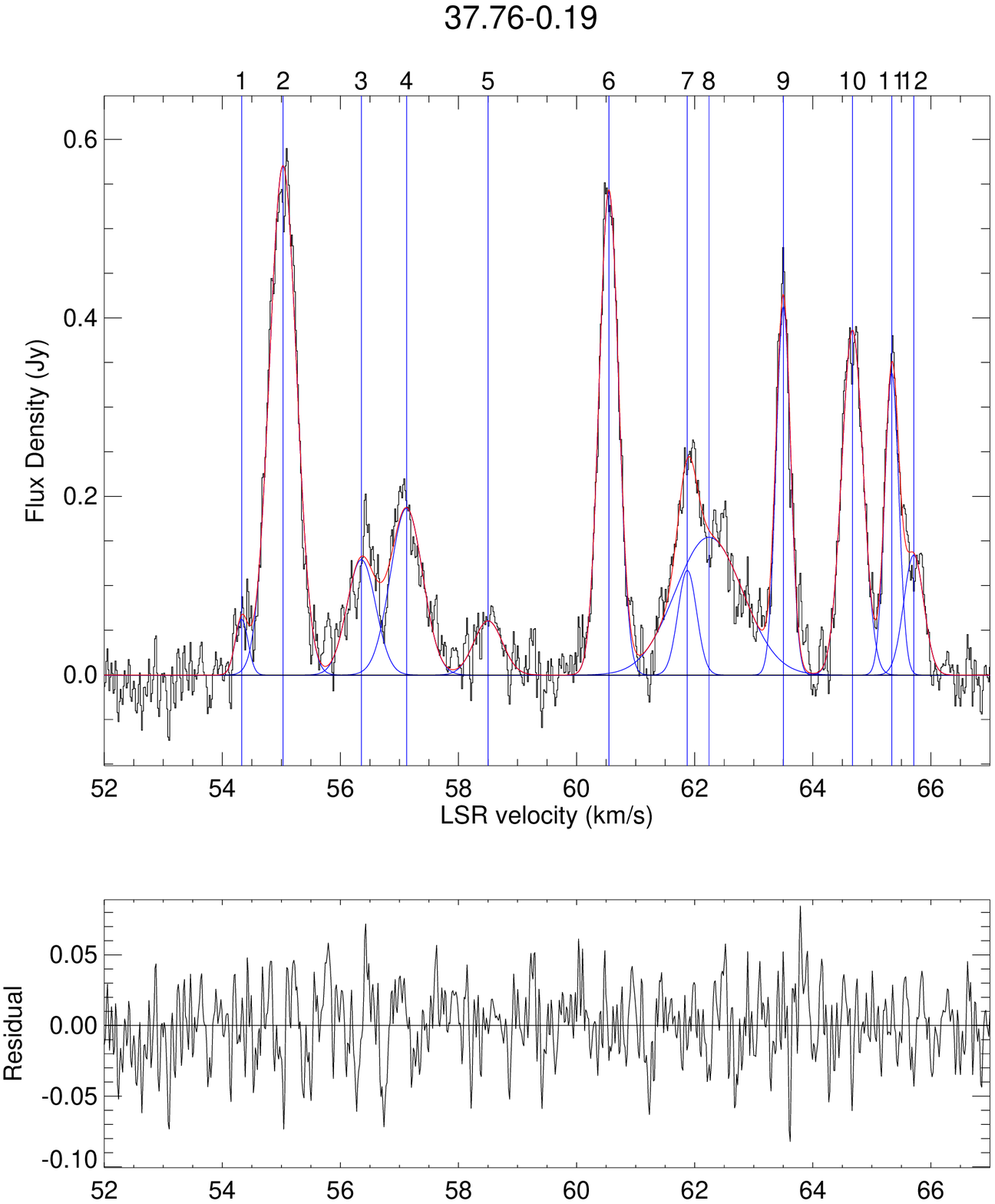}}\\
\centerline{Fig. 8. --- Continued.}
{\includegraphics[width=0.9\textwidth]{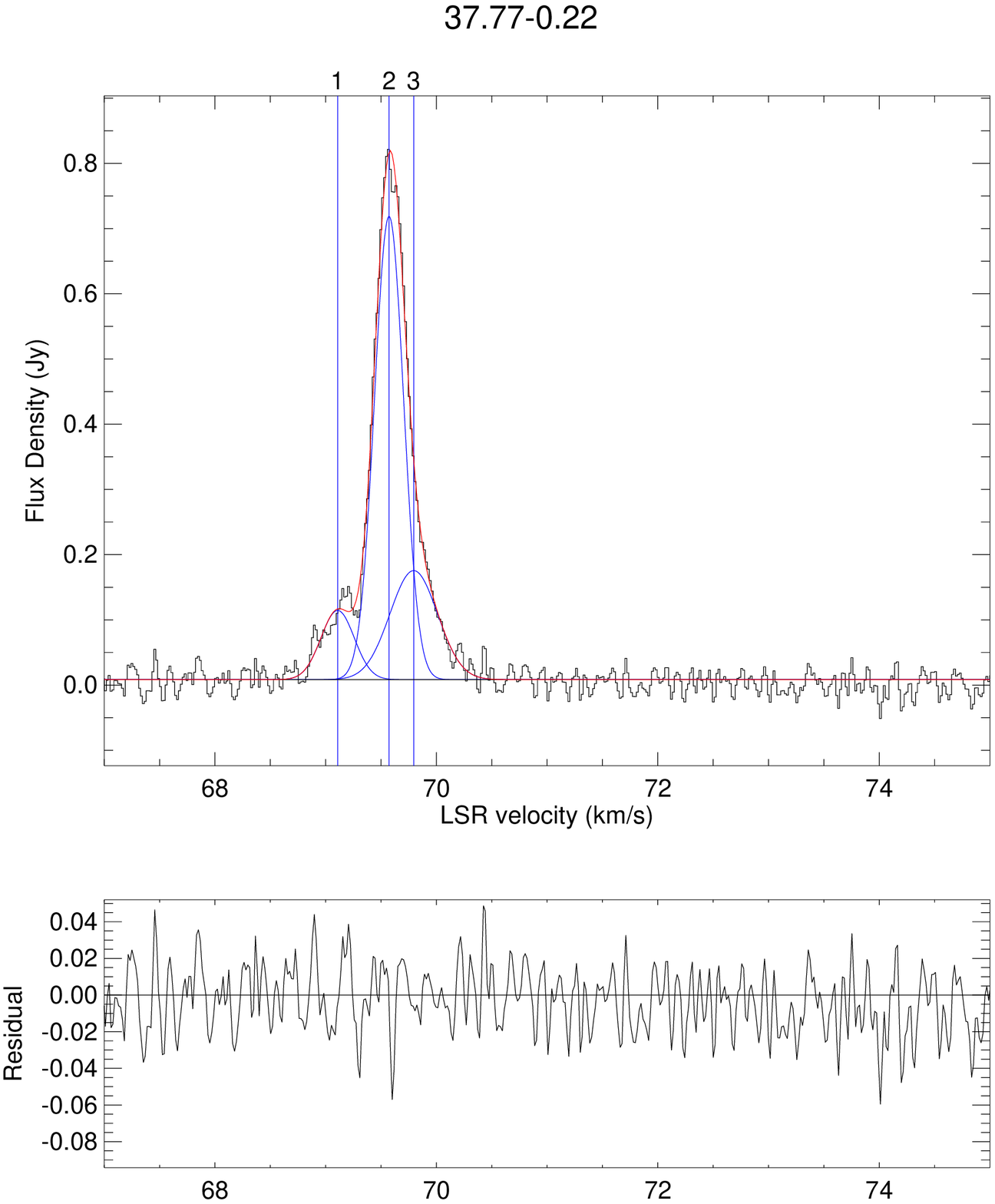}}\\
\centerline{Fig. 8. --- Continued.}
\clearpage
{\includegraphics[width=0.9\textwidth]{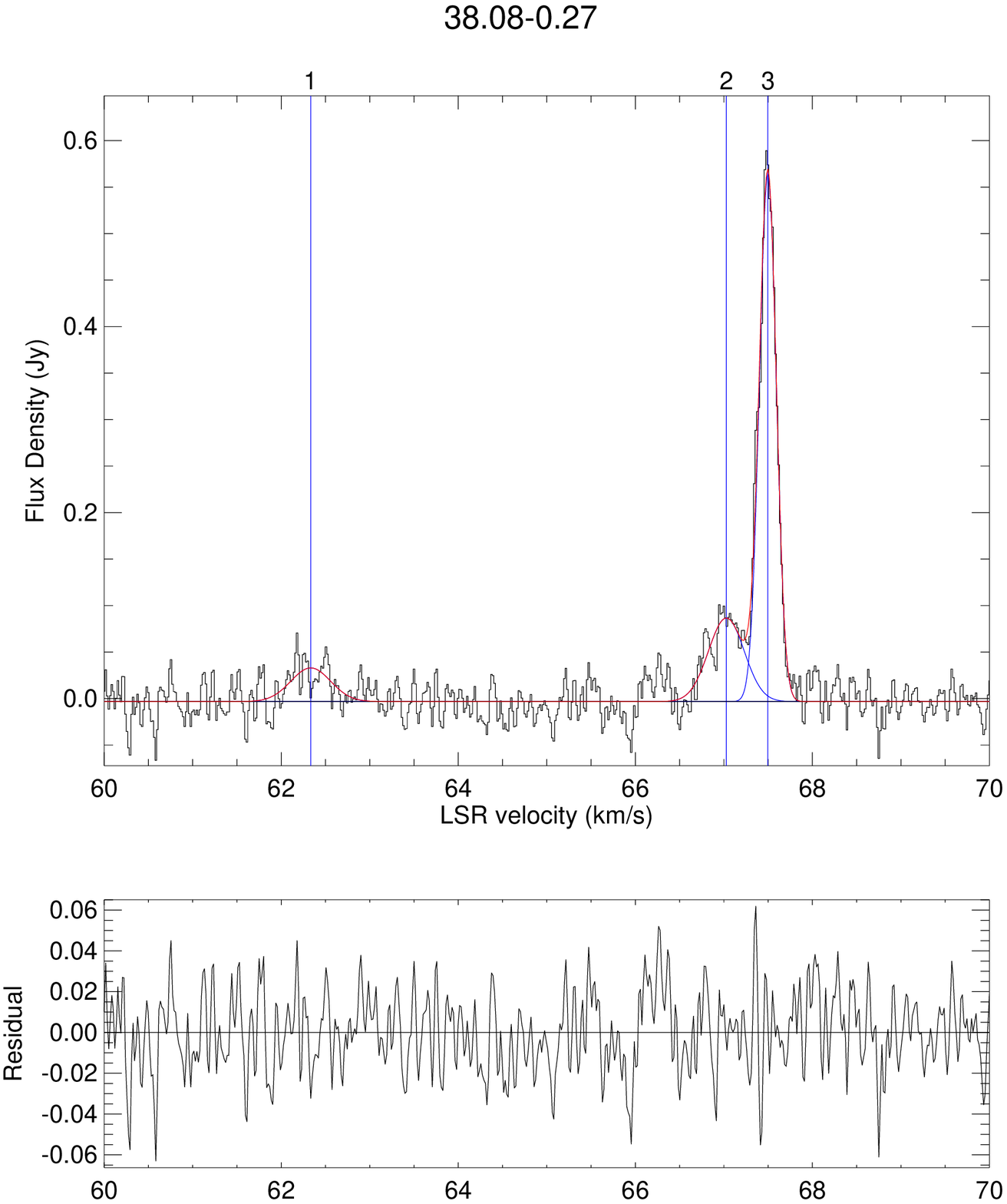}}\\
\centerline{Fig. 8. --- Continued.}
{\includegraphics[width=0.9\textwidth]{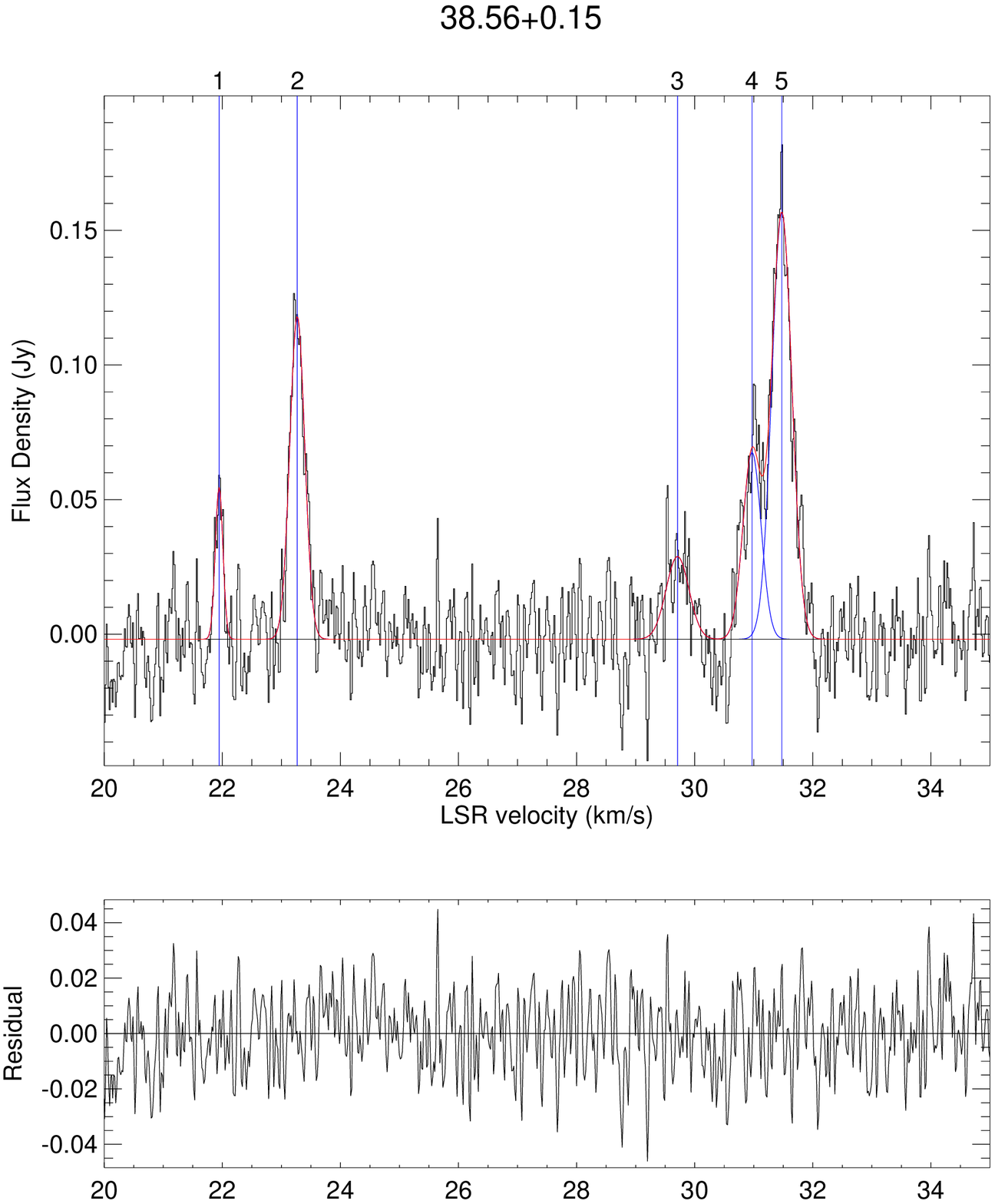}}\\
\centerline{Fig. 8. --- Continued.}
{\includegraphics[width=0.9\textwidth]{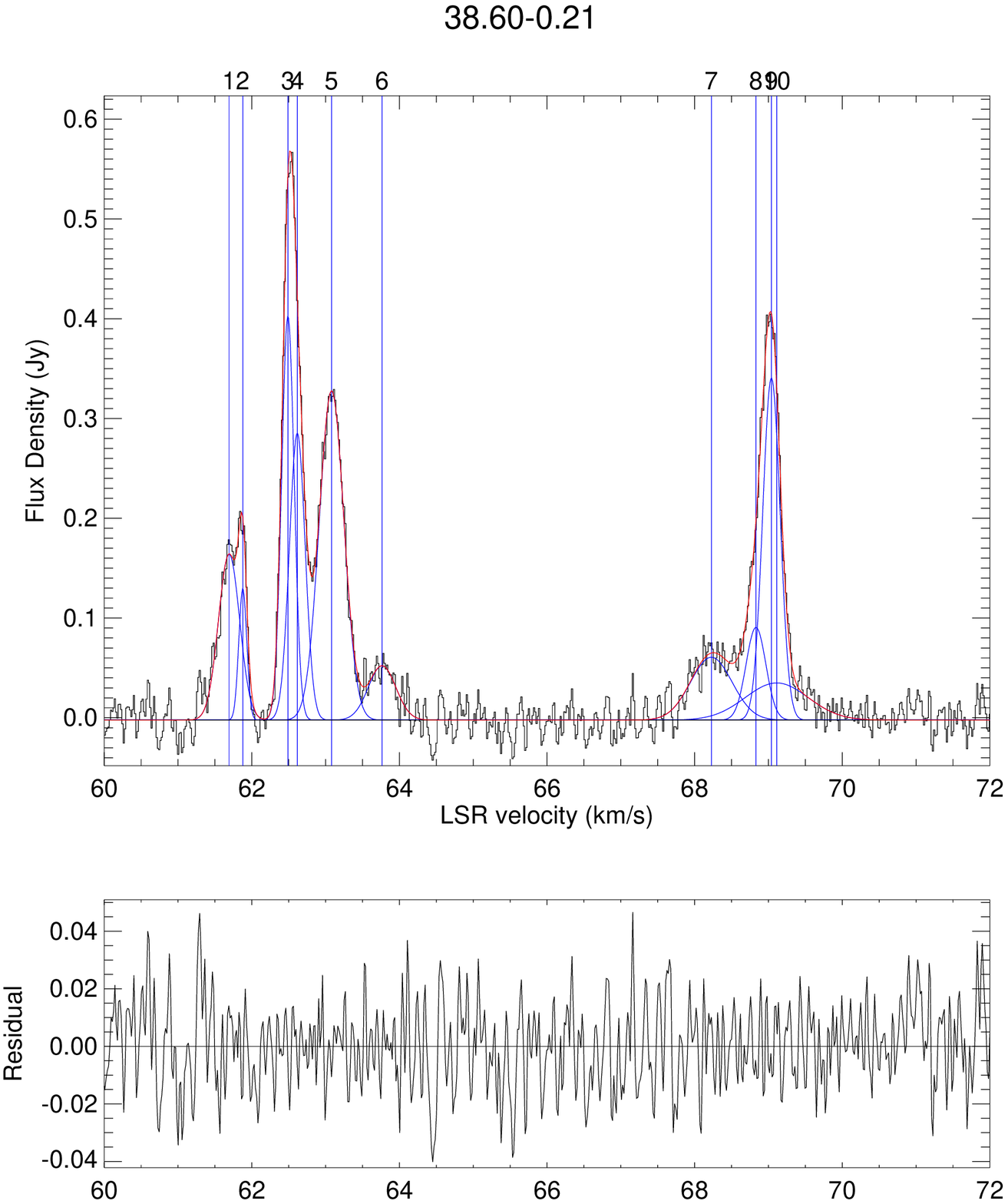}}\\
\centerline{Fig. 8. --- Continued.}
{\includegraphics[width=0.9\textwidth]{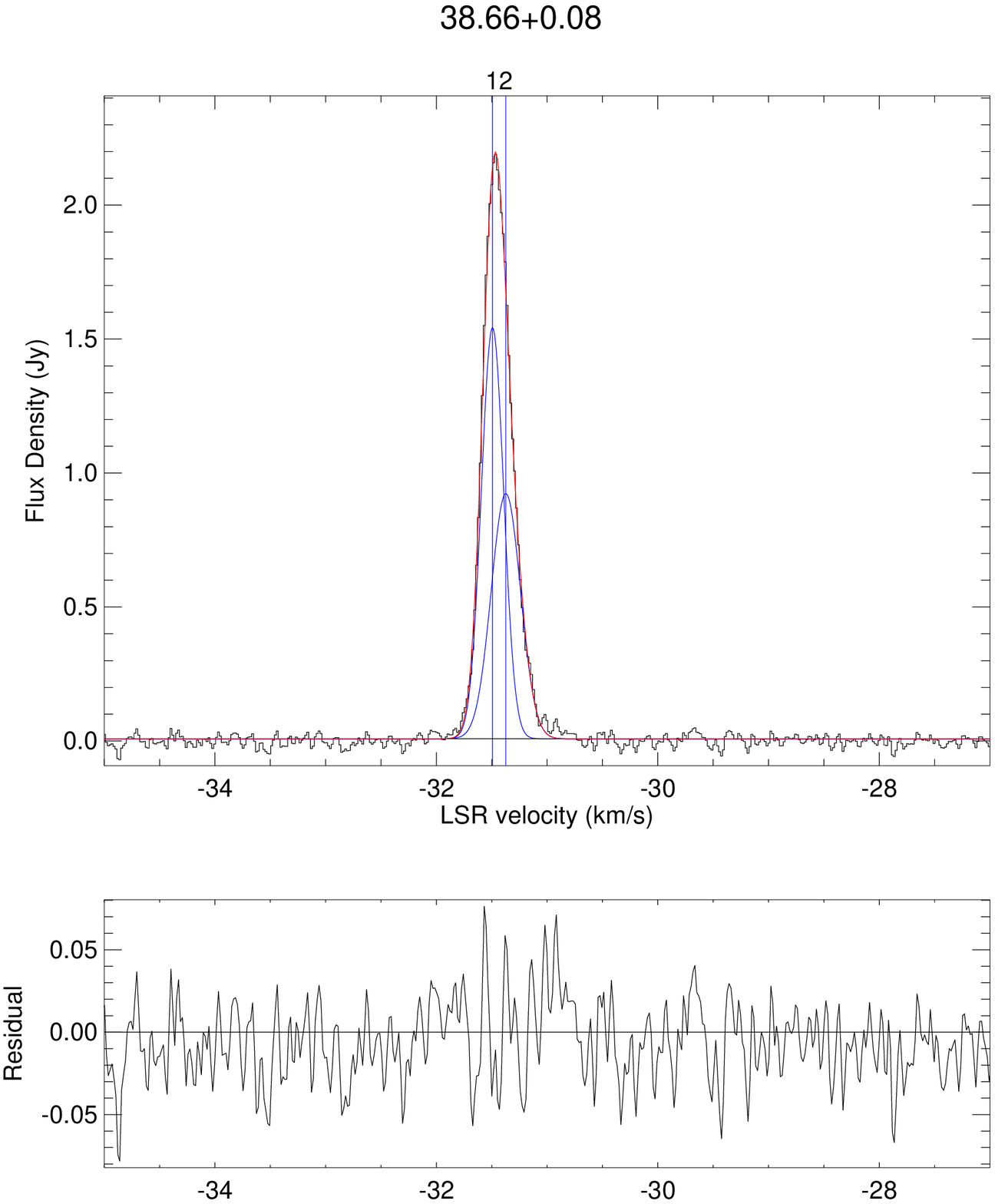}}\\
\centerline{Fig. 8. --- Continued.}
{\includegraphics[width=0.9\textwidth]{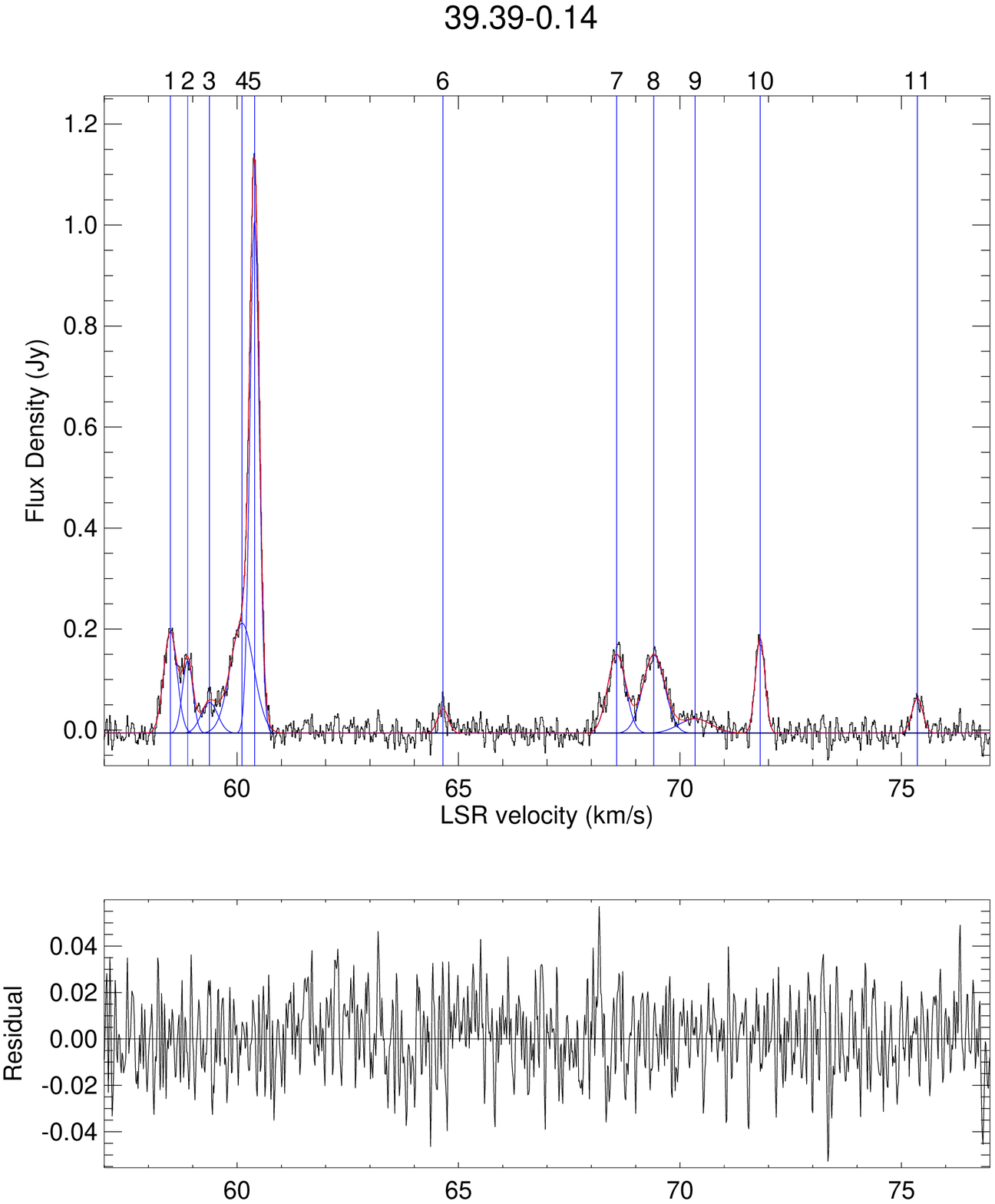}}\\
\centerline{Fig. 8. --- Continued.}
{\includegraphics[width=0.9\textwidth]{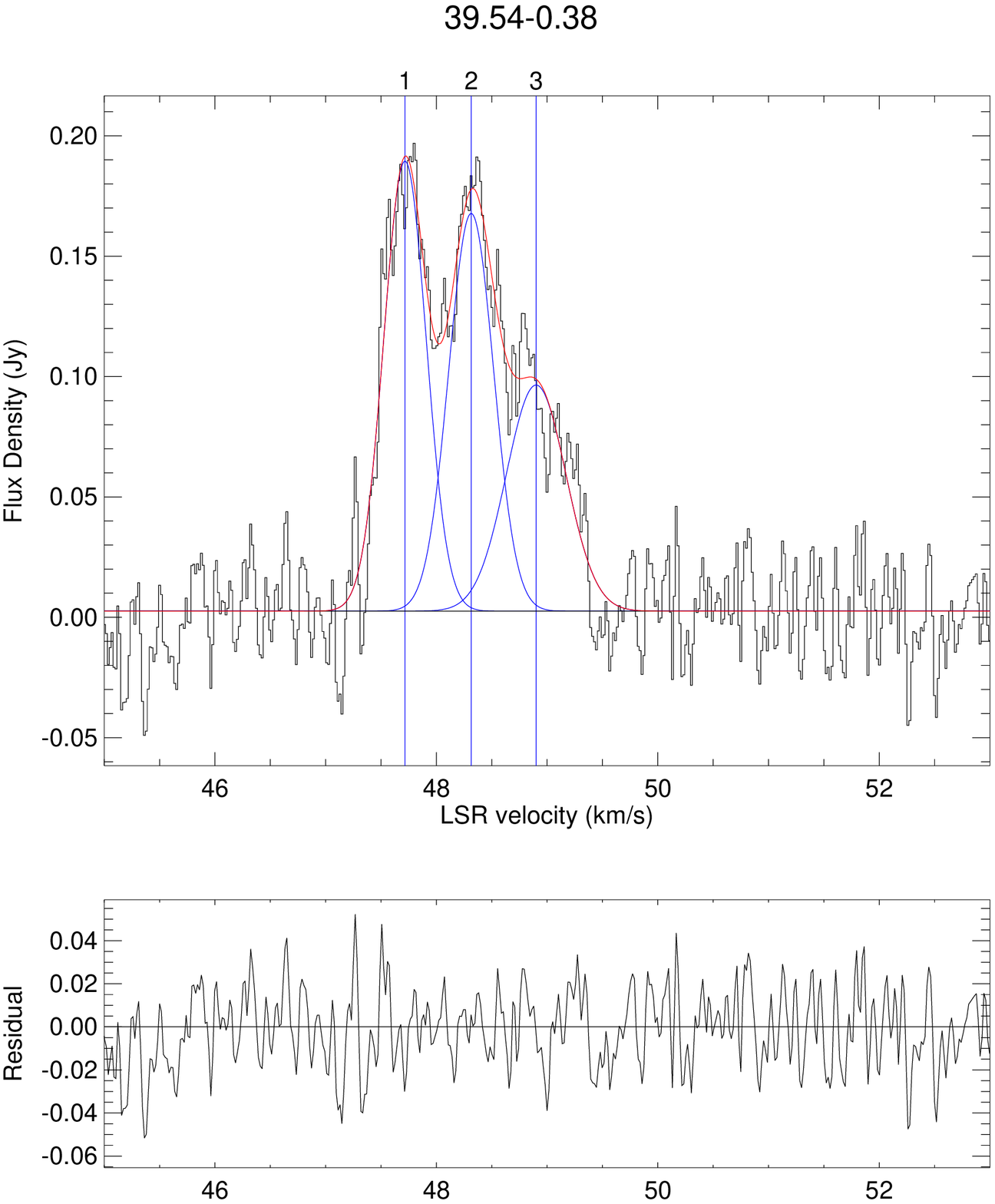}}\\
\centerline{Fig. 8. --- Continued.}
{\includegraphics[width=0.9\textwidth]{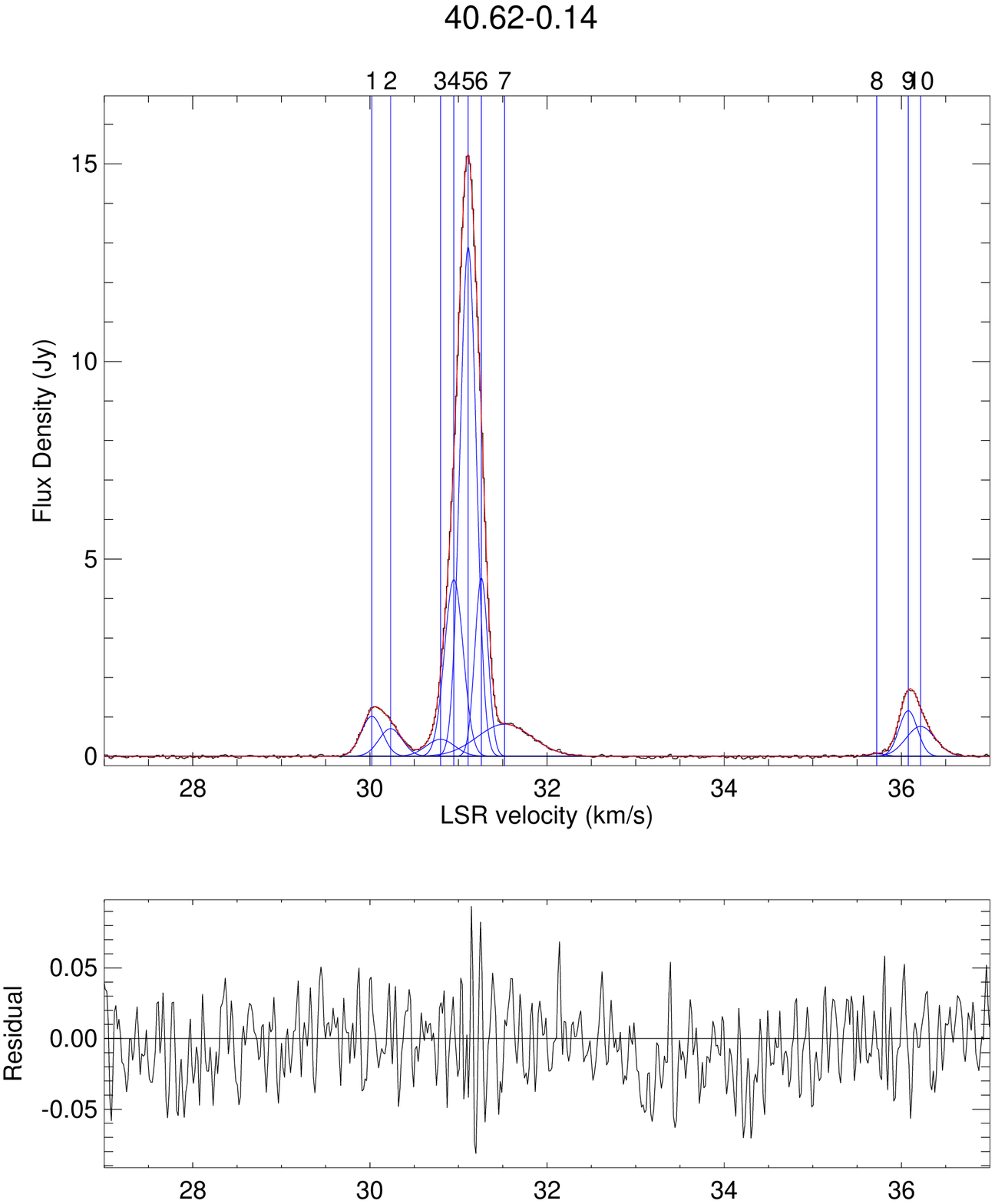}}\\
\centerline{Fig. 8. --- Continued.}
{\includegraphics[width=0.9\textwidth]{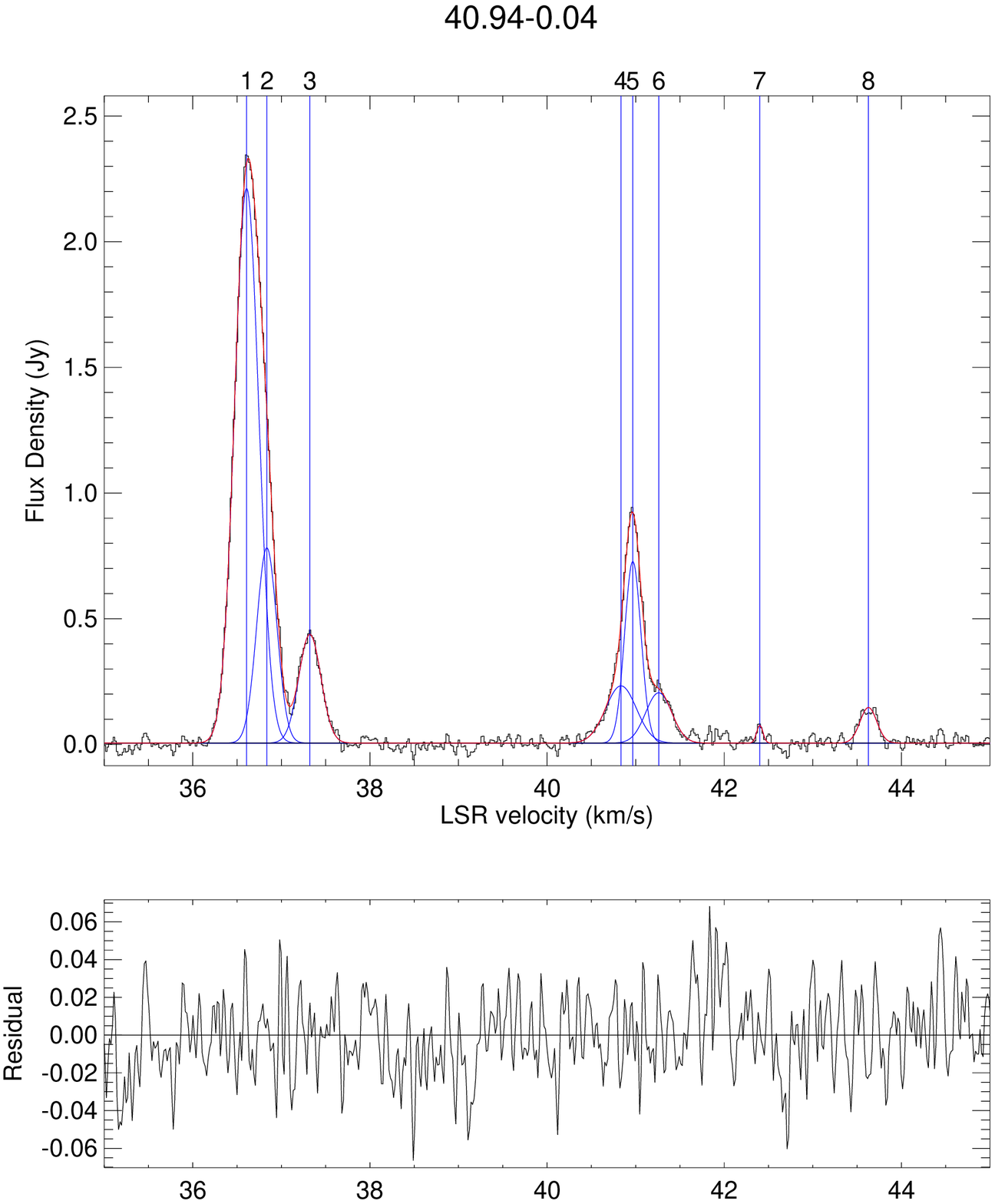}}\\
\centerline{Fig. 8. --- Continued.}
{\includegraphics[width=0.9\textwidth]{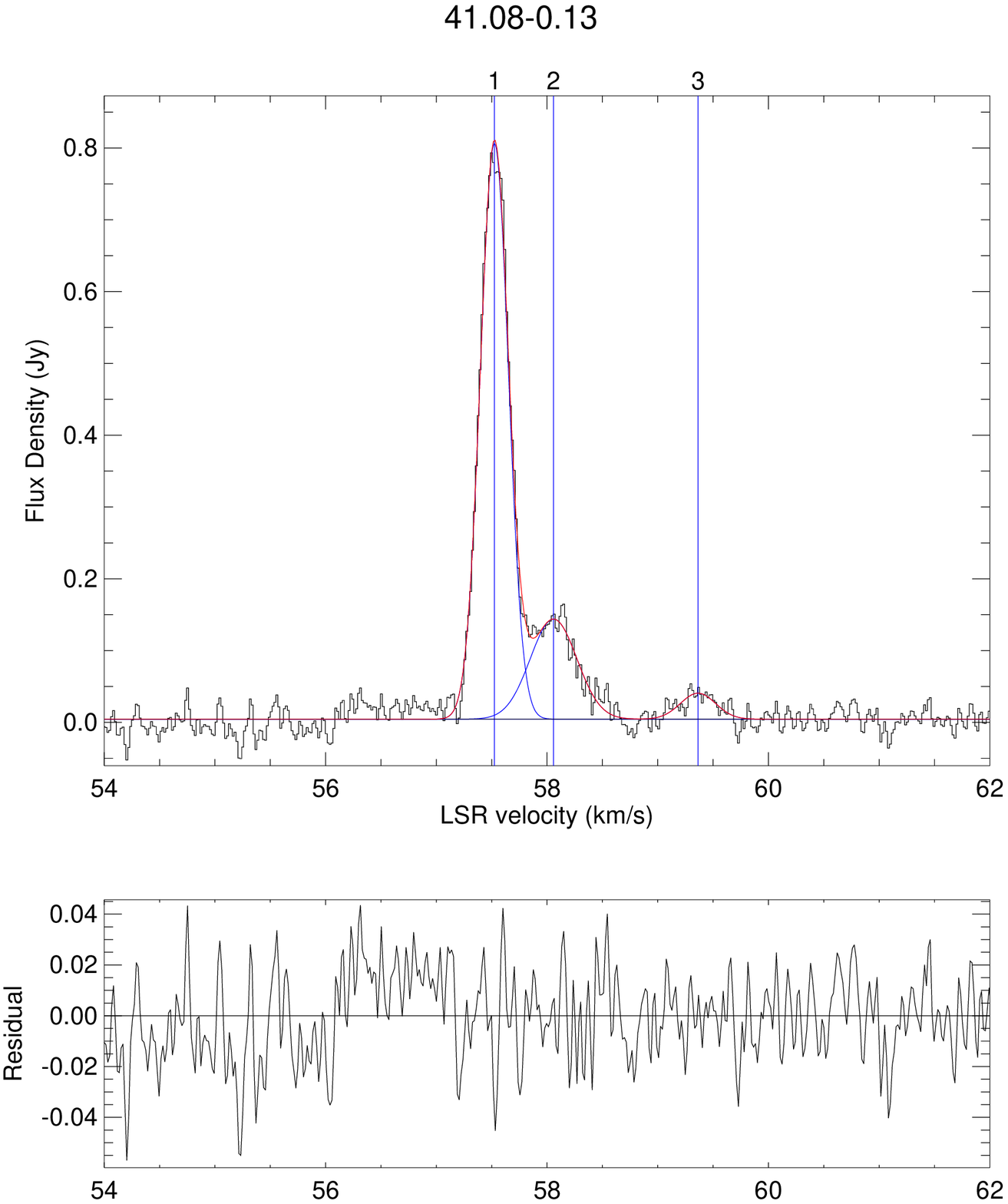}}\\
\centerline{Fig. 8. --- Continued.}
{\includegraphics[width=0.9\textwidth]{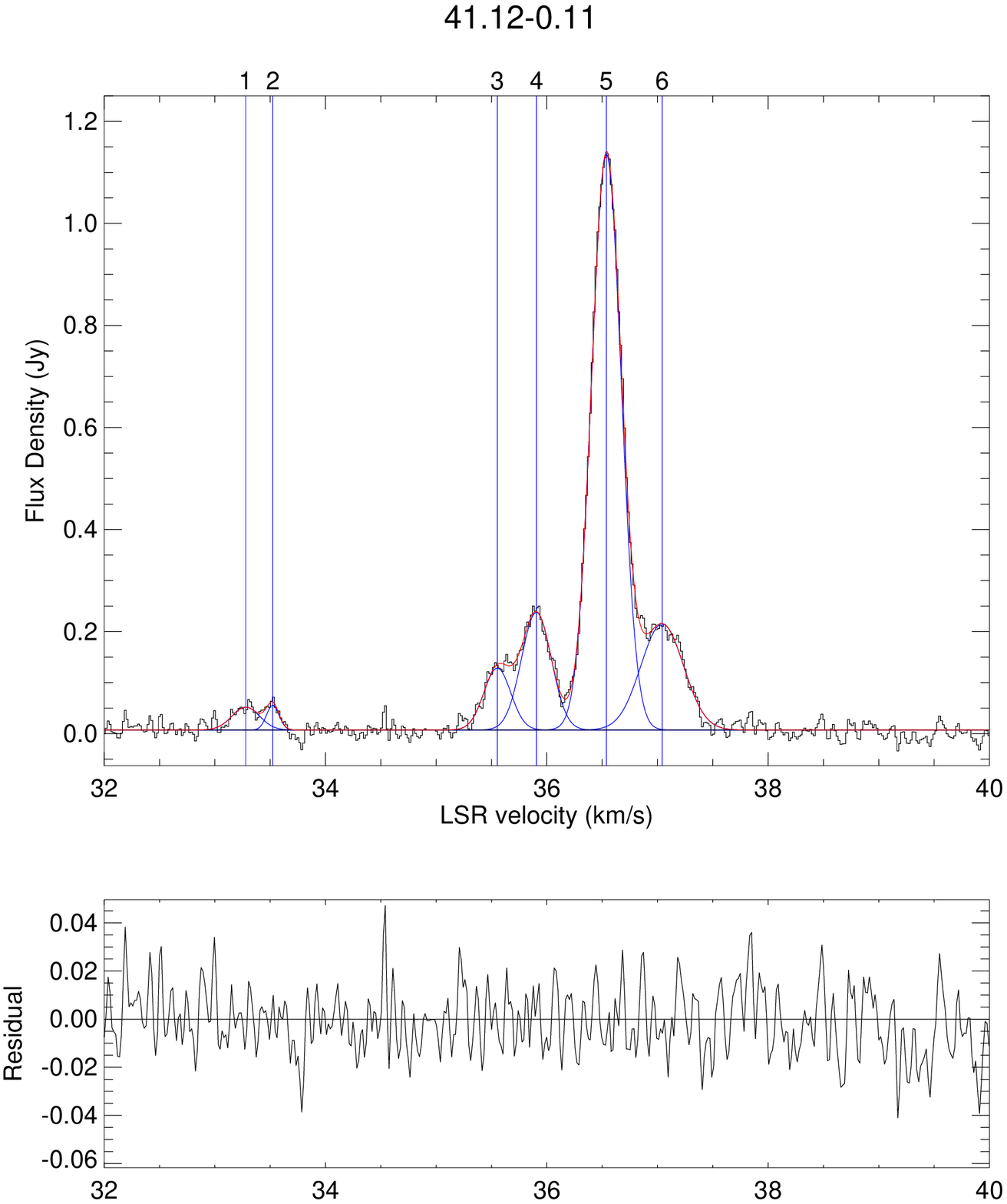}}\\
\centerline{Fig. 8. --- Continued.}
\clearpage
{\includegraphics[width=0.9\textwidth]{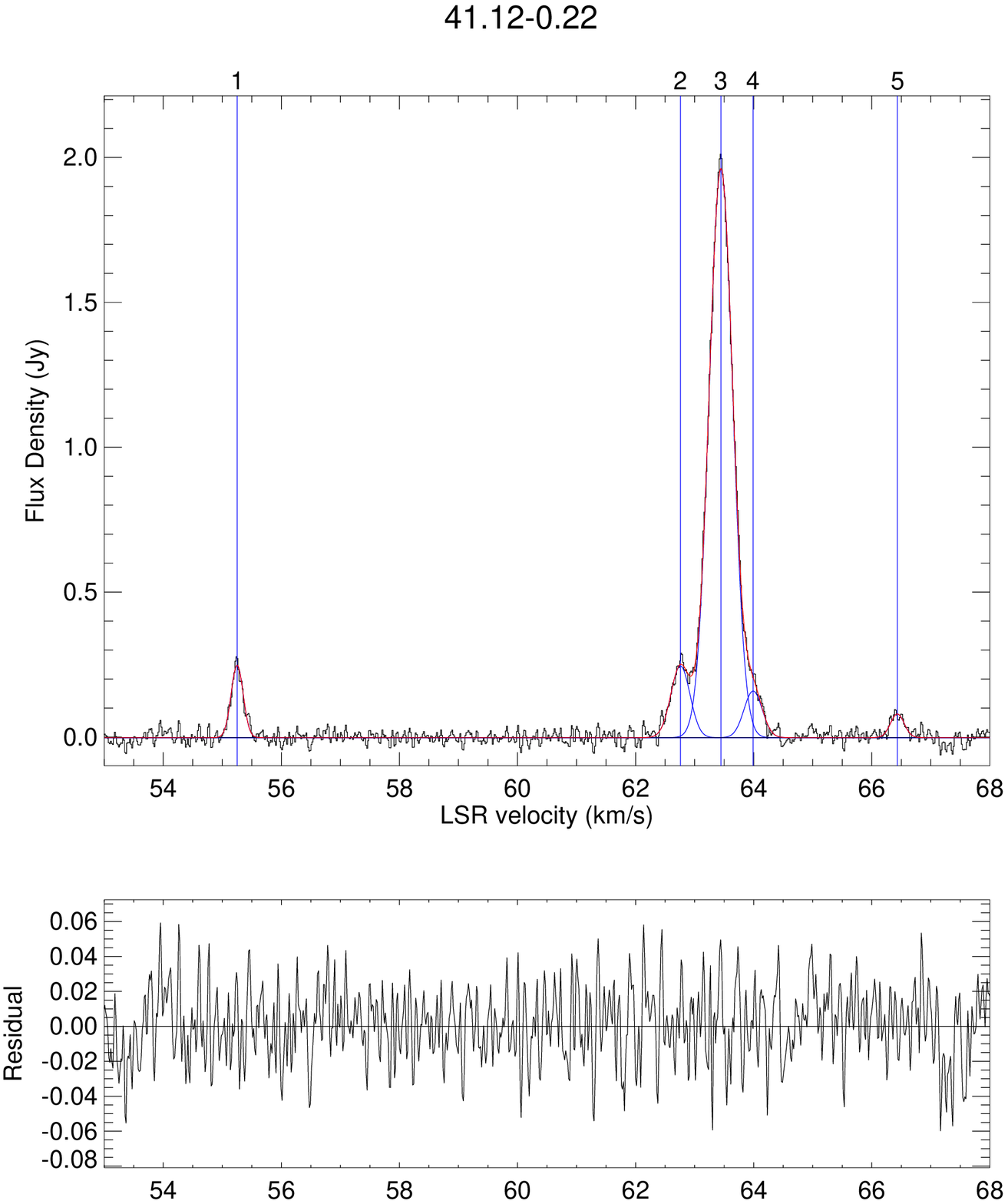}}\\
\centerline{Fig. 8. --- Continued.}
{\includegraphics[width=0.9\textwidth]{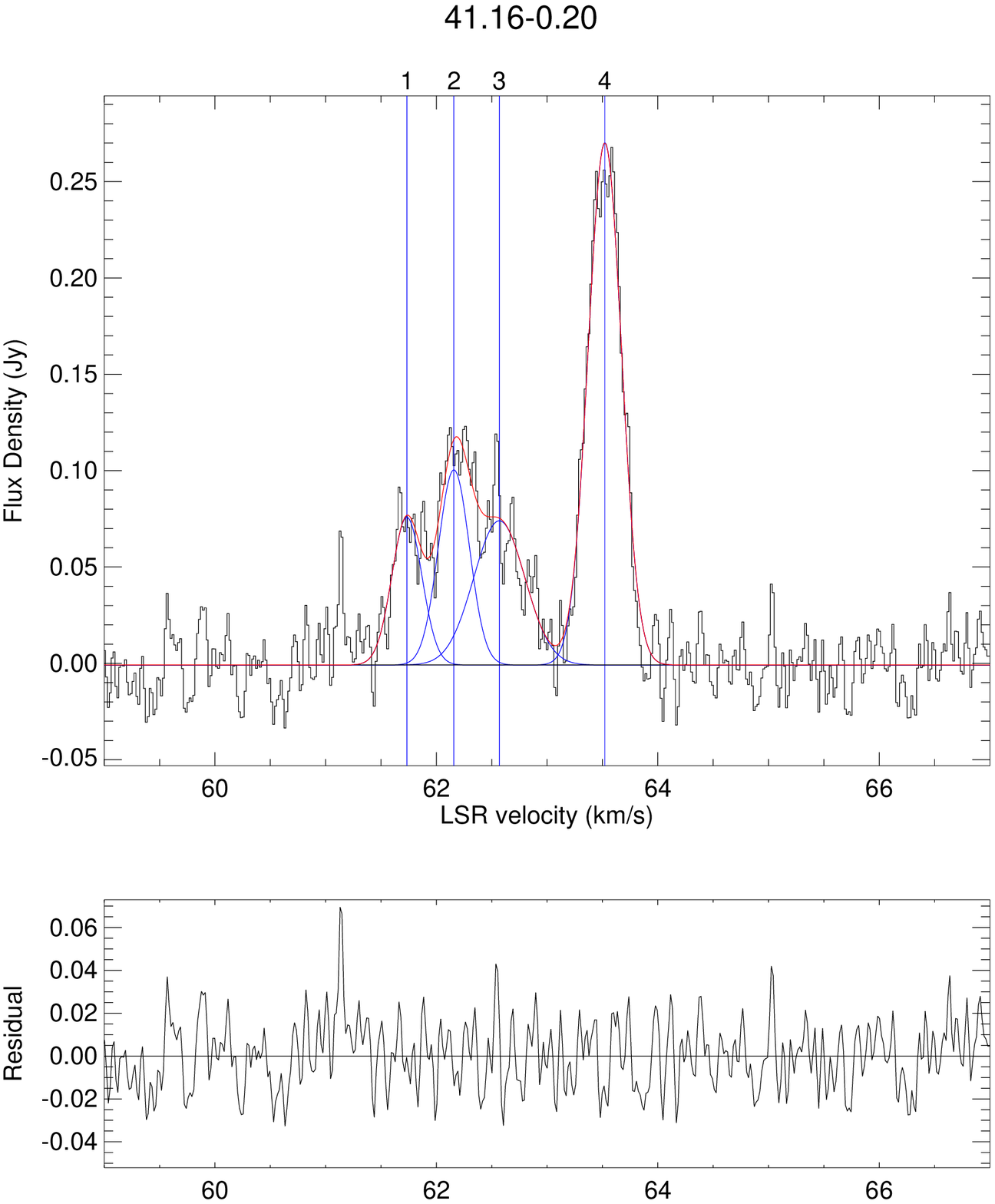}}\\
\centerline{Fig. 8. --- Continued.}
{\includegraphics[width=0.9\textwidth]{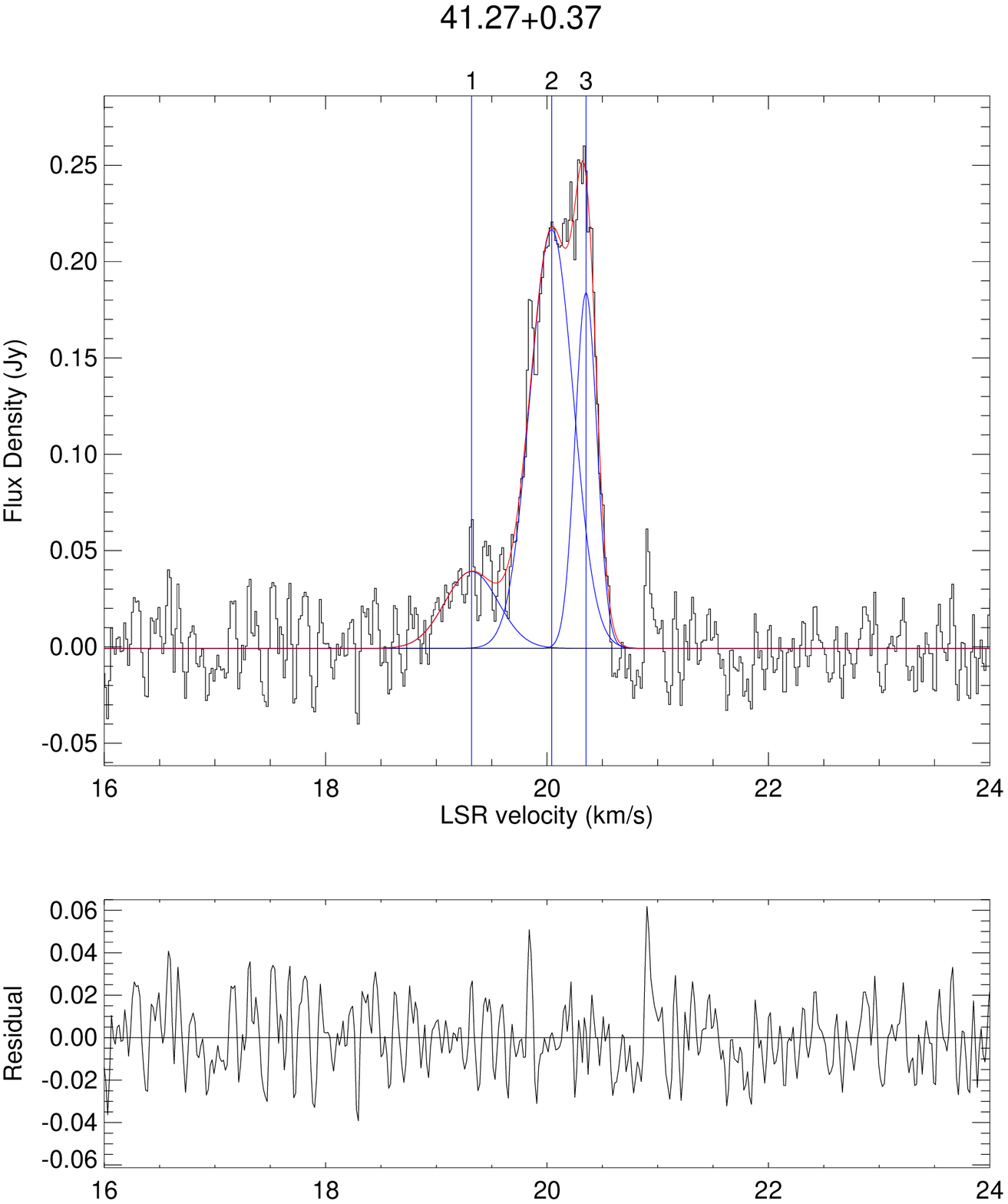}}\\
\centerline{Fig. 8. --- Continued.}
{\includegraphics[width=0.9\textwidth]{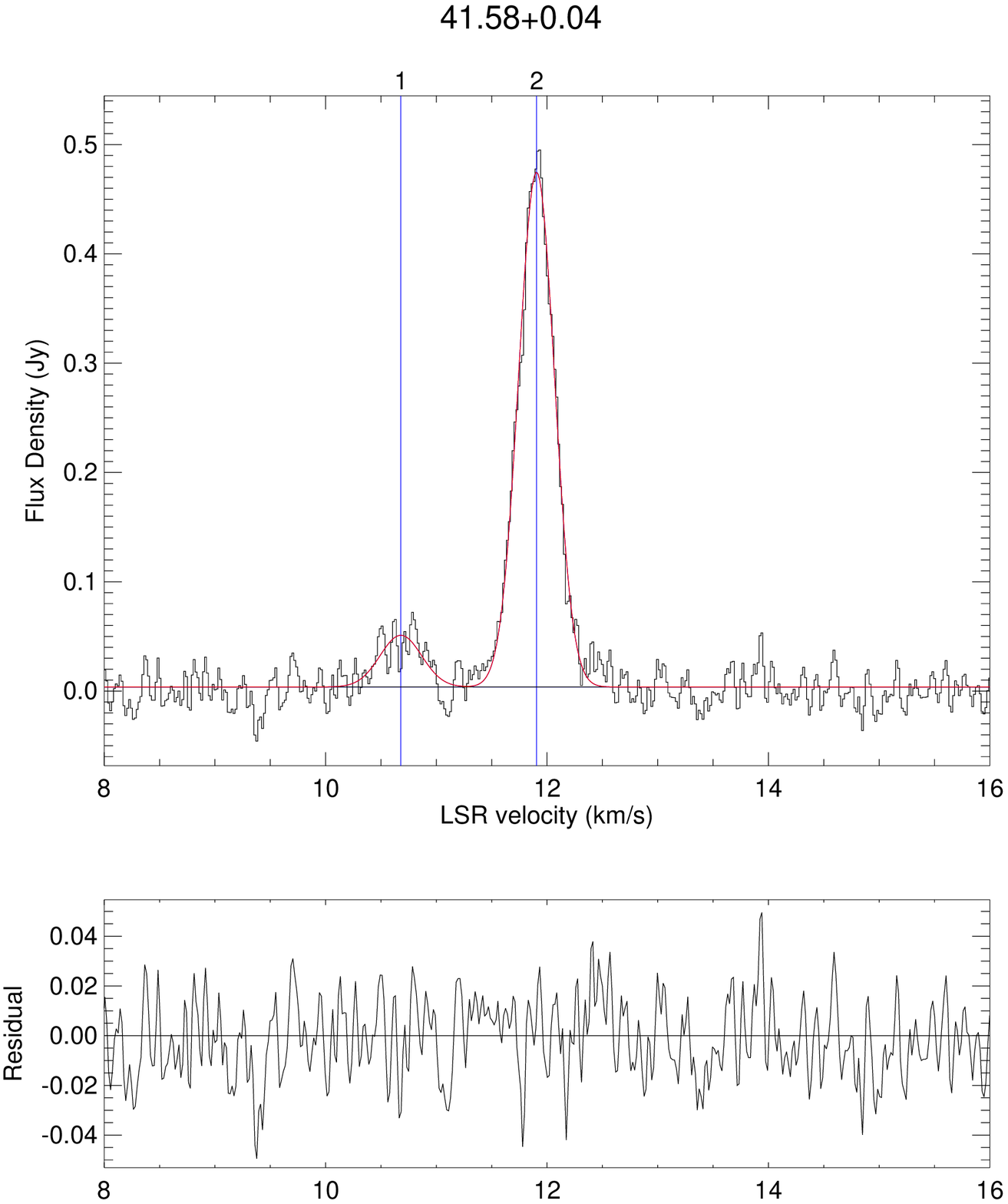}}\\
\centerline{Fig. 8. --- Continued.}
{\includegraphics[width=0.9\textwidth]{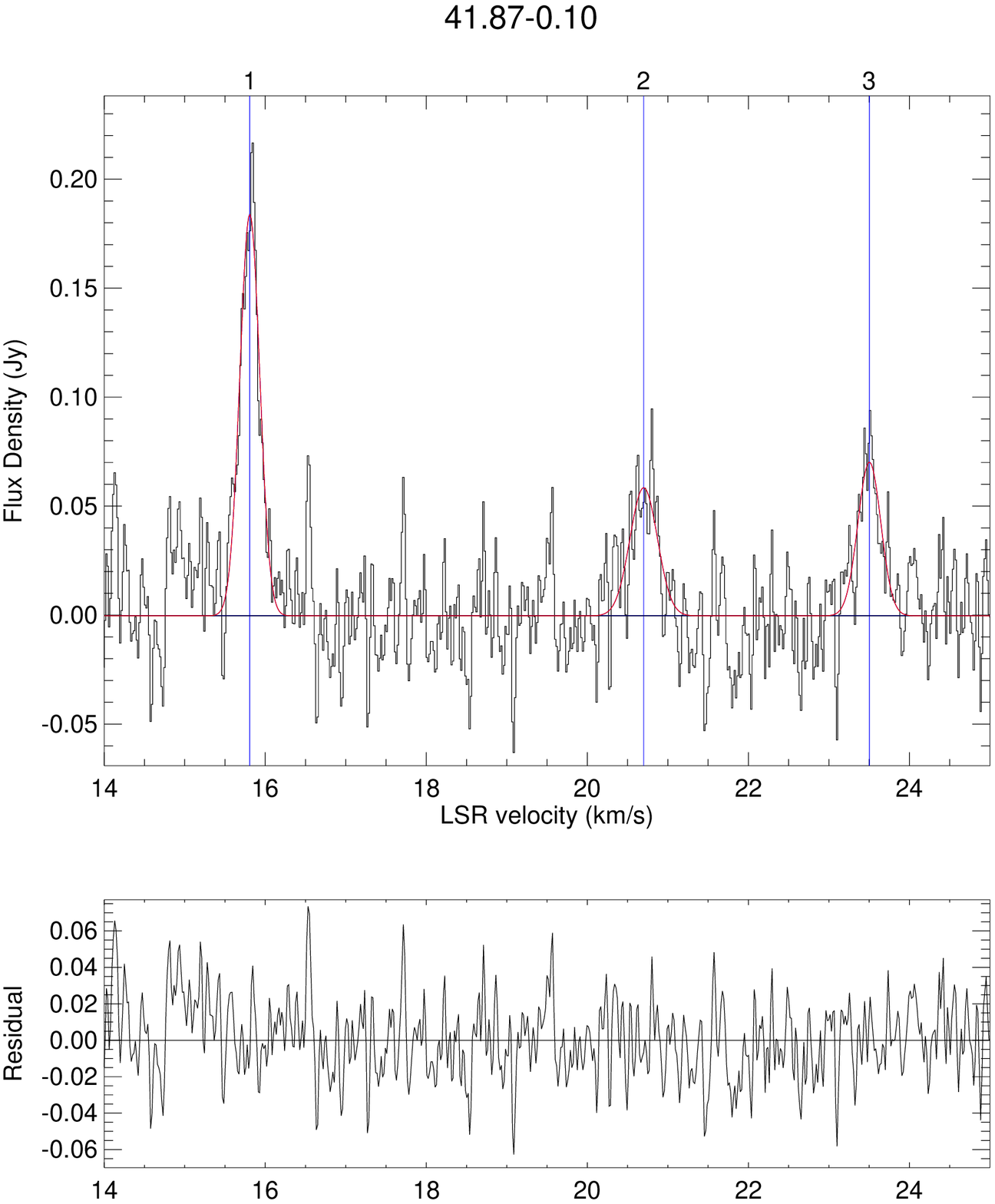}}\\
\centerline{Fig. 8. --- Continued.}
{\includegraphics[width=0.9\textwidth]{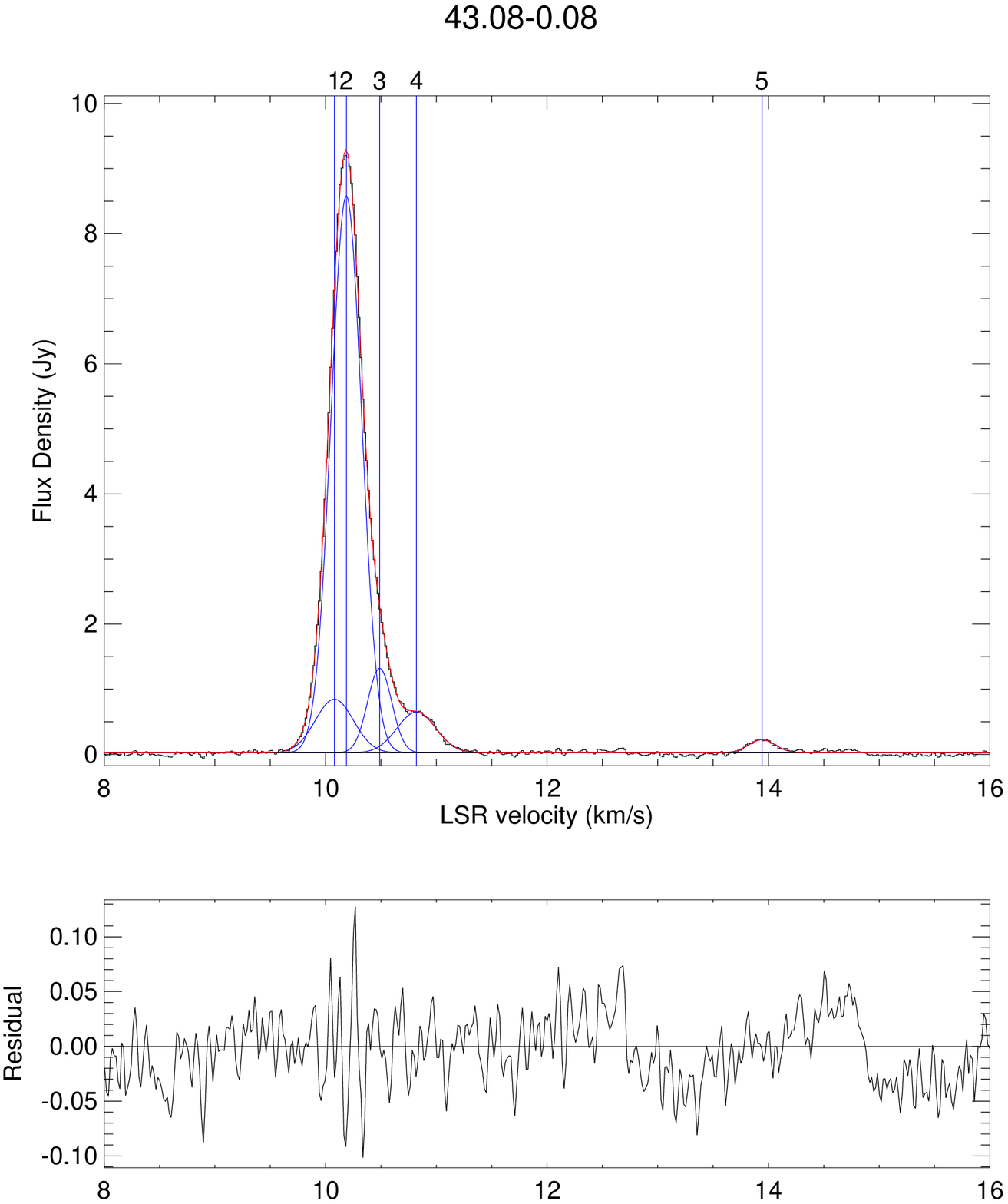}}\\
\centerline{Fig. 8. --- Continued.}
{\includegraphics[width=0.9\textwidth]{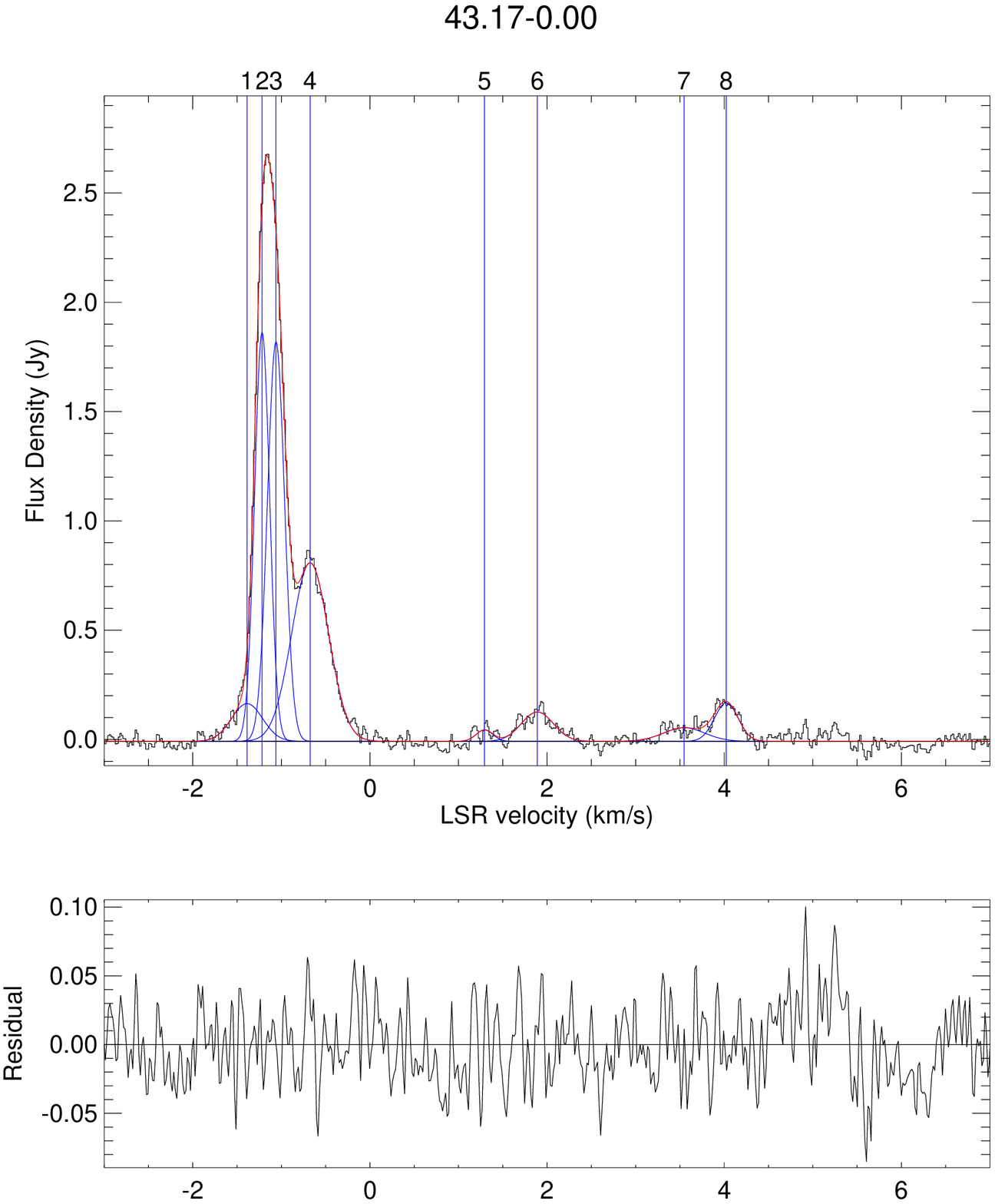}}\\
\centerline{Fig. 8. --- Continued.}
{\includegraphics[width=0.9\textwidth]{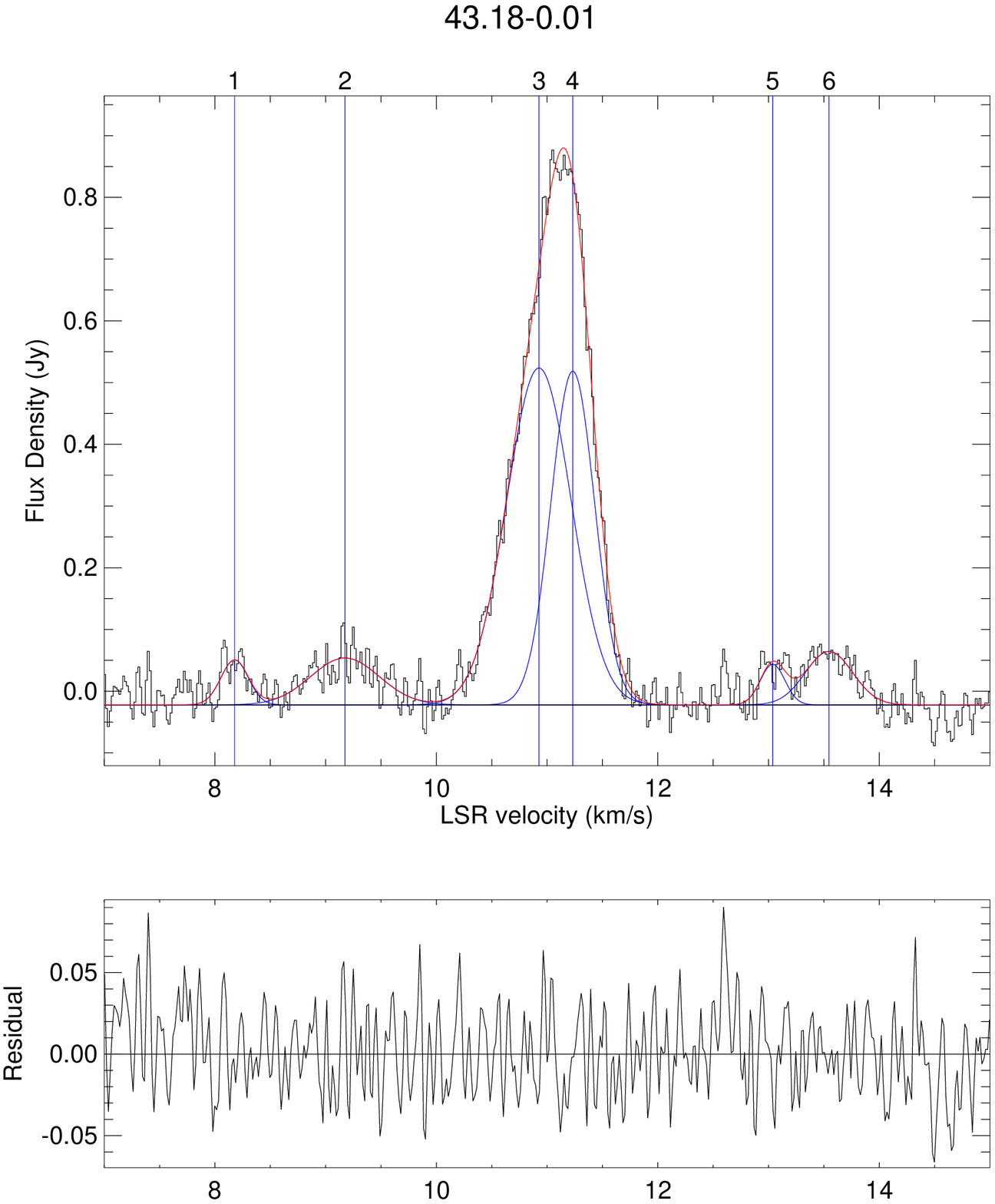}}\\
\centerline{Fig. 8. --- Continued.}
{\includegraphics[width=0.9\textwidth]{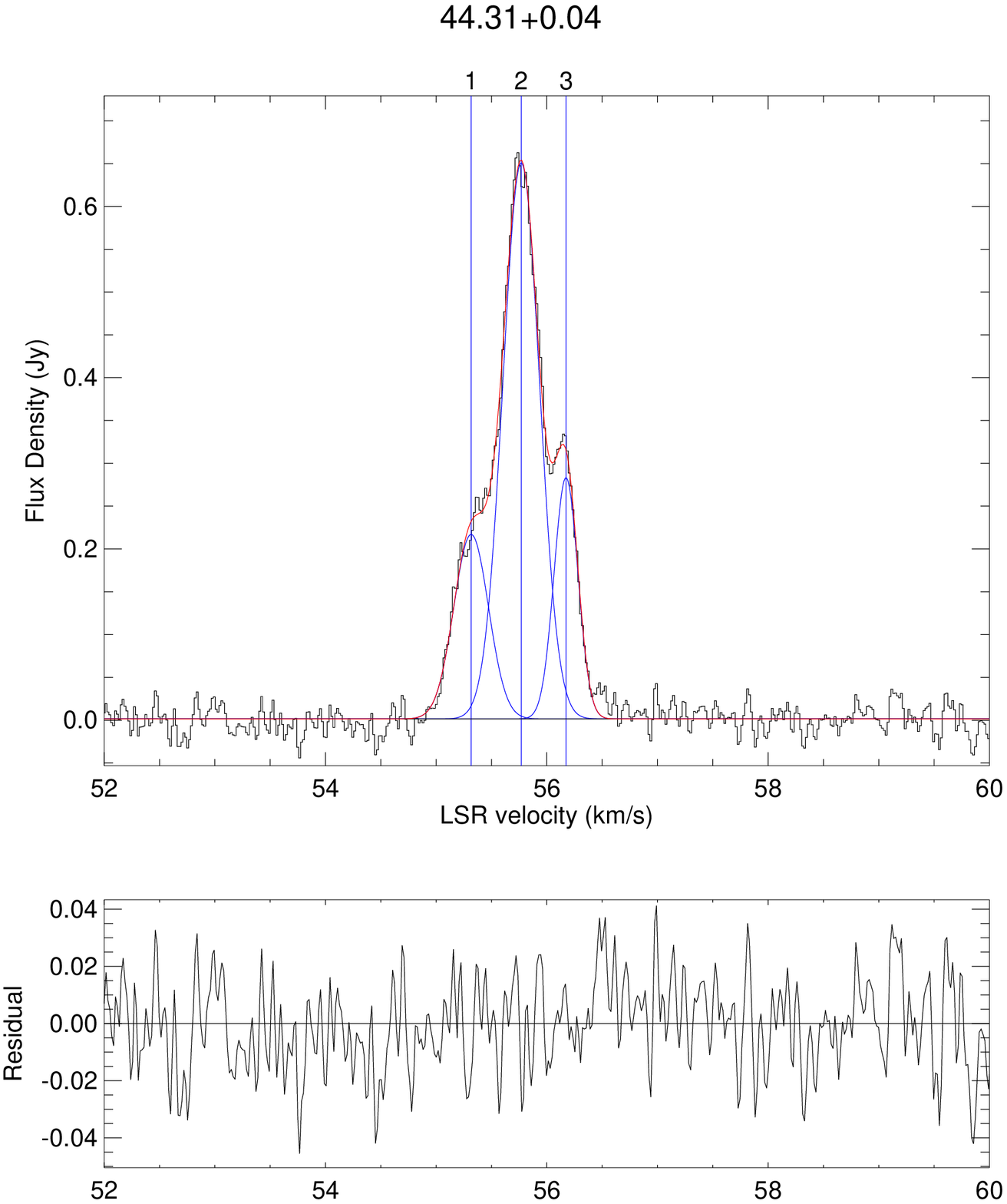}}\\
\centerline{Fig. 8. --- Continued.}
{\includegraphics[width=0.9\textwidth]{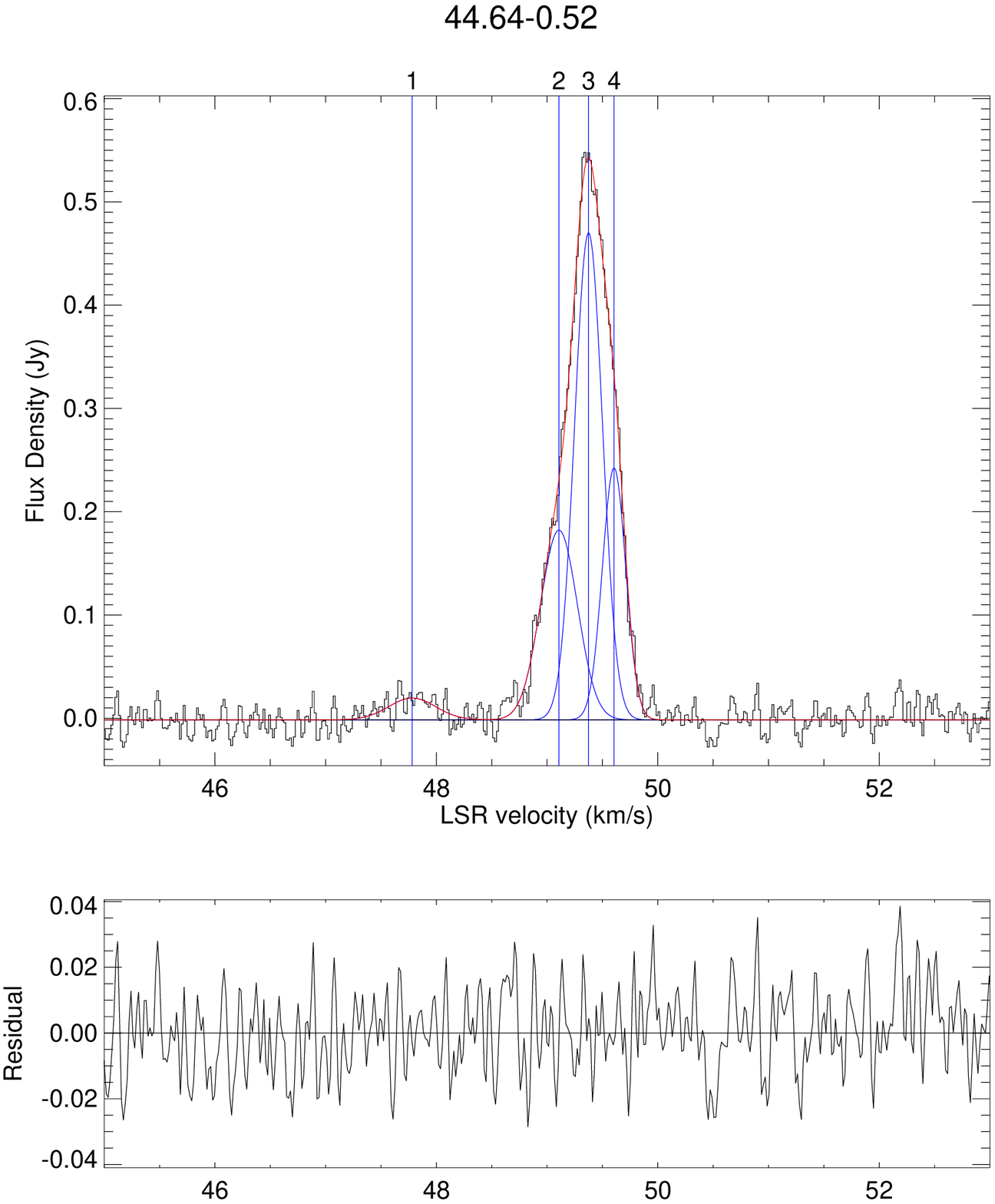}}\\
\centerline{Fig. 8. --- Continued.}
\clearpage
{\includegraphics[width=0.9\textwidth]{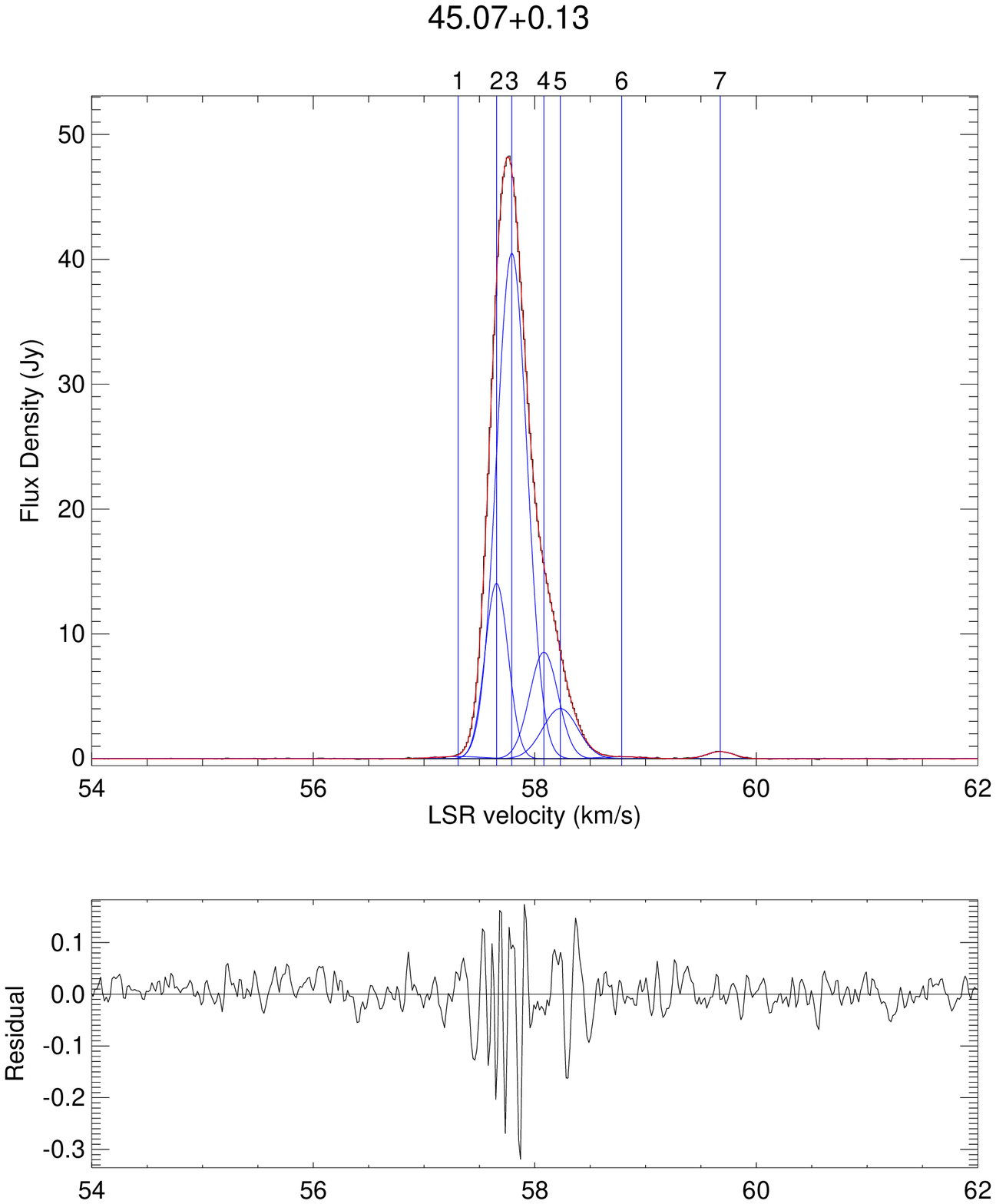}}\\
\centerline{Fig. 8. --- Continued.}
{\includegraphics[width=0.9\textwidth]{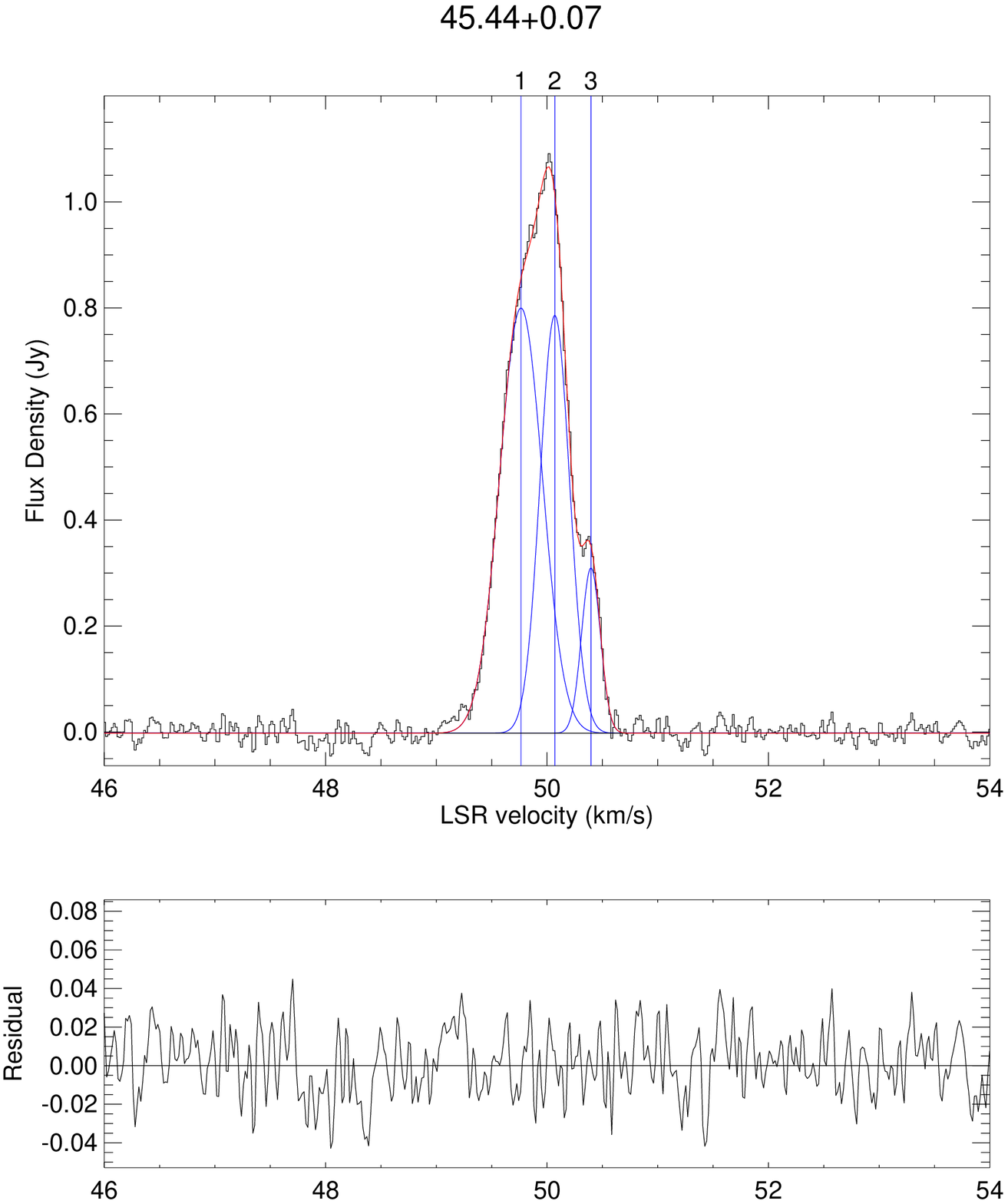}}\\
\centerline{Fig. 8. --- Continued.}
{\includegraphics[width=0.9\textwidth]{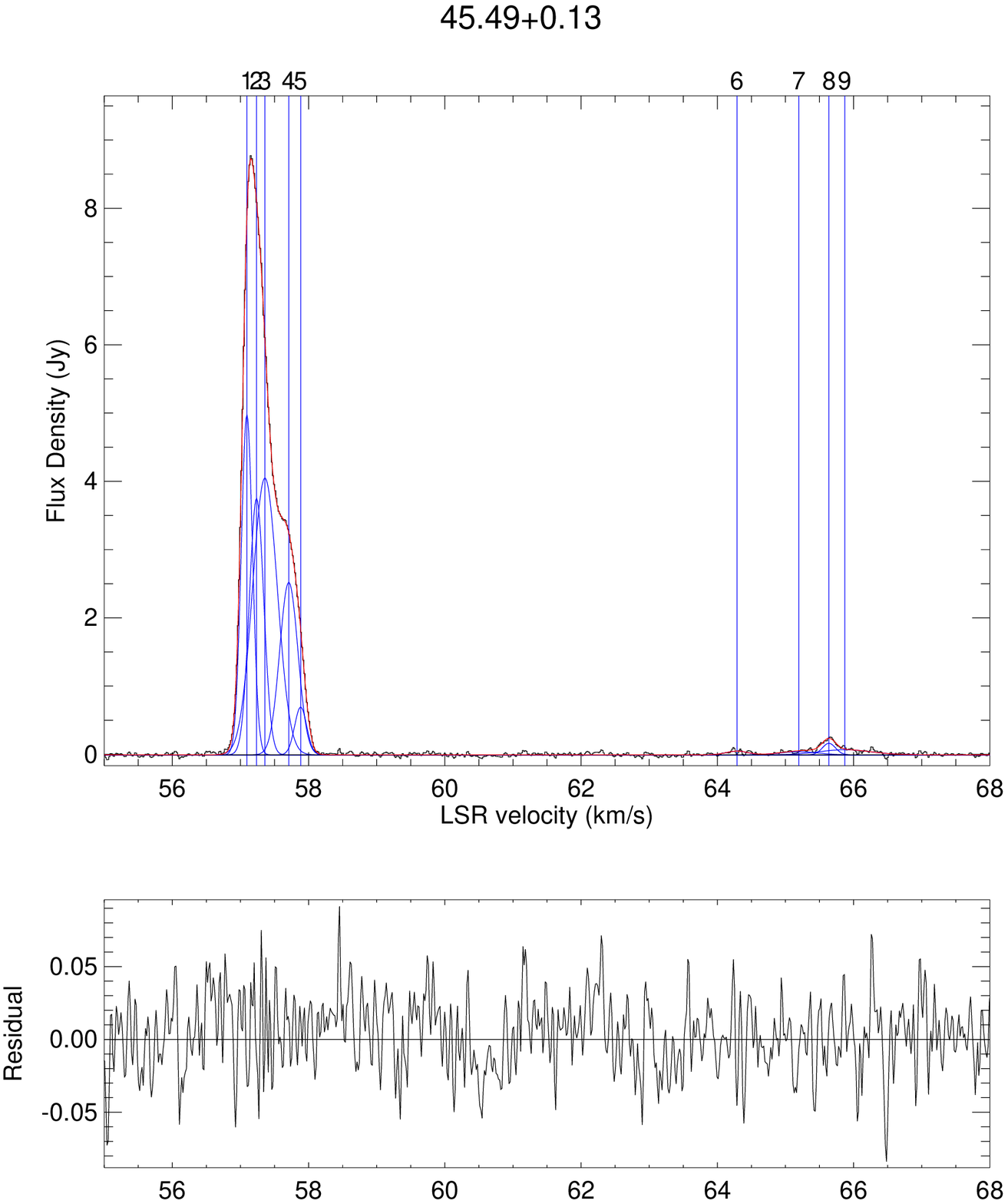}}\\
\centerline{Fig. 8. --- Continued.}
{\includegraphics[width=0.9\textwidth]{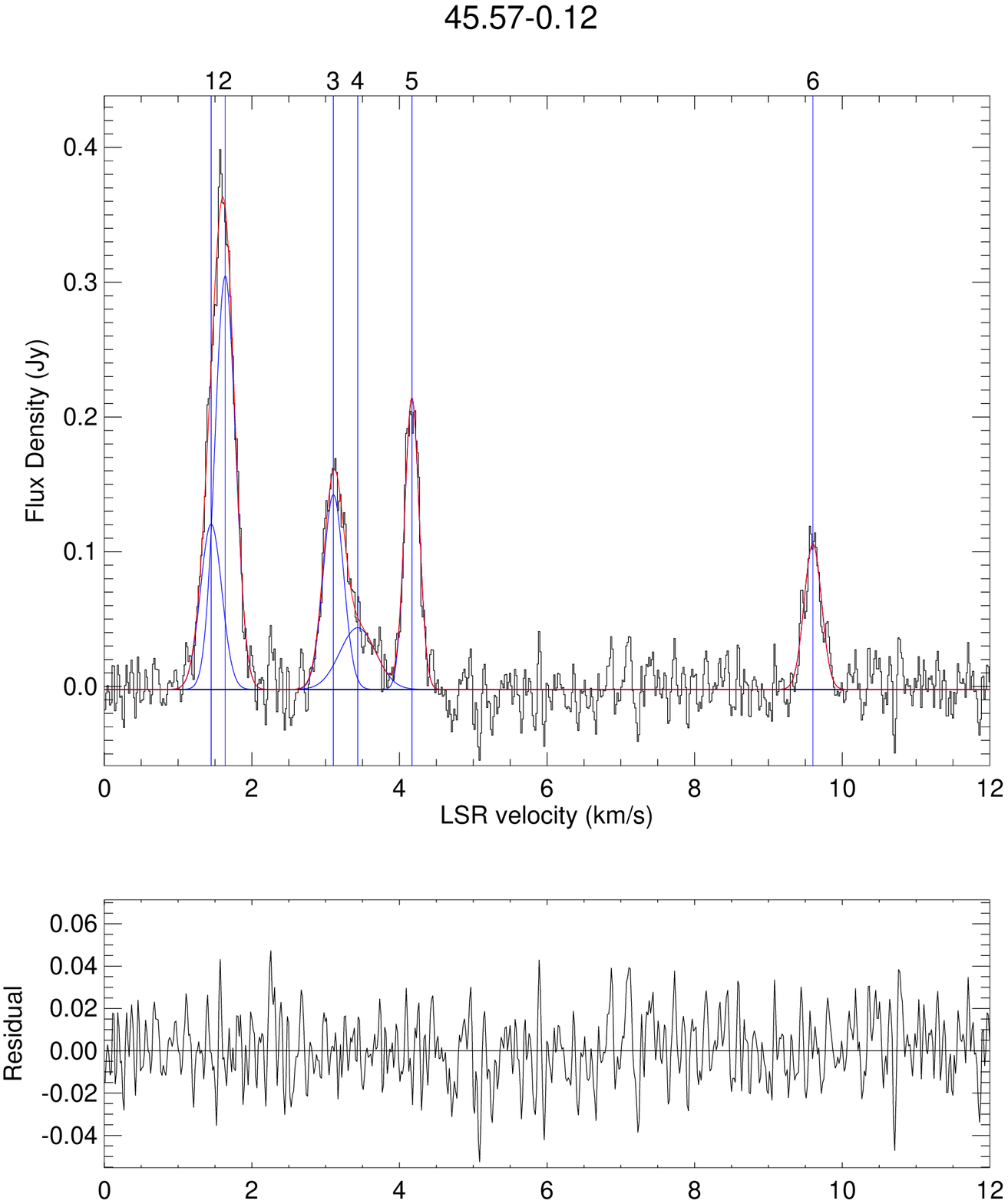}}\\
\centerline{Fig. 8. --- Continued.}
{\includegraphics[width=0.9\textwidth]{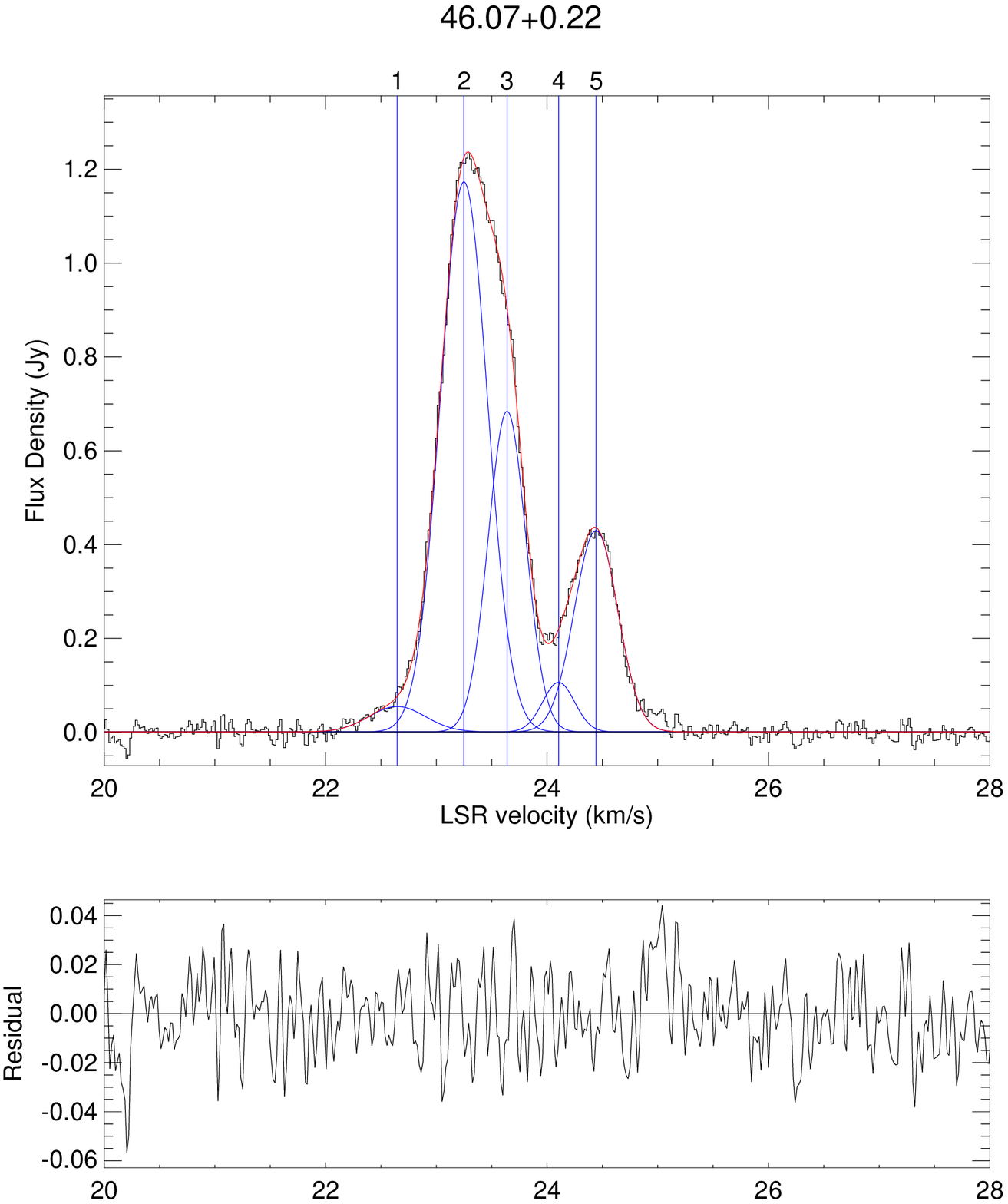}}\\
\centerline{Fig. 8. --- Continued.}
{\includegraphics[width=0.9\textwidth]{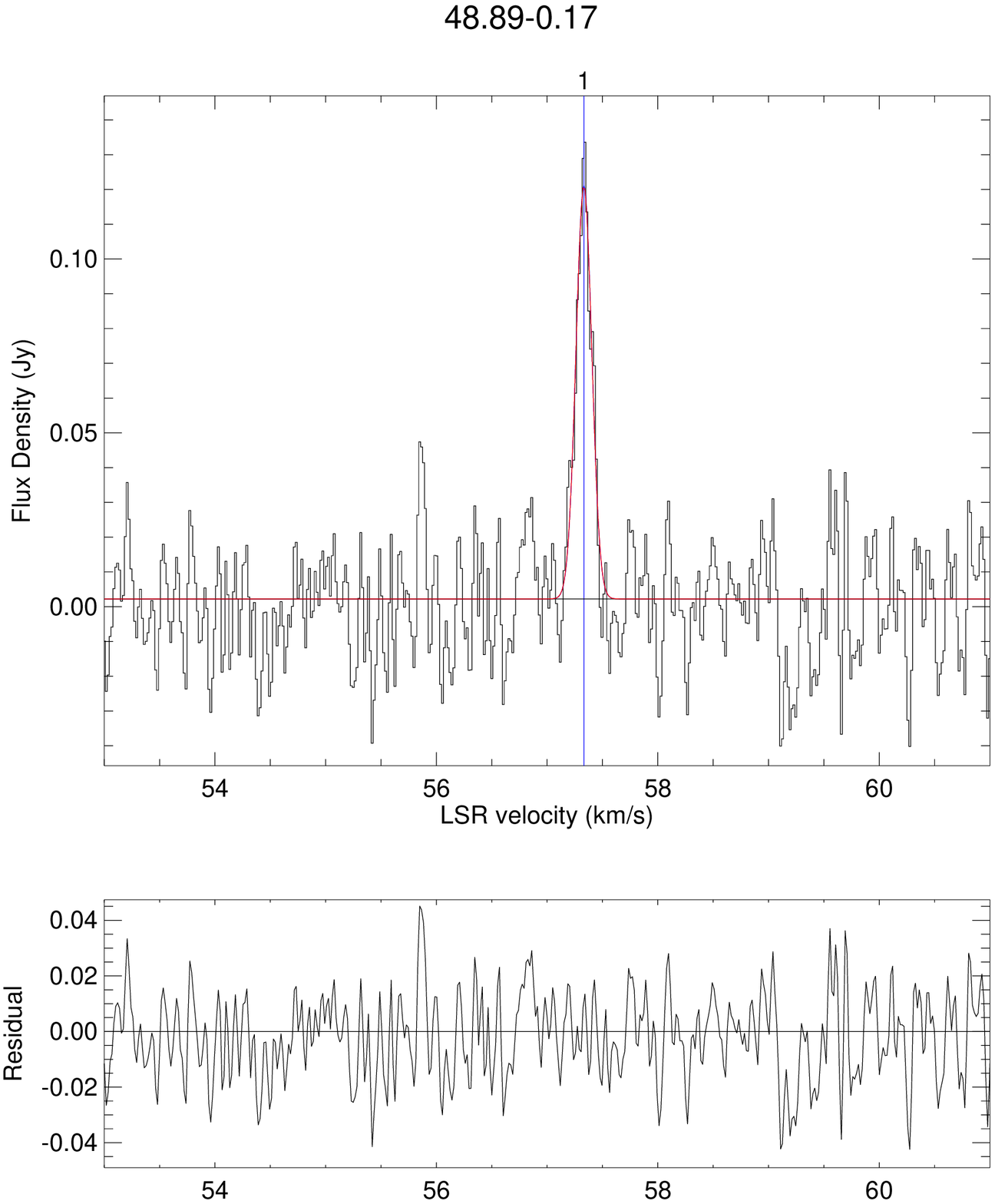}}\\
\centerline{Fig. 8. --- Continued.}
{\includegraphics[width=0.9\textwidth]{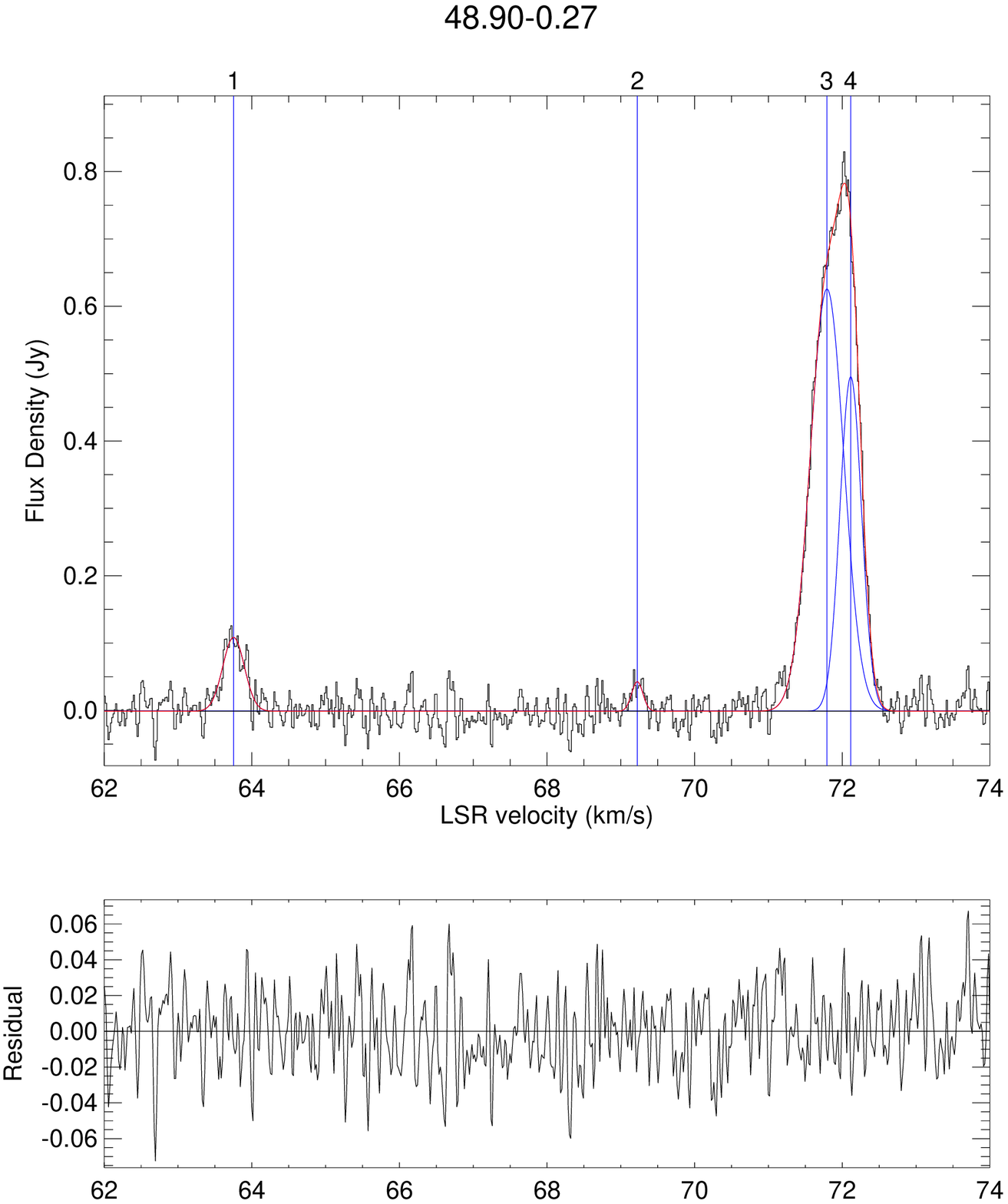}}\\
\centerline{Fig. 8. --- Continued.}
{\includegraphics[width=0.9\textwidth]{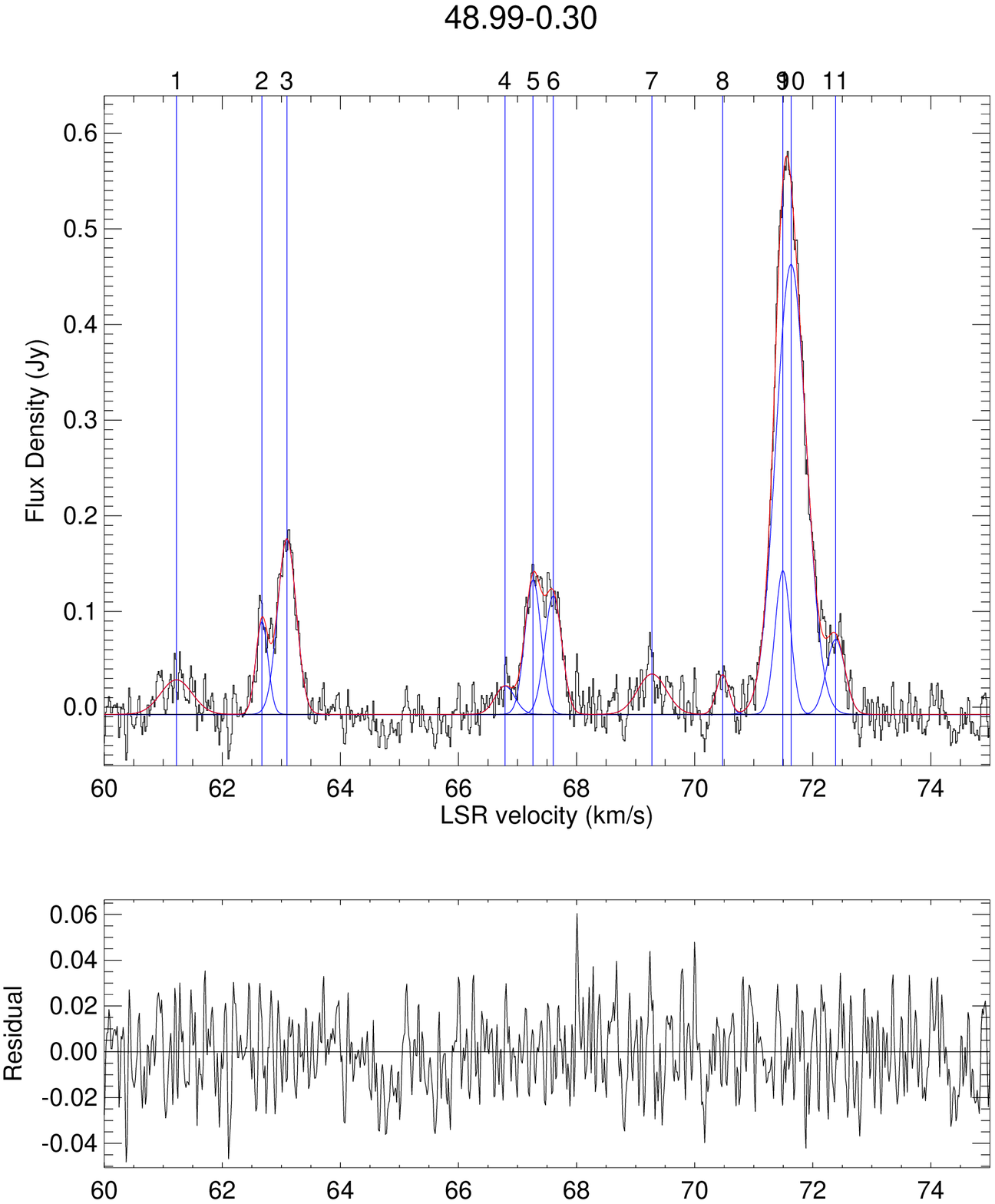}}\\
\centerline{Fig. 8. --- Continued.}
{\includegraphics[width=0.9\textwidth]{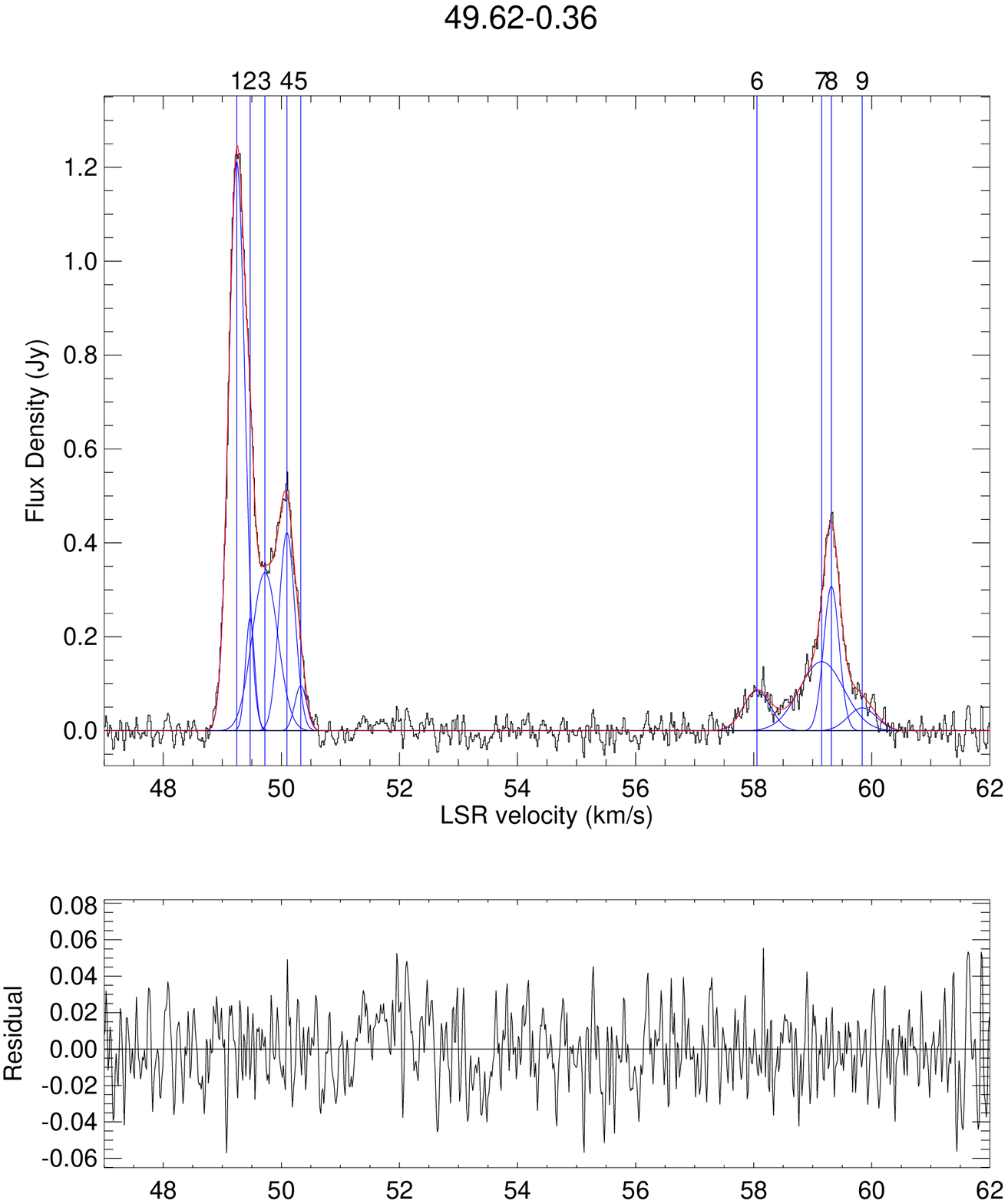}}\\
\centerline{Fig. 8. --- Continued.}
{\includegraphics[width=0.9\textwidth]{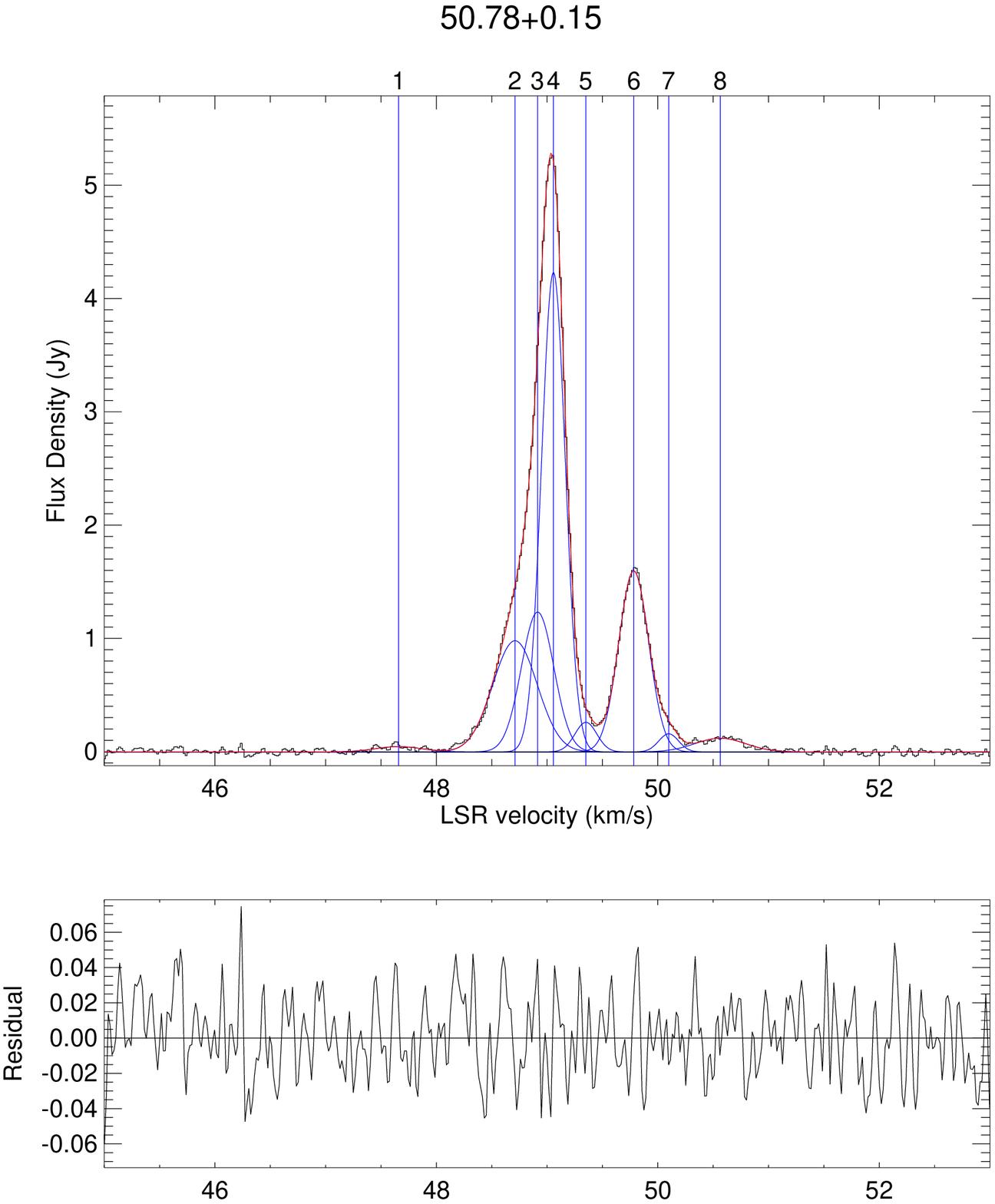}}\\
\centerline{Fig. 8. --- Continued.}
\clearpage
{\includegraphics[width=0.9\textwidth]{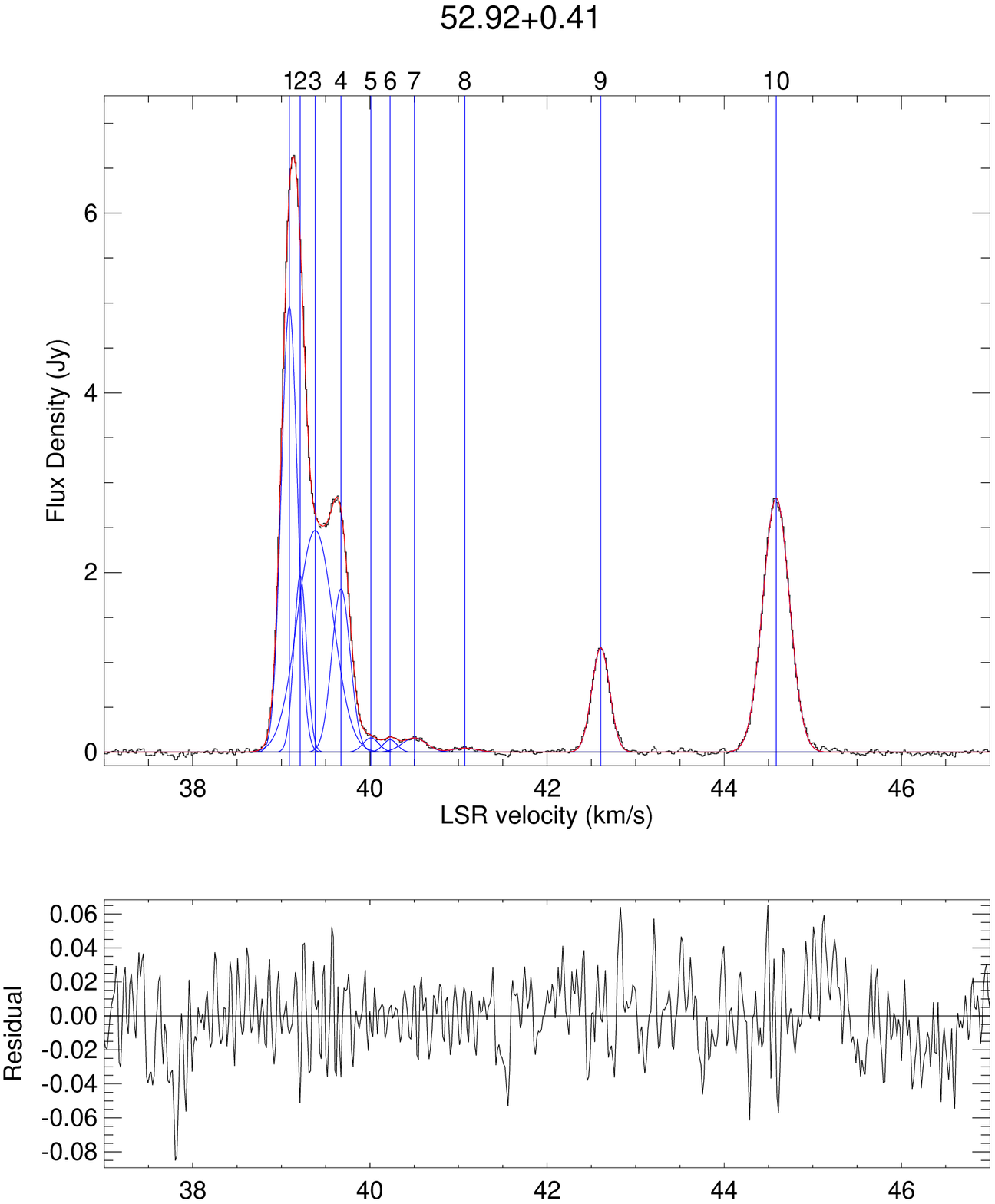}}\\
\centerline{Fig. 8. --- Continued.}
{\includegraphics[width=0.9\textwidth]{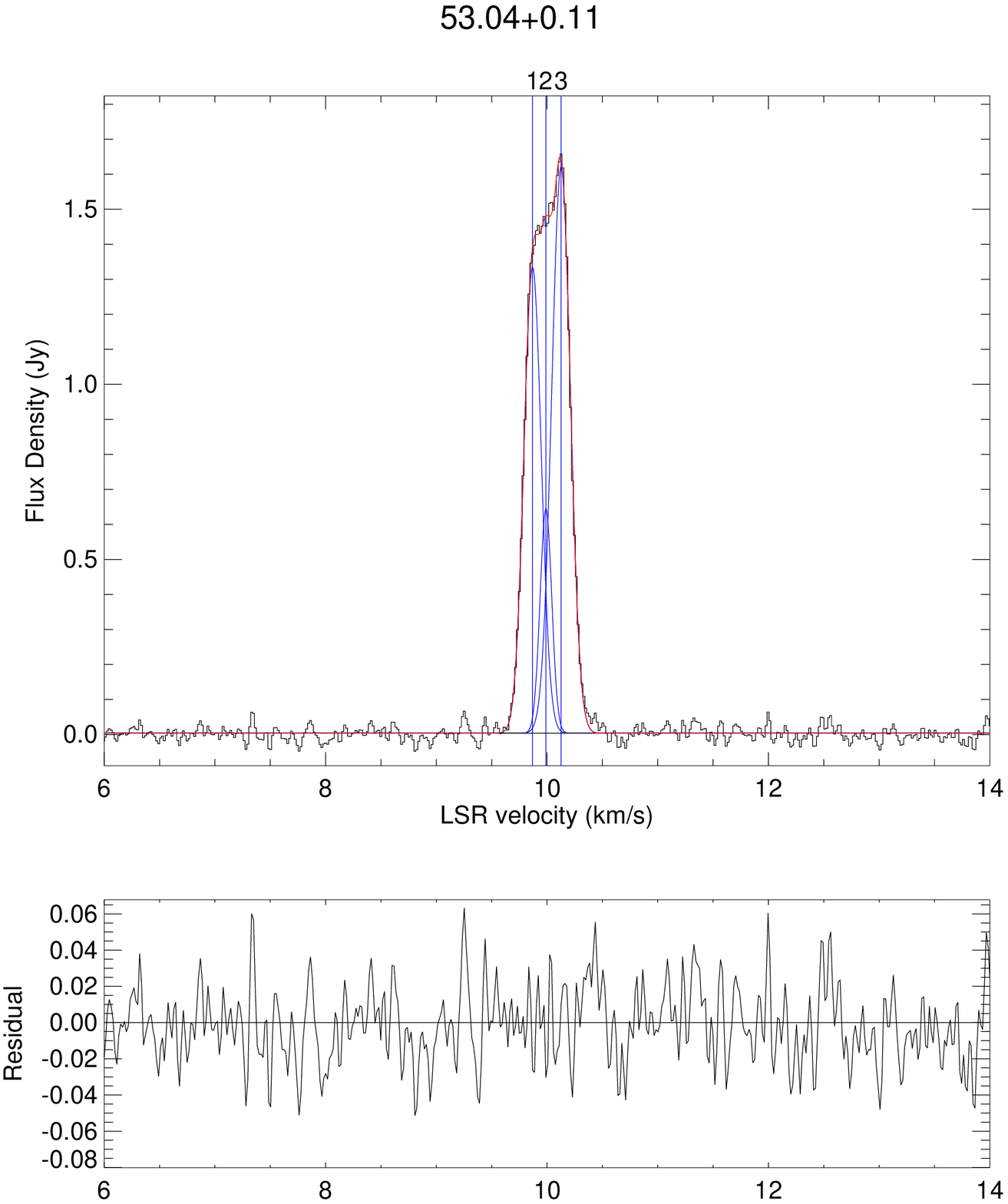}}\\
\centerline{Fig. 8. --- Continued.}
{\includegraphics[width=0.9\textwidth]{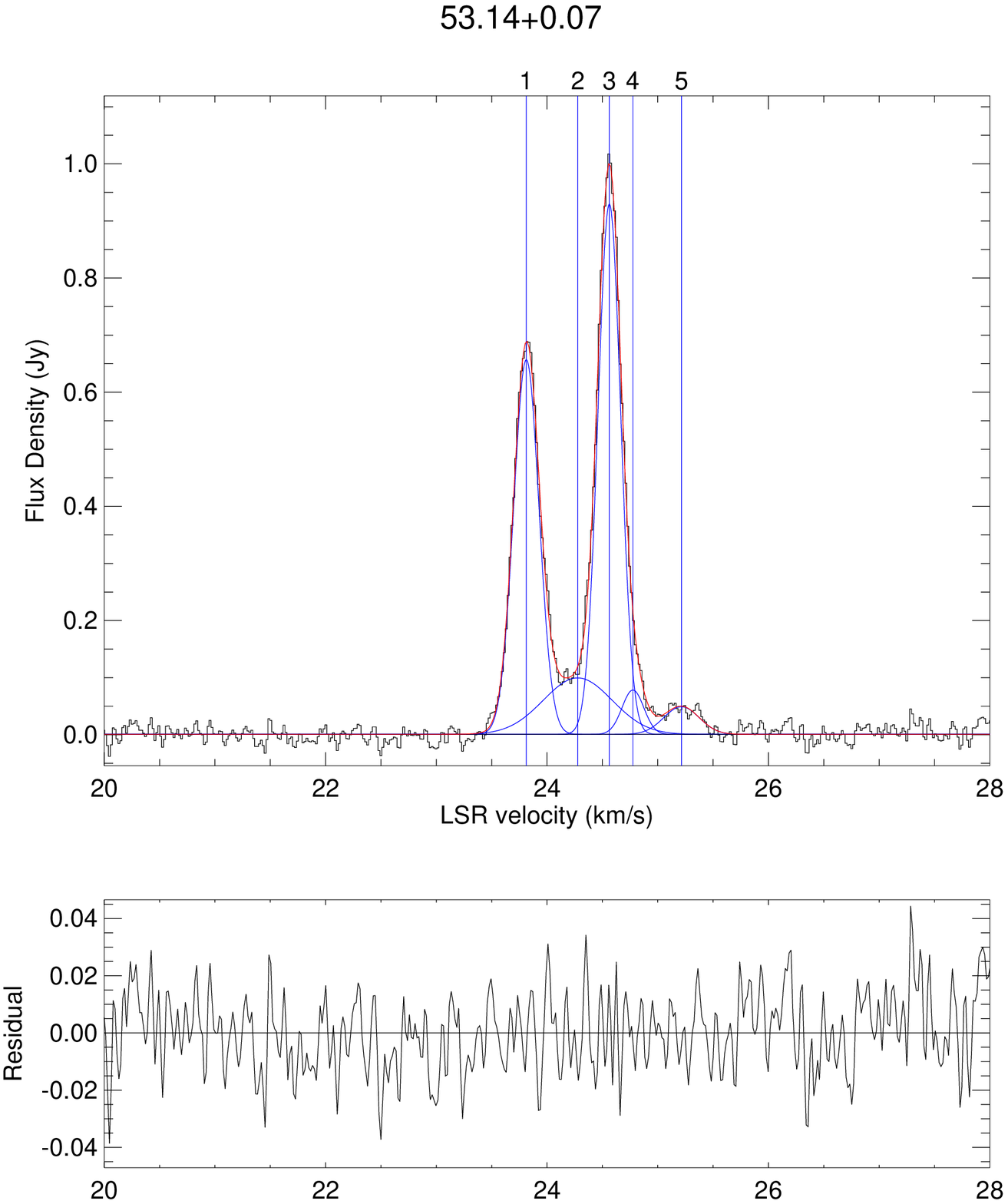}}\\
\centerline{Fig. 8. --- Continued.}
{\includegraphics[width=0.9\textwidth]{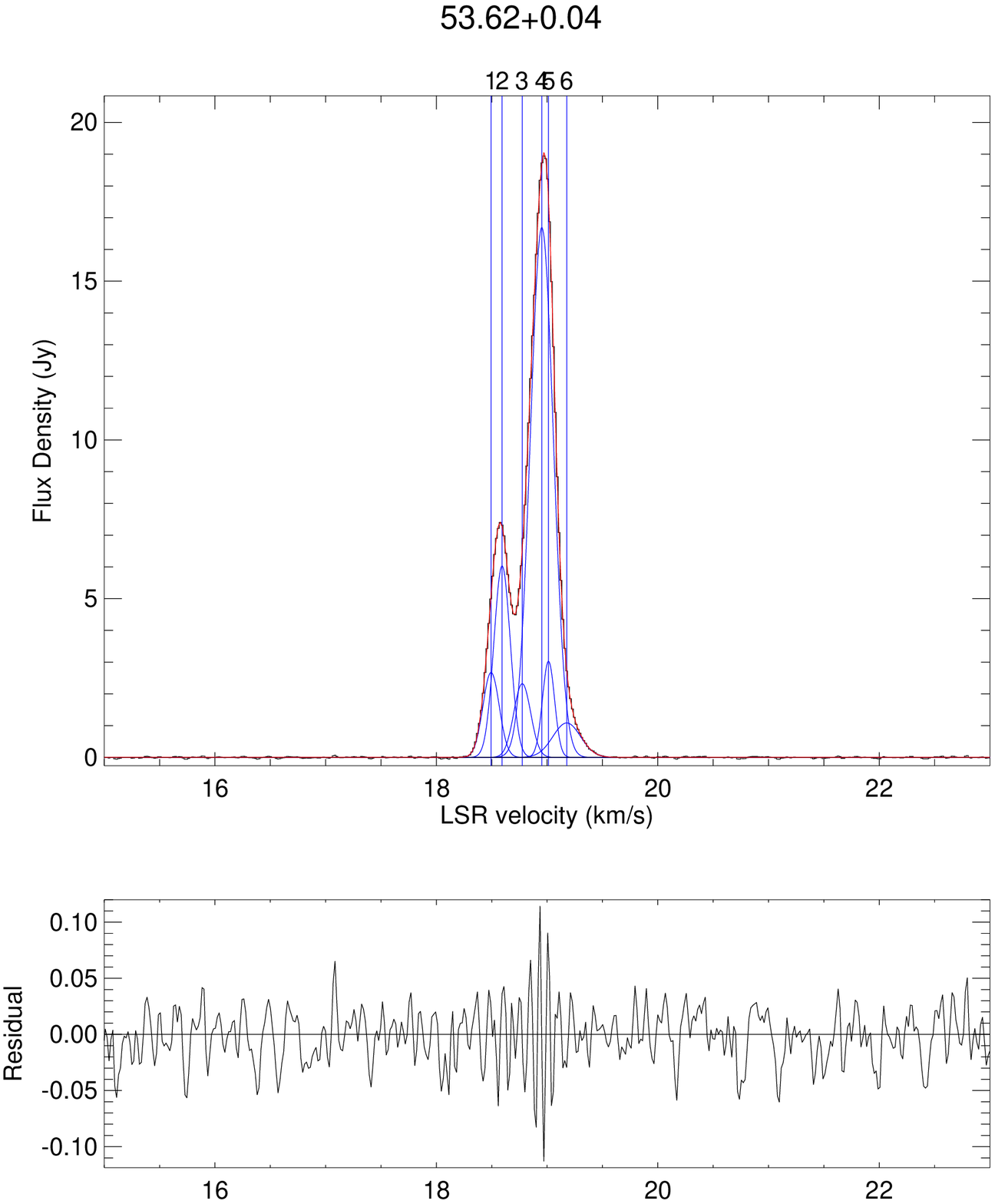}}\\
\centerline{Fig. 8. --- Continued.}
\clearpage

\end{document}